\documentclass[twocolumn,superscriptaddress,amsmath, amssymb, amsfonts,preprintnumbers,aps,prd,longbibliography,nofootinbib]{revtex4-1}

\renewcommand{\sec}[1]{\textit{#1. --- }}

\usepackage{graphicx}
\usepackage{dcolumn}
\usepackage{bm}
 \usepackage{amstext}
 \usepackage{amssymb}
 \usepackage{amsmath}
 \usepackage{graphicx}
 \usepackage{color}
 \usepackage{bbold}
 \usepackage{delimset} 
 \usepackage[caption=false,justification=justified]{subfig}
\usepackage[colorlinks=true]{hyperref}

\usepackage{tikz}
\usetikzlibrary{decorations.pathmorphing}
\usetikzlibrary{decorations.markings}
\usetikzlibrary{positioning, shapes, snakes, arrows}

\tikzset{
	graviton/.style={line width=.8pt, -latex,decorate, decoration={snake, segment length=4pt,amplitude=1.8pt, pre length=.1cm, post length=.25cm}},
	worldline/.style={gray, line width=1pt},
	worldlineBold/.style={black, line width=.6pt},
	zUndirected/.style={line width=1pt},
	zParticle/.style={line width=1pt,postaction={decorate},decoration={markings,mark=at position .6 with {\arrow[#1]{latex}}}},
	zParticleF/.style={line width=1pt,postaction={decorate}},
	cscalar/.style={line width=1pt,postaction={decorate},decoration={markings,mark=at position .6 with {\arrow[#1]{latex}}}},
	cscalar2/.style={line width=1pt,postaction={decorate},decoration={markings,mark=at position .8 with {\arrow[#1]{latex}}}},
	photon/.style={line width =.8pt, decorate, decoration={snake, segment length=4pt, amplitude=1.8pt,  pre length=.1cm, post length=.1cm}}
}

\makeatletter
\newlength{\apb@width}
\newcommand{\autoparbox}[2][c]{\settowidth{\apb@width}{#2}\parbox[#1]{\apb@width}{#2}}

\makeatother

\makeatletter
\def\mr@ignsp#1 {\ifx\:#1\@empty\else #1\expandafter\mr@ignsp\fi}%
\newcommand{\multiref}[1]{\begingroup
\xdef\mr@no@sparg{\expandafter\mr@ignsp#1 \: }%
\def\mr@comma{}%
\@for\mr@refs:=\mr@no@sparg\do{\mr@comma\def\mr@comma{,}\ref{\mr@refs}}%
\endgroup}
\makeatother

\renewcommand{\eqref}[1]{(\multiref{#1})}

\newcommand{\sfrac}[2]{{\textstyle\frac{#1}{#2}}}

\newcommand{\vev}[1]{\langle #1\rangle}

\newcommand{\be}{\begin{equation}}
\newcommand{\ee}{\end{equation}}
\newcommand{\ba}{\begin{align}}
\newcommand{\ea}{\end{align}}
\newcommand{\eqn}[1]{(\ref{#1})}

\def\eqn#1{eq.~\eqref{#1}}

\def\Rcite#1{Ref.~\cite{#1}}
\def\Rcites#1{Refs.~\cite{#1}}

\newcommand{\nn}{\nonumber}
\def\dd{\delta\!\!\!{}^-\!}
\newcommand{\eps}{\varepsilon}

\usepackage{feynmp-auto}
\usepackage{feynmp} 
\newcommand{\hut}{}

\def\tbmo{|\tilde{\mathbf{b}}|_1}
\def\d{\mathrm{d}}
\def\eps{\epsilon}

\def\cN{\mathcal{N}}
\def\cS{\mathcal{S}}
\def\cO{\mathcal{O}}

\def\cM{\mathcal{M}}

\def\eps{\epsilon}

\def\ltb{{\tilde b}}

\def\bv{\hat{\mathbf e}}

\def\jty{\mathcal{J}}
\def\ity{\mathcal{I}}

\begin{document}

\preprint{
HU-EP-21/15-RTG
}

\title{Gravitational Bremsstrahlung and Hidden Supersymmetry of Spinning Bodies}
\author{Gustav Uhre Jakobsen} 
\email{gustav.uhre.jakobsen@physik.hu-berlin.de}
\affiliation{%
Institut f\"ur Physik und IRIS Adlershof, Humboldt-Universit\"at zu Berlin,
Zum Gro{\ss}en Windkanal 2, 12489 Berlin, Germany
}
 \affiliation{Max Planck Institute for Gravitational Physics (Albert Einstein Institute), Am M\"uhlenberg 1, 14476 Potsdam, Germany}

\author{Gustav Mogull}
\email{gustav.mogull@aei.mpg.de} 
\affiliation{%
Institut f\"ur Physik und IRIS Adlershof, Humboldt-Universit\"at zu Berlin,
Zum Gro{\ss}en Windkanal 2, 12489 Berlin, Germany
}
 \affiliation{Max Planck Institute for Gravitational Physics (Albert Einstein Institute), Am M\"uhlenberg 1, 14476 Potsdam, Germany}

\author{Jan Plefka} 
\email{jan.plefka@hu-berlin.de}
\affiliation{%
Institut f\"ur Physik und IRIS Adlershof, Humboldt-Universit\"at zu Berlin,
Zum Gro{\ss}en Windkanal 2, 12489 Berlin, Germany
}

\author{Jan Steinhoff} 
\email{jan.steinhoff@aei.mpg.de}
 \affiliation{Max Planck Institute for Gravitational Physics (Albert Einstein Institute), Am M\"uhlenberg 1, 14476 Potsdam, Germany}


\begin{abstract}
The recently established formalism of a worldline quantum field theory,
which describes the classical scattering of massive bodies
(black holes, neutron stars or stars) in Einstein gravity,
is generalized up to quadratic order in spin,
revealing an alternative $\cN=2$ supersymmetric description
of the symmetries inherent in spinning bodies.
The far-field time-domain waveform of the gravitational waves produced in such a
spinning encounter is computed at leading order in the post-Minkowskian
(weak field, but generic velocity) expansion, and exhibits this supersymmetry.
From the waveform we extract the
leading-order total radiated angular momentum in a generic reference frame,
and the total radiated energy in the center-of-mass frame
to leading order in a low-velocity approximation.
\end{abstract}

\maketitle

The rise of gravitational wave (GW) astronomy~\cite{Abbott:2016blz,*LIGOScientific:2018mvr,*Abbott:2020niy} offers new paths to explore our universe, including black hole (BH) population and formation studies~\cite{Abbott:2020gyp}, tests of gravity in the strong-field regime~\cite{Abbott:2020jks}, measurements of the Hubble constant~\cite{Abbott:2019yzh},
and investigations of strongly interacting matter inside neutron stars~\cite{Abbott:2018exr}.
This form of astronomy relies heavily on Bayesian methods to infer
probability distributions for theoretical GW predictions (templates),
depending on a source's parameters, to match the measured strain on detectors.
With the network of GW observatories steadily increasing in sensitivity~\cite{TheLIGOScientific:2014jea,*TheVirgo:2014hva,*Aso:2013eba}, theoretical GW predictions need to keep pace with the accuracy requirements placed on templates~\cite{Purrer:2019jcp}.
For the inspiral and merger phases of a binary
an important strategy is to synergistically combine approximate and numerical relativity predictions~\cite{Buonanno:1998gg,*Ajith:2007qp},
each applicable only to a corner of the parameter space~\cite{vandeMeent:2020xgc}.

In this Letter we calculate gravitational waveforms ---
the primary observables of GW detectors ---
produced in the parameter-space region of highly eccentric
(scattering) \emph{spinning} BHs and neutron stars (NSs),
to leading order in the weak-field, or post-Minkowskian (PM), approximation.
Following the above strategy, this is a valuable input for future eccentric waveform models.
Indeed, the extension of contemporary quasi-circular (non-eccentric) waveform models for spinning binaries to eccentric orbits (including scattering) is under active investigation~\cite{Ramos-Buades:2019uvh,*Chiaramello:2020ehz,*Nagar:2021gss,*Liu:2021pkr,*Khalil:2021txt,*Hinderer:2017jcs,*Islam:2021mha}.
This is motivated, for instance, by the potential insight gained on the formation channels or astrophysical environments of binary BHs (BBHs) through measurements of eccentricity~\cite{Samsing:2017xmd,*Rodriguez:2017pec,*Gondan:2020svr} and spins~\cite{LIGOScientific:2018jsj}, or the search for scattering BHs~\cite{Kocsis:2006hq,*Mukherjee:2020hnm,*Zevin:2018kzq,*Gamba:2021gap} in our universe.

Accurate predictions for GWs from BBHs should crucially also account
for the BHs' spins~\cite{Zackay:2019tzo,*Huang:2020ysn},
and this is an important aspect of the present work.
The gravitational waveforms presented here are valid up to quadratic order
in angular momenta (spins) of the compact stars;
that is, we extend Crowley, Kovacs and Thorne's seminal non-spinning result
\cite{1975ApJ...200..245T,*1977ApJ...215..624C,*Kovacs:1977uw,*Kovacs:1978eu}.
We also improve on our earlier reproduction of the non-spinning result~\cite{Jakobsen:2021smu} by presenting results in a compact Lorentz-covariant form,
using an improved integration strategy.

To obtain these results we generalize the recently introduced
worldline quantum field theory (WQFT) formalism
\cite{Mogull:2020sak,Jakobsen:2021smu}
to spinning particles on the worldline.
This is achieved by including anticommuting worldline fields
carrying the spin degrees of freedom,
building upon Refs.~\cite{Howe:1988ft,Gibbons:1993ap,Bastianelli:2005vk,*Bastianelli:2005uy}.
Our formalism manifests an $\mathcal{N}=2$ extended worldline supersymmetry (SUSY)
which holds up to the desired quadratic order in spin.
The SUSY implies conservation of the covariant spin-supplementary condition (SSC),
and thus represents an alternative formulation of the symmetries inherent to spinning bodies.
It also operates on the spinning waveform.

The spinning WQFT innovates over previous approaches to classical spin based on corotating-frame variables~\cite{Porto:2005ac,Levi:2015msa} in the effective field theory (EFT) of compact objects~\cite{Goldberger:2004jt,*Goldberger:2006bd,*Goldberger:2009qd,Porto:2016pyg,*Levi:2018nxp} ---
see Refs.~\cite{Goldberger:2017ogt,*Goldberger:2016iau,*Shen:2018ebu} for the construction of PM integrands and Refs.~\cite{Liu:2021zxr,Kalin:2020mvi} for worldline and spin deflections
(in agreement with scattering amplitude results~\cite{Bern:2020buy,Kosmopoulos:2021zoq}).
The worldline EFT was applied to radiation also in the weak-field and slow-motion,
i.e.~post-Newtonian (PN),
approximation~\cite{Porto:2010zg,*Porto:2012as,*Maia:2017gxn,*Maia:2017yok,*Cho:2021mqw}--- see Refs.~\cite{Mishra:2016whh,*Buonanno:2012rv} for more traditional methods.
Other approaches to PM spin effects can be found in Refs.~\cite{Vines:2017hyw,*Bini:2017xzy,*Bini:2018ywr,*Guevara:2017csg,*Vines:2018gqi,*Guevara:2018wpp,*Chung:2018kqs,*Guevara:2019fsj,*Chung:2019duq,*Damgaard:2019lfh,*Aoude:2020onz,*Guevara:2020xjx}. 

\sec{Spinning Worldline Quantum Field Theory} 
It has been known since the 1980s  \cite{Howe:1988ft} that the relativistic wave equation
for a massless or massive spin-$\cN/2$ field in flat spacetime
(generalizing the Klein-Gordon, Dirac and Maxwell or Proca equations)
may be obtained by quantization of an extended supersymmetric particle model
where one augments the bosonic trajectory $x^{\mu}(\tau)$ by $\cN$ anticommuting,
real worldline fields.
Generalizing this to a curved background spacetime
comes with consistency problems beyond $\cN=2$.
Yet the situation for spins up to one is well understood
\cite{Bastianelli:2005vk,Bastianelli:2005uy},
and sufficient for our purposes of describing 
two-body scattering up to quadratic order in spin.

We therefore augment the worldline trajectories $x_i^{\mu}(\tau_{i})$ ($i=1,2$)
of our two massive bodies
by anticommuting \emph{complex} Grassmann fields $\psi_i^{a}(\tau_{i})$.
These are vectors in the flat tangent Minkowski spacetime connected to
the curved spacetime via the vierbein $e^{a}_{\mu}(x)$.
The worldline action in the massive case for each body
takes the form (suppressing the $i$ subscripts)
\cite{Bastianelli:2005uy,Jakobsen:2021zvh}
\begin{align}\label{N=2act}
	S= -m\int\!\d\tau \Bigl [&\sfrac{1}{2}g_{\mu\nu}\dot x^{\mu}\dot x^{\nu}
	\!+\! i\bar\psi_a\sfrac{D\psi^a}{D\tau}\!+\!\sfrac{1}{2}
	R_{abcd}\bar\psi^{a}\psi^{b}\bar\psi^{c}\psi^{d}\Bigr ]\,,
\end{align}
where $g_{\mu\nu}=e^{a}_{\mu}e^{b}_{\nu}\eta_{ab}$ is the metric in mostly minus signature,
$\sfrac{D\psi^a}{D\tau}=\dot\psi^{a}+\dot x^{\mu}{{\omega_\mu}^a}_b\psi^{b}$
includes the spin connection $\omega_{\mu ab}$
and the Riemann tensor is
$R_{\mu\nu ab}=e^c_\mu e^d_\nu R_{abcd}=
2(\partial_{[\mu}\omega_{\nu]ab}+\omega_{[\mu\,a}{}^{c}\omega_{\nu]cb})$.
This theory enjoys a global $\cN=2$ SUSY:
it is invariant under
\be
\label{N=2SUSY}
\delta x^{\mu} = i\bar\epsilon \psi^{\mu} + i\epsilon \bar\psi^{\mu} \, ,
\quad
\delta \psi^{a}= -\epsilon e^{a}_{\mu}{\dot x}^{\mu} -\delta x^{\mu}\, \omega_{\mu}{}^{a}{}_{b}\psi^{b}\, ,
\ee
with constant SUSY parameters $\epsilon$ and $\bar \epsilon = \epsilon^{\dagger}$.

The connection to a traditional description of spinning bodies in general relativity,
using the spin field $S^{\mu\nu}$ and the Lorentz body-fixed frame
$\Lambda^{A}_{\mu}$~\cite{Vines:2016unv,Porto:2008jj,Porto:2005ac,Levi:2015msa,Porto:2016pyg,*Levi:2018nxp},
comes about upon identifying the spin field $S^{\mu\nu}(\tau)$ with the Grassmann bilinear:
\begin{equation}
	S^{\mu\nu}=-2i e^{\mu}_{a}e^{\nu}_{b}\,\bar{\psi}^{[a}\psi^{b]}\,.
\end{equation}
One can easily show that $S^{ab}$ obeys the Lorentz algebra under Poisson brackets $\{\psi^a,\bar\psi^b\}_{\text{P.B.}}=-i\eta^{ab}$. 
In fact, the spin-supplementary condition (SSC) and preservation of spin length
may be related to $\cN=2$ SUSY-related constraints~\cite{Jakobsen:2021zvh}.
Finally, by deriving the classical equations of motion from the action
these can be shown to match the Mathisson-Papapetrou equations~\cite{Mathisson:1937zz,*Papapetrou:1951pa,*Dixon:1970zza} at quadratic spin order.
This indicates a hidden $\cN=2$ SUSY in the
actions of Refs.~\cite{Porto:2008jj,Vines:2016unv,Levi:2015msa}.

The actions of Refs.~\cite{Porto:2008jj,Vines:2016unv,Levi:2015msa}
also carry a first spin-induced \emph{multipole moment term}
at quadratic order in spins with an undertermined Wilson
coefficient $C_{E}$,
where here $C_E=0$ for a Kerr BH.
Translating it to our formalism this term reads
\be
S_{ES^{2}}:= -m\int\!\d\tau\, C_E E_{ab}  \bar\psi^a \psi^b\, \bar\psi\cdot\psi\,,
\ee
where $E_{ab}:= R_{a \mu b\nu} \dot x^{\mu} \dot x^{\nu}$ is the ``electric'' part of the Riemann tensor.
The $\cN=2$ SUSY is now maintained only in an approximate sense~\cite{Jakobsen:2021zvh}:
it survives in the action for terms up to $\cO(\psi^5)$,
i.e.~quadratic order in spin.

In order to describe a scattering scenario we expand the worldline fields
about solutions of the equations of motion along straight-line trajectories:
\begin{align}\label{bgexp}
\begin{aligned}
x^\mu_{i}(\tau_{i})&=b^\mu_{i}+v^\mu_{i}\tau_i +z^\mu_{i}(\tau_{i}) \, ,\\
\psi^{a}_{i}(\tau_{i}) & = \Psi_{i}^{a} +\psi_{i}^{\prime a}(\tau_{i})\,,
\end{aligned}
\end{align}
where $\cS_{i}^{\mu\nu}:=-2i\bar\Psi_{i}^{[\mu}\Psi_{i}^{\nu]}$
captures the initial spin of the two massive objects.
The weak gravity expansion of the vierbein reads
\be
e^{a}_{\mu} = \eta^{a\nu}\left(\eta_{\mu\nu}+ \frac{\kappa}{2}h_{\mu\nu} - 
\frac{\kappa^2}{8}h_{\mu\rho}{h^\rho}_\nu  + \cO(\kappa^3) \right)\, ,
\ee
introducing the graviton field $h_{\mu\nu}(x)$
and the gravitational coupling $\kappa^{2}= 32\pi G$. 
Note that in this perturbative framework the
distinction between curved $\mu,\nu,\ldots$ and tangent
$a,b,\ldots$ indices necessarily drops.

The spinning WQFT has the partition function
\begin{align}\label{ZWQFTdef}
\mathcal{Z}_{\text{WQFT}}
&:= \text{const} \times
\int \!\!D[h_{\mu\nu}]
 \,  e^{i (S_{\rm EH}+S_{\rm gf})}\\ &\quad
\times\int \prod_{i=1}^{2} D [z_i^\mu] D[{\psi_{i}^{\prime}}^\mu]
\exp\Bigl[i\sum_{i=1}^{2} S^{(i)}+S_{ES^{2}}^{(i)}\Bigr ],\nn
\end{align}
where $S_{\rm EH}$ is the Einstein-Hilbert action
and the gauge-fixing term $S_{\rm gf}$ enforces de Donder gauge.
The SUSY variations \eqref{N=2SUSY}  leave an
imprint on the free energy (or eikonal)
$F_{\text{WQFT}}(b_{i},v_{i},\cS_{i}) := -i\log \mathcal{Z}_{\text{WQFT}}$:
after integrating out the fluctuations $h_{\mu\nu}$,
$z^{\mu}$ and $\psi'^\mu$ in the path integral \eqref{ZWQFTdef},
the SUSY variations of the background trajectories \eqref{bgexp} remain intact
in an asymptotically flat spacetime.
That is, the transformations
\begin{align}\label{susybg}
\begin{aligned}
\delta b^{\mu}_{i}&= i \bar\epsilon \Psi^{\mu}_{i} + i \epsilon\bar{\Psi}^{\mu}_{i}\, ,
\quad \delta v^{\mu}_{i}=0 \, ,
\quad \delta\Psi_{i}^\mu=-\epsilon v_{i}^{\mu}\, \\
\Rightarrow&\quad \delta \cS_{i}^{\mu\nu} = v_{i}^{\mu}\, \delta b_{i}^{\nu} 
-v_{i}^{\nu}\, \delta b_{i}^{\mu}
\end{aligned}
\end{align}
are a symmetry of $F_{\text{WQFT}}(b_{i},v_{i},\cS_{i})$
(only up to quadratic spin order when the Wilson coefficients $C_{E,i}$ are included).
As we shall see, this is also a symmetry of the waveform.
Using a suitable shift of the proper times $\tau_i$ we may choose $b\cdot v_i=0$,
where $b^\mu=b_2^\mu-b_1^\mu$ is the relative impact parameter;
by gauge fixing the SUSY transformations \eqref{susybg}
we impose $v_{i,\mu}\cS_i^{\mu\nu}=0$ (the covariant SSC).



\sec{Feynman rules}
As the Feynman rules for the Einstein-Hilbert action are conventional we will not dwell on them;
the only subtlety is our use of a \emph{retarded} graviton propagator:
\begin{align}
	\begin{tikzpicture}[baseline={(current bounding box.center)}]
	\coordinate (x) at (-.7,0);
	\coordinate (y) at (0.5,0);
	\draw [photon] (x) -- (y) node [midway, below] {$k$};
	\draw [fill] (x) circle (.08) node [above] {$\mu\nu$};
	\draw [fill] (y) circle (.08) node [above] {$\rho\sigma$};
	\end{tikzpicture}&=i\frac{P_{\mu\nu;\rho\sigma}}{(k^{0}+i\epsilon)^{2}-\mathbf{k}^2}\,,
\end{align}
with $P_{\mu\nu;\rho\sigma}:=\eta_{\mu(\rho}\eta_{\sigma)\nu}-
\sfrac12\eta_{\mu\nu}\eta_{\rho\sigma}$.
On the worldline we work in one-dimensional energy (frequency) space:
the propagators for the fluctuations $z^\mu(\omega)$
and anti-commuting vectors $\psi^{\prime\mu}(\omega)$ are respectively
\begin{subequations}\label{eq:Propagators}
	\begin{align}
	\begin{tikzpicture}[baseline={(current bounding box.center)}]
	\coordinate (in) at (-0.6,0);
	\coordinate (out) at (1.4,0);
	\coordinate (x) at (-.2,0);
	\coordinate (y) at (1.0,0);
	\draw [zUndirected] (x) -- (y) node [midway, below] {$\omega$};
	\draw [dotted] (in) -- (x);
	\draw [dotted] (y) -- (out);
	\draw [fill] (x) circle (.08) node [above] {$\mu$};
	\draw [fill] (y) circle (.08) node [above] {$\nu$};
	\end{tikzpicture}&=-i\frac{\eta^{\mu\nu}}{m\,(\omega+i\eps)^2}\,, \\
	\begin{tikzpicture}[baseline={(current bounding box.center)}]
		\coordinate (in) at (-0.6,0);
		\coordinate (out) at (1.4,0);
		\coordinate (x) at (-.2,0);
		\coordinate (y) at (1.0,0);
		\draw [zParticle] (x) -- (y) node [midway, below] {$\omega$};
		\draw [dotted] (in) -- (x);
		\draw [dotted] (y) -- (out);
		\draw [fill] (x) circle (.08) node [above] {$\mu$};
		\draw [fill] (y) circle (.08) node [above] {$\nu$};
		\end{tikzpicture}&=-i\frac{\eta^{\mu\nu}}{m\,(\omega+i\eps)}\,,
	\end{align}
\end{subequations}
which also both involve a retarded $i\eps$ prescription.
The former was already used in \Rcites{Mogull:2020sak,Jakobsen:2021smu}.

\begin{figure}[t!]
	\label{fig:1a}
	\includegraphics[width=6.0cm]{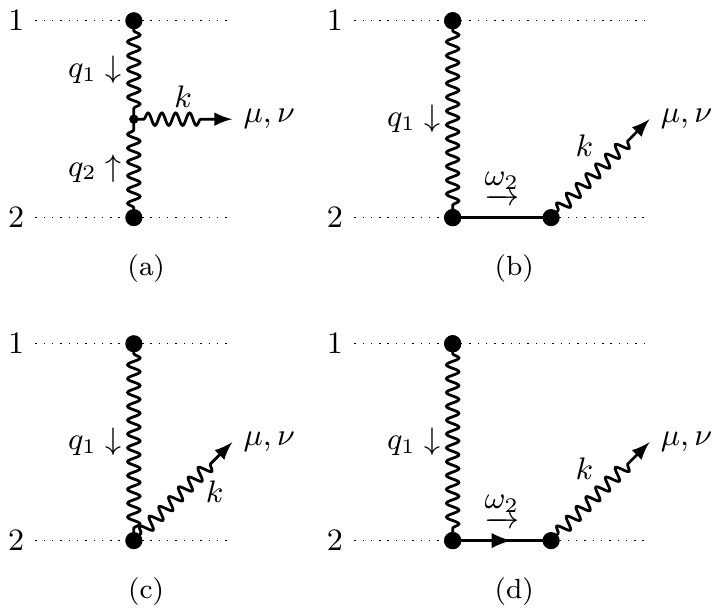}
	\caption{The four diagram topologies contributing to the 2PM Bremsstrahlung up to $\cO(\cS^2)$,
		where $\omega_i=k\cdot v_i$ by energy conservation at the worldline vertices.
		For diagrams (b)--(d) we also include the corresponding
		flipped topologies with massive bodies 1$\leftrightarrow$2;
		for diagram (d) (which includes the propagating fermion $\psi_2^{\prime\mu}$)
		we also include the graph with the arrow reversed.}
	\label{fig:1}
\end{figure}

Next we consider the worldline vertices.
The simplest of these is the single-graviton emission vertex:
\begin{align}\label{eq:vertexH}
	\begin{tikzpicture}[baseline={(current bounding box.center)}]
	\coordinate (in) at (-1,0);
	\coordinate (out) at (1,0);
	\coordinate (x) at (0,0);
	\node (k) at (0,-1.3) {$h_{\mu\nu}(k)$};
	\draw [dotted] (in) -- (x);
	\draw [dotted] (x) -- (out);
	\draw [graviton] (x) -- (k);
	\draw [fill] (x) circle (.08);
	\end{tikzpicture}&=
	-i\frac{m\kappa }{2}e^{ik\cdot b}\dd(k\cdot v)
	\bigg(v^\mu v^\nu+ik_\rho\cS^{\rho(\mu}v^{\nu)} \nn \\[-10pt]
  &+\frac12k_\rho k_\sigma\cS^{\rho\mu}\cS^{\nu\sigma}
  +\frac{C_E}{2}v^\mu v^\nu(k\cdot\cS\cdot\cS\cdot k)\bigg)\,,
\end{align}
where $\dd(\omega):=(2\pi)\delta(\omega)$ and we have used
$\cS^{\mu\nu}=-2i\bar{\Psi}^{[\mu}\Psi^{\nu]}$.
The other worldline-based vertices required for the 2PM Bremsstrahlung all appear
in Fig.~\ref{fig:1}:
the two-point interaction between a graviton and a single $z^\mu$ mode in (b),
the two-graviton emission vertex in (c),
and the two-point interaction between a graviton and ${\psi'}^\mu$ in (d).
Full expressions for these vertices are provided in the Supplementary Material.

\sec{Waveform from WQFT}
To describe the Bremsstrahlung at 2PM order including spin effects
we compute the expectation value
$k^2\langle h_{\mu\nu}(k)\rangle_{\rm WQFT}$.
This requires us to compute four kinds of Feynman graphs,
illustrated in Fig.~\ref{fig:1}.
Explicit expressions for the first two graphs (a) and (b)
were given in the non-spinning case \cite{Jakobsen:2021smu};
these are now modified by terms up to $\cO({\cal S}^2)$.
Graphs (c) and (d) are unique to the spinning case ---
for the latter we sum over both routings of the fermion line.

From this result we seek to obtain the waveform in spacetime in the \emph{wave zone},
where the distance to the observer $|\mathbf{x}|=r$ is large compared to all other lengths.
Following Ref.~\cite{Jakobsen:2021smu} the gauge-invariant \emph{frequency-domain waveform} $4G\,\eps^{\mu\nu}{S}_{\mu\nu}(k^{\mu}=\Omega\,(1,\hat{\bf x}))$
is extracted from the WQFT via
\be
{S}_{\mu\nu}(k)= \frac{2}{\kappa} k^{2} \vev{h_{\mu\nu}(k)}_{\rm WQFT}\, ,\ee
where $\Omega$ is the GW frequency and $\mathbf{\hat x}=\mathbf{x}/r$ points towards the observer. 
However, it is advantageous to study the \emph{time-domain waveform}
$f(u,\mathbf{\hat{x}})$  which is given by a Fourier transform: 
\be\label{eq:startingpoint}
\kappa \epsilon^{\mu\nu} h_{\mu\nu}=\frac{f(u,\mathbf{\hat{x}})}{r} = \frac{4G}{r}
\int_\Omega e^{-i k\cdot x} \,\epsilon^{\mu\nu}\,
S_{\mu\nu}(k) \Bigr |_{k^{\mu}=\Omega\, \rho^{\mu}}\, .
\ee
We have contracted with a polarization tensor
$\epsilon^{\mu\nu}=\sfrac12\epsilon^{\mu}\epsilon^{\nu}$,
$\int_\Omega:=\int_{-\infty}^\infty\sfrac{{\rm d}\Omega}{2\pi}$, and
$\rho^{\mu}=(1,\hat{\bf x})$;
in a PM decomposition $f=\sum_nG^nf^{(n)}$
we seek the 2PM component $f^{(2)}$.
Note that $k\cdot x = \Omega (t-r)$ yields the retarded time $u=t-r$,
and $\epsilon\cdot\epsilon=\epsilon\cdot\rho=0$.

\sec{Integration}
Our integration procedure follows closely
that used for the non-spinning calculation in Ref.~\cite{Jakobsen:2021smu},
the main difference being that we maintain four-dimensional Lorentz covariance.
Each diagram contributing to
$k^2\langle h_{\mu\nu}(k)\rangle_{\rm WQFT}$ carries
the overall factor
\begin{equation}\label{intmeasureGold}
	\hut{\mu}_{1,2}(k)=
	e^{i(q_1\cdot\hut{b}_1+q_2\cdot\hut{b}_2)}
	\dd(q_1\cdot\hut{v}_1)\dd(q_2\cdot\hut{v}_2)
	\dd(k-q_1-q_2)\,.
\end{equation}
We integrate over $q_i$, the momentum emitted from each worldline
(see Fig.~\ref{fig:1}).
When we also integrate over $\Omega$ ---
as in Eq.~\eqref{eq:startingpoint} ---
the full integration measure becomes
\begin{equation}\label{eq:measure}
	\int_{\Omega,q_1,q_2}\mu_{1,2}(k)e^{-ik\cdot x}=
	\frac1{\rho\cdot v_2}\int_{q_1}\dd(q_1\cdot v_1)e^{-iq_1\cdot\tilde{b}}\,,
\end{equation}
where $\int_{q_i}:=\int\!\sfrac{\d^4 q_i}{(2\pi)^4}$;
the delta function constraints give
$\Omega=\sfrac{q_1\cdot v_2}{\rho\cdot v_2}$ and $q_2=k-q_1$.
The shifted impact parameter,
\begin{align}\label{eq:shifted}
	\tilde{b}^\mu=\tilde{b}_2^\mu-\tilde{b}_1^\mu\,, &&
	\tilde{b}_i^\mu=b_i^\mu+u_iv_i^\mu\,,
\end{align}
extends the original impact parameter $b^\mu=b_2^\mu-b_1^\mu$
along the undeflected trajectories of the two bodies.
Finally, $u_i$ is the retarded time in the $i$'th rest frame:
\begin{equation}
	u_i=\frac{\rho\cdot(x-b_i)}{\rho\cdot v_i}\,,
\end{equation}
This implies $\rho\cdot\tilde{b}_i=\rho\cdot x=u$, so $\rho\cdot\tilde{b}=0$.

Rewriting the integral measure as in Eq.~\eqref{eq:measure}
is convenient for performing the integrals of diagrams (b)--(d),
in the rest frame of body 1.
The mirrored counterparts to these diagrams are easily recovered after
integration using the $1\leftrightarrow2$ symmetry of the waveform.
To integrate diagram (a) we insert the partial-fraction identity
$q_1^{-2}q_2^{-2}=-q_1^{-2}(2k\cdot q_1)^{-1}-q_2^{-2}(2k\cdot q_2)^{-1}$
(which is valid for $k$ on-shell)
and focus on the first term.

The full 2PM waveform is then written schematically as
(dropping the subscript on $q_1$)
\begin{align}\label{eq:schematic}
	\frac{f^{(2)}}{m_1 m_2}
	&=4\pi \int_{q}\dd(q\cdot v_1)
	  \frac{e^{-iq\cdot\tilde{b}}}{q^2}\left(
	  \frac{\mathcal{N}(q)}{q\cdot v_2+i\eps}+
	  \frac{\mathcal{M}(q)}{(q\cdot v_2)(q\cdot\rho)}
	  \right)\,,\nn\\
&\qquad+(1\leftrightarrow2)\,,
\end{align}
the $\cN$- and $\cM$-contributions corresponding to diagrams
(b)--(d) and (a) in Fig.~\ref{fig:1} respectively.
The numerators $\mathcal{N}(q)$ and $\mathcal{M}(q)$ have a uniform
power counting in $q$ for each spin order:
\begin{align}
	\mathcal{N}(q)&=
	\mathcal{N}_\mu q^\mu+\mathcal{N}_{\mu\nu} q^\mu q^\nu+\mathcal{N}_{\mu\nu\rho} q^\mu q^\nu q^\rho\,,\\
	\mathcal{M}(q)&=
	\mathcal{M}_{\mu\nu} q^\mu q^\nu+\mathcal{M}_{\mu\nu\rho}q^\mu q^\nu q^\rho+
	\mathcal{M}_{\mu\nu\rho\sigma}q^\mu q^\nu q^\rho q^\sigma\,,\nn
\end{align}
and the non-spinning result involves only $\cN_\mu$ and $\cM_{\mu\nu}$.
We present full expressions for $\cN$ and $\cM$ in the
ancillary file attached to the \texttt{arXiv} submission of this Letter.

To lowest order in $q^\mu$, the first integral in \eqn{eq:schematic} is
\begin{align}\label{eq:firstIntegral}
\begin{aligned}
	&4\pi \int_{q}\dd(q\cdot v_1)
	\frac{e^{-iq\cdot\tilde{b}}}{q^2}
	\frac{q^\mu}{q\cdot v_2+i\eps}\\
	&\qquad=
	\frac{P_1^{\mu\nu}v_{2,\nu}}{(\gamma^2-1)|\tilde{\mathbf{b}}|_1}-
	\frac{b^\mu}{|b|^2}\!
	\left(\frac1{\sqrt{\gamma^2-1}}+\frac{u_2}{|\tilde{\mathbf{b}}|_1}\right)\,,
\end{aligned}
\end{align}
where $P_i^{\mu\nu}:=\eta^{\mu\nu}-v_i^\mu v_i^\nu$ is a projector
into the rest frame of the $i$'th body,
$|b|=-\sqrt{b^\mu b_\mu}$ (the impact parameter is spacelike) and
\begin{align}\label{eq:modBDef}
	|\tilde{\mathbf{b}}|_{1,2}&:=\sqrt{-\tilde{b}_\mu P_{1,2}^{\mu\nu}\tilde{b}_\nu}
	=\sqrt{|b|^2+(\gamma^2-1)u_{2,1}^2}
	\end{align}
are the lengths of the shifted impact parameter $\tilde{b}^\mu$
\eqref{eq:shifted} in the two rest frames.
The second integral in \eqn{eq:schematic} is
\begin{align}\label{eq:secondIntegral}
\begin{aligned}
    &
    4\pi \int_{q}\dd(q\cdot v_1)
    \frac{e^{-iq\cdot\tilde{b}}}{q^2}
    \frac{q^\mu q^\nu}{q\cdot v_2\ q\cdot\rho}
    \\
    &\qquad
    =
    \frac{
      K_1^{\mu\nu}\ v_2 \cdot K_1 \cdot \rho
      -
      2 (v_{2}\cdot K_1)^{(\mu} (\rho\cdot K_1)^{\nu)}
    }{(\gamma^2-1)\ (\rho\cdot v_1)^2\  |b|^2\  |\tilde{b}|^2\ |\tilde{\mathbf{b}}|_1}\,,
\end{aligned}
\end{align}
where we have introduced the symmetric tensor
\begin{equation}\label{eq:kTensor}
	K_i^{\mu\nu}:=P_i^{\mu\nu}|\tilde{\mathbf{b}}|_i^2 
	+ (P_i\cdot\ltb)^\mu (P_i\cdot\ltb)^\nu\,,
\end{equation}
with the property that $K_i^{\mu\nu}v_{i,\nu}=K_i^{\mu\nu}\tilde{b}_\nu=0$.
Both integrals are derived in the Supplementary Material;
one generalizes to higher powers of $q^\mu$ in the numerators
by taking derivatives with respect to $\tilde{b}^\mu$.

\sec{Results}
The 2PM waveform takes the schematic form
\begin{align}\label{eq:result}
	\frac{f^{(2)}}{m_1 m_2} & = 
	\sum_{s=0}^2\frac1{|\tilde{\mathbf{b}}|_1^{2s+1}}\left[
	\alpha_1^{(s)}+
	\frac{\beta_1^{(s)}}{|\tilde{b}|^{2s+2}}
	\right]+(1\leftrightarrow2)\,,
\end{align}
where the coefficients $\alpha_i^{(s)}$, $\beta_i^{(s)}$,
provided in the ancillary file,
are associated with the $\cN$- and $\cM$-type contributions
in Eq.~\eqref{eq:schematic} respectively;
they are functions of $u_i$, $b^\mu$, $v_i^\mu$, $\rho^\mu$, and $\cS_i^{\mu\nu}$
and bi-linear in $\epsilon^\mu$.
The waveform $f$ is invariant under
the SUSY transformations in Eq.~\eqref{susybg}
to quadratic order in spin regardless of the values of $C_{E,i}$.
To see this we expand the waveform at all PM orders in powers of spin:
\begin{align}
	f & = f_{0}
	\!+\sum_{i=1}^2\cS_{i,\mu\nu}f^{\mu\nu}_{i} 
		+\sum_{i,j=1}^2\!\!\cS_{i,\mu\nu}\cS_{j,\rho\sigma} f_{ij}^{\mu\nu;\rho\sigma}
		+\cO(\cS^{3})\,,
\end{align}
where $f_i^{\mu\nu}$ and $f_{ij}^{\mu\nu;\rho\sigma}$ are defined
modulo terms that vanish on support of $v_{i,\mu}\cS_i^{\mu\nu}=0$.
The SUSY links higher-spin to lower-spin terms:
\begin{align}\label{eq:susyrel}
\frac{1}{2}\frac{\partial f_{0}}{\partial b_{i,\mu}} &=
v_{i,\nu}\, f_{i}^{[\mu\nu]}\, ,
&
\frac{1}{4} \frac{\partial f_{i}^{\mu\nu}}{\partial b_{j,\rho}}  &=
v_{j,\sigma}\, f_{ij}^{\mu\nu;[\rho\sigma]}\, ,
\end{align}
and these identities are satisfied by the waveform \eqref{eq:result}.

To illustrate the waveform we consider the \emph{gravitational wave memory}
$\Delta f(\mathbf{\hat{x}}):=f(+\infty,\mathbf{\hat{x}})-f(-\infty,\mathbf{\hat{x}})$.
The constant spin tensors are decomposed in terms of the
Pauli-Lubanski vectors $a_i^\mu$ as
$\cS_i^{\mu\nu}={\eps^{\mu\nu}}_{\rho\sigma}v_i^\rho a_i^\sigma$,
the latter satisfying $a_i\cdot v_i=0$.
In the aligned-spin case $a_i\cdot b=a_i\cdot v_j=0$,
i.e.~the spin vectors are orthogonal to the plane of scattering. Writing
$|a_i|=\sqrt{-a_{i}^{2}}$ the wave memory is then
proportional to the non-spinning result:
\begin{align}
	&\Delta f^{(2)}=\left(1+\frac{2v|a_3|}{b(1+v^2)}+\frac{|a_3|^2}{|b|^2}-
	\sum_{i=1}^2\frac{C_{E,i}|a_i|^2}{|b|^2}\right)\!
	\Delta f^{(2)}_{\cS=0} ,\nn\\
	&\frac{\Delta f^{(2)}_{\cS=0}}{m_1m_2}=
	\frac{4(2\gamma^2-1)\eps\cdot v_1(2b\cdot\eps\,\rho\cdot v_1-b\cdot\rho\,\eps\cdot v_1)}
	{|b|^2\sqrt{\gamma^2-1}(\rho\cdot v_1)^2}\nn\\
	&\qquad\qquad\qquad+(1\leftrightarrow2)\,,
\end{align}
where $a_3^\mu=a_1^\mu+a_2^\mu$.
For two Kerr black holes ($C_{E,i}=0$) with equal-and-opposite spins
($a_1^\mu=-a_2^\mu$) we see that $\Delta f^{(2)}=\Delta f^{(2)}_{\cS=0}$,
which we observe also when the spins are mis-aligned to the plane of scattering.

\begin{figure}[t!]
 \includegraphics[width=8.5cm]{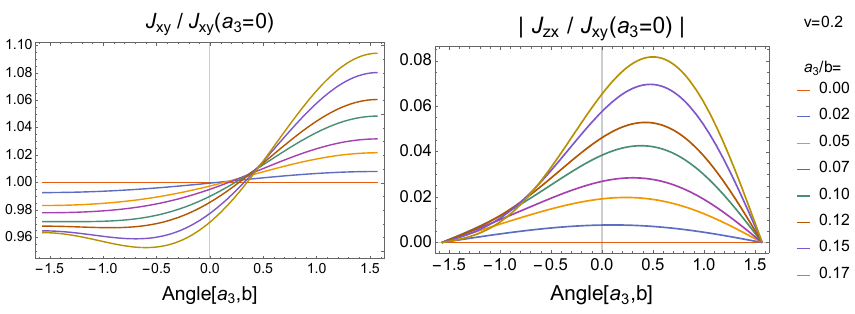}
   \caption{Total radiated angular momenta for the scattering of two Kerr-BHs with $v=0.2$ 
   as a function of the angle between the total initial spins 
   $\mathbf{a_{3}}=\mathbf{a_{1}} + \mathbf{a_{2}}$ and 
   $\mathbf{b}$ (with $\mathbf{a_{i}}\cdot\mathbf{v_{i}}=0$) for a range of ratios
   $|\mathbf{a_{3}}|/|\mathbf{b}|$. We show the normalized ratio of angular momenta emitted orthogonal to the $\mathbf{b},\mathbf{v}$ plane (left plot) and in the $\mathbf{b}$ direction (right plot), normalization is w.r.t.~angular momentum emitted in the spinless case. }
  \label{fig:2}
\end{figure}

There is also a 1PM (non-radiating) contribution to the waveform
consisting of single-graviton emission from either massive body:
\begin{equation}\label{eq:res1PM}
	f^{(1)}(\mathbf{\hat{x}})=\frac{2m_1}{\rho\cdot v_1}(\eps\cdot v_1)^2+
	\frac{2m_2}{\rho\cdot v_2}(\eps\cdot v_2)^2\,.
\end{equation}
At 1PM order there is manifestly no dependence on either the spins $\cS_i^{\mu\nu}$
or impact parameters $b_i^\mu$,
so the SUSY identities in Eq.~\eqref{eq:susyrel} are trivially satisfied.

Finally, the wave memory and 1PM part of the waveform
contribute to the total radiated angular momentum $J_{ij}^{\rm rad}$.
Using three-dimensional Cartesian basis vectors $\bv_i$,
we choose a frame of reference with the initial velocities $v_i^\mu$
restricted to the $t$--$x$ plane;
$\mathbf{b}=|b|\,\bv_2$ is orthogonal to these.
Then we find two non-zero components of $J_{ij}^{\rm rad}$:
$J_{xy}^{\rm rad}$ and $J_{zx}^{\rm rad}$,
which are conveniently arranged into
\begin{align}\label{eq:radAngMom}
\begin{aligned}
	&\frac{J^{\text{rad}}_{xy}+iJ^{\text{rad}}_{zx}}{\left.J^{\rm init}_{xy}\right|_{\cS=0}}=
	\frac{4G^2m_1m_2}{|b|^2}\frac{(2\gamma^2-1)}{\sqrt{\gamma^2-1}}{\cal I}(v)\\
	&\times\left(1-\frac{2iv\,\mathbf{a}_3\cdot\mathbf{l}}{|b|(1+v^2)}
	-\frac{(\mathbf{a}_3\cdot\mathbf{l})^2}{|b|^2}
	+\sum_{i=1}^2\frac{C_{E,i}}{|b|^2}(\mathbf{a}_i\cdot\mathbf{l})^2
	\right)\\
	&\qquad+\cO(G^3)\,.
\end{aligned}
\end{align}
We normalize with respect to $\left.J^{\rm init}_{xy}\right|_{\cS=0}$,
the initial angular momentum in the non-spinning case.
The spin vectors $\mathbf{a}_1$ and $\mathbf{a}_2$
are taken in the rest frame of each massive body;
$\mathbf{a}_3=\mathbf{a}_1+\mathbf{a}_2$, $\mathbf{l}=\bv_2+i\bv_3$, and
\begin{align}
	{\cal I}(v)=-\frac83+\frac1{v^2}+\frac{(3v^2-1)}{v^3}{\rm arctanh}(v)
\end{align}
is a universal prefactor.
Eq.~\eqref{eq:radAngMom} holds in the
rest frame of either body or the center-of-mass (c.o.m.) frame;
see Fig.~\ref{fig:2} for plots. For a derivation we refer the reader to the Supplementary Material.
There we also compute the total radiated energy in the c.o.m.~frame.
Due to the multi-scale nature of the waveform it is difficult
to perform the necessary time and solid
angle-integrals, so we performed a low velocity expansion.
For terms up to $\cO(v^{2})$ we find
\begin{widetext}
  \begin{align}\label{eq:radEnergy}
    &E^{\text{rad,LO}}_{\text{CoM}}
    =
    \frac{v G^3 m_1^2 m_2^2 \pi}{|b|^3}
    \bigg[
    \frac{37}{15}
    +\frac{
      v 
      (65 m_1+69 m_2) (\mathbf{a}_1\! \!\cdot \! \mathbf{\hat{e}_{3}})
    }{
      10 |b| (m_1+m_2)
    }
     +\frac{
    1503 (\mathbf{a}_1\!\! \cdot \!\mathbf{{\hat e}_1})( \mathbf{a}_2\!\! \cdot \!\mathbf{{\hat e}_1})
    -
    3559 (\mathbf{a}_1\!\! \cdot \!\mathbf{{\hat e}_2})( \mathbf{a}_2\!\! \cdot \!\mathbf{{\hat e}_2})
    +
    1816 (\mathbf{a}_1\!\! \cdot \!\mathbf{{\hat e}_3})( \mathbf{a}_2\!\! \cdot \!\mathbf{{\hat e}_3})
    }{320 |b|^2}
    \nn\\&\quad
    +\frac{9   (185- 176 C_{E,1}) (\mathbf{a}_1\!\! \cdot \!\mathbf{{\hat e}_1})^2-
      (3385-3472 C_{E,1})
    (\mathbf{a}_1\! \!\cdot \!\mathbf{{\hat e}_2})^2+8(245- 236 C_{E,1}) (\mathbf{a}_1\!\! \cdot \!{\mathbf{{\hat e}_3}})^2}{320 |b|^2}
   + (1\leftrightarrow 2)
   +\mathcal{O}\left(v ^2\right) 
    \bigg] ,
  \end{align}
\end{widetext}
where the swap $(1\leftrightarrow 2)$ does not affect the basis vectors $\mathbf{{\hat e}_i}$ or the constant term $\frac{37}{15}$.
It is straighforward to extend this result to higher orders in $v$.

\sec{Conclusions}
In this Letter we extended the WQFT to describe spinning compact bodies to quadratic order in spin, and calculated the leading-PM order waveform for highly eccentric (scattering) orbits.
Our accompanying work~\cite{Jakobsen:2021zvh} presents an application to further observables such as the spin kick and deflection \cite{Liu:2021zxr,Kosmopoulos:2021zoq} at 2PM order
and gives details on the approximate SUSY and its relation to the SSC.
The radiated energy~\eqref{eq:radEnergy} should also be
particularly useful for future studies.
In Refs.~\cite{Herrmann:2021lqe,Herrmann:2021tct} the $\cO(G^3)$ energy loss
from a scattering of non-spinning black holes was recently computed to all
orders in velocity using the KMOC formalism \cite{Kosower:2018adc}
(see also Ref.~\cite{Maybee:2019jus});
a similar result could conceivably be obtained at $\cO(\cS^2)$,
and then checked against Eq.~\eqref{eq:radEnergy} in the low-velocity limit.
Similarly, the remarkably simple result for radiated angular momentum
\eqref{eq:radAngMom} at 2PM order is intriguing; it may be important for understanding the high-energy limit, see \Rcite{DiVecchia:2020ymx,Damour:2020tta} for the non-spinning case.

The application of modern on-shell and integration techniques to compute scattering amplitudes~\cite{Bern:1994zx,*Bern:1994cg,*Britto:2004nc,Bern:2008qj,*Bern:2010ue,*Bern:2012uf,*Bern:2017ucb,*Bern:2018jmv,*Bern:2019prr,Bern:2019crd,Parra-Martinez:2020dzs,Bern:2021dqo,Herrmann:2021lqe} holds great promise for pushing calculations to higher PM orders.
This is demonstrated by the impressive calculation of the 4PM conservative dynamics in the potential region~\cite{Bern:2021dqo,Dlapa:2021npj} --- see also Refs.~\cite{Bern:2019nnu,Bern:2019crd,Cheung:2020gyp,*Kalin:2020fhe,DiVecchia:2020ymx,Damour:2020tta,DiVecchia:2021bdo,*Bjerrum-Bohr:2021din,Bini:2020flp,*Cheung:2020sdj,*Haddad:2020que,*Kalin:2020lmz,*Brandhuber:2019qpg,*Huber:2019ugz,*AccettulliHuber:2020oou,*AccettulliHuber:2020oou,*Bern:2020uwk,*Cheung:2020gbf,*Aoude:2020ygw,Amati:1990xe,*DiVecchia:2019myk,*DiVecchia:2019kta,*Bern:2020gjj,*Huber:2020xny,*DiVecchia:2021ndb,*Bautista:2019tdr,*Laddha:2018rle,*Laddha:2018myi,*Sahoo:2018lxl,*Laddha:2019yaj,*Saha:2019tub,*A:2020lub,*Sahoo:2020ryf,*Sahoo:2018lxl}.
The connection between amplitudes and classical physics was studied in Refs.~\cite{Kosower:2018adc,Maybee:2019jus,Damour:2019lcq,*Bjerrum-Bohr:2019kec,*Bjerrum-Bohr:2021vuf}, and Refs.~\cite{Kalin:2020mvi,*Kalin:2019rwq,*Kalin:2019inp} discussed the connection to bound orbits.
Our WQFT framework~\cite{Mogull:2020sak,Jakobsen:2021smu} provides an efficient,
rather intuitive way to connect amplitude and (classical) worldline EFT calculations.
It may therefore benefit from modern amplitude techniques at higher PM orders in future work, building on the compact Lorentz-covariant master integrals provided here.

\medskip

\sec{Acknowledgments}
We would like to thank F.~Bautista, R.~Bonezzi, A.~Buonanno, P.~Pichini  and J.~Vines for very helpful discussions.
We are also grateful for use of G.~K\"alin's C\texttt{++} graph library.
GUJ's and GM's research is funded by the Deutsche Forschungsgemeinschaft (DFG, German Research Foundation) Projektnummer 417533893/GRK2575 ``Rethinking Quantum Field Theory''.

\appendix
\section*{Supplementary Material}

\sec{Feynman rules}
Here we give explicit expressions for the worldline Feynman rules used in the main calculation,
the single-graviton emission vertex having already been given in Eq.~\eqref{eq:vertexH}.
Adding an outgoing $z^\mu$ line we have:
\begin{widetext}
\begin{align}\label{eq:vertexHZ}
\begin{aligned}
\begin{tikzpicture}[baseline={(current bounding box.center)}]
\coordinate (in) at (-1,0);
\coordinate (out) at (1,0);
\coordinate (x) at (0,0);
\node (k) at (0,-1.3) {$h_{\mu\nu}(k)$};
\draw (out) node [right] {$z^\rho(\omega)$};
\draw [dotted] (in) -- (x);
\draw [zUndirected] (x) -- (out);
\draw [graviton] (x) -- (k);
\draw [fill] (x) circle (.08);
\end{tikzpicture}\!\!=
\frac{m\kappa }{2}e^{ik\cdot b}\dd(k\cdot v+\omega)
\bigg(2\omega v^{(\mu}\delta^{\nu)}_\rho+v^\mu v^\nu k_\rho
+i(k\cdot\cS)^{(\mu}(k_\rho v^{\nu)}+\omega\delta_\rho^{\nu)})+
\frac12k_\rho(k\cdot\cS)^{\mu}(\cS\cdot k)^{\nu}&\\[-15pt]
+\frac{C_E}{2}\Big(\left(2\omega v^{(\mu}\delta^{\nu)}_\rho+v^\mu v^\nu k_\rho\right)(k\cdot\cS\cdot\cS\cdot k)-
\omega^2k_\rho(\cS\cdot\cS)^{\mu\nu}+
2\omega^2(k\cdot\cS\cdot \cS)^{(\mu}\delta^{\nu)}_\rho\Big)\bigg)\,,&
\end{aligned}
\end{align}
where we have adopted the shorthands $(k\cdot\cS)^\mu=k_\nu\cS^{\nu\mu}$, $(\cS\cdot\cS)^{\mu\nu}=\cS^{\mu\rho}\cS_{\rho}{}^{\nu}$
and $(\cS\cdot k)^\mu=\cS^{\mu\nu}k_\nu$.
Both this and the single-graviton emission vertex appear in the non-spinning case,
and by setting $\cS^{\mu\nu}=0$ we recover the corresponding
expressions from \Rcites{Mogull:2020sak,Jakobsen:2021smu}.
New to the spinning case is the coupling with  $\psi^{\prime\mu}$:
\begin{align}\label{eq:vertexHS}
  \begin{aligned}
	\begin{tikzpicture}[baseline={(current bounding box.center)}]
	\coordinate (in) at (-1,0);
	\coordinate (out) at (1,0);
	\coordinate (x) at (0,0);
	\node (k) at (0,-1.3) {$h_{\mu\nu}(k)$};
	\draw (out) node [right] {$\psi^{\prime\rho}(\omega)$};
	\draw [dotted] (in) -- (x);
	\draw [zParticle] (x) -- (out);
	\draw [graviton] (x) -- (k);
	\draw [fill] (x) circle (.08);
	\end{tikzpicture}\!\!\!\!\!&=
	-im\kappa e^{ik\cdot b}\dd(k\cdot v+\omega)
	\bigg(k_{[\rho}\delta_{\sigma]}^{(\mu}\!\left(v^{\nu)}-i(\cS\cdot k)^{\nu)}\right)+
	iC_E\!\left(v^{(\mu}k_\lambda+\omega\delta^{(\mu}_\lambda\right)\!
	\left(v^{\nu)}k_{[\rho}+\omega\delta^{\nu)}_{[\rho}\right)\!{\cS^\lambda}_{\sigma]}
	\bigg)\bar{\Psi}^\sigma.
  \end{aligned}
\end{align}
The vertex with $\bar{\psi}^{\prime\mu}(\omega)$ on an outgoing
line is identical, except with $\bar{\Psi}^\mu\to\Psi^\mu$.
Finally, starting at linear order in spin there is also the two-graviton emission vertex:
\begin{align}
  \begin{tikzpicture}[baseline={(current bounding box.center)}]
  \coordinate (in) at (-1,0);
  \coordinate (out) at (1,0);
  \coordinate (x) at (0,0);
  \node (k1) at (-.9,-1.3) {$h_{\mu_1\nu_1}(k_1)$};
  \node (k2) at (.9,-1.3) {$h_{\mu_2\nu_2}(k_2)$};
  \draw [dotted] (in) -- (x);
  \draw [dotted] (x) -- (out);
  \draw [graviton] (x) -- (k1);
  \draw [graviton] (x) -- (k2);
  \draw [fill] (x) circle (.08);
  \end{tikzpicture}&=
  -\frac{m\kappa^2}{4}e^{i(k_1+k_2)\cdot b}\dd((k_1+k_2)\cdot v)
  \big(
	(k_1\cdot\cS)^{\mu_2}v^{\mu_1}\eta^{\nu_1\nu_2}-
	\cS^{\mu_1\mu_2}\!\left(v^{\nu_1}k_1^{\nu_2}-
	\sfrac12k_1\cdot v\eta^{\nu_1\nu_2}\right)\nn\\[-15pt]
  &\quad
  +i\left((\cS\cdot k_1)^{\mu_1}(\cS\cdot k_1)^{\mu_2}
  +\sfrac12(\cS\cdot k_2)^{\mu_1}(\cS\cdot k_1)^{\mu_2}
  -\sfrac12\cS^{\mu_1\mu_2}(k_1\cdot\cS\cdot k_2)\right)\eta^{\nu_1\nu_2}\nn\\
  &\quad
  +\sfrac{i}4k_1\cdot k_2\cS^{\mu_1\nu_2}\cS^{\mu_2\nu_1}
  -ik_1^{\nu_2}(\cS\cdot(k_1+k_2))^{\mu_1}\cS^{\mu_2\nu_1}\nn\\
  &\quad
  +i\,C_E\left(
	2k_1\cdot v(\cS\cdot\cS\cdot(k_1+k_2))^{\mu_2}v^{\mu_1}
  	-\sfrac12(k_1\cdot v)^2(\cS\cdot\cS)^{\mu_1\mu_2}
	-\sfrac12(k_1\cdot\cS\cdot\cS\cdot k_2)v^{\mu_1}v^{\mu_2}
  \right)\eta^{\nu_1\nu_2}\nn\\
  &\quad
  +i\,C_E\big(
	-\sfrac12k_1\cdot k_2(\cS\cdot\cS)^{\nu_1\nu_2}v^{\mu_1}v^{\mu_2}
	+k_1^{\nu_2}(\cS\cdot\cS\cdot k_2)^{\nu_1}v^{\mu_1}v^{\mu_2}
	-k_1^{\nu_2}(\cS\cdot\cS\cdot k_1)^{\mu_2}v^{\mu_1}v^{\nu_1}\nn\\
  &\quad
  	-k_1^{\nu_2}(\cS\cdot\cS\cdot k_2)^{\mu_2}v^{\mu_1}v^{\nu_1}
	-(\cS\cdot\cS)^{\mu_2\nu_2}\left(k_1\cdot vk_2^{\nu_1}-\sfrac12k_1\cdot k_2v^{\nu_1}\right)v^{\mu_1}
  \big)\big)+(1\leftrightarrow2)\,,
\end{align}
with implicit symmetrization on $(\mu_1,\nu_1)$ and $(\mu_2,\nu_2)$.
\end{widetext}

\sec{Integration}
To compute the 2PM waveform we require explicit results
for the following integrals:
\begin{align}
  \mathcal{J}^{
    \mu_1\mu_2...\mu_n
  }
  &=
  4\pi \int_{q}\dd(q\cdot v_1)
  \frac{e^{-iq\cdot\tilde{b}}}{q^2}
  \frac{
    q^{\mu_1} q^{\mu_2} ... q^{\mu_n}
  }{
    q\cdot v_2+i\eps}
  \ ,\label{eq:jInts}
  \\
  \mathcal{I}^{
    \mu_1\mu_2...\mu_n
  }
  &=
  4\pi \int_{q}\dd(q\cdot v_1)
  \frac{e^{-iq\cdot\tilde{b}}}{q^2}
  \frac{
    q^{\mu_1} q^{\mu_2} ... q^{\mu_n}
  }{
    q\cdot v_2\ q\cdot\rho}
  \ ,\label{eq:iInts}
\end{align}
with $n=1,2,3$ for the $\jty$-integrals and $n=2,3,4$ for the $\ity$-integrals.
Expressions for $\jty^\mu$ and $\ity^{\mu\nu}$ were presented in
Eqs.~\eqref{eq:firstIntegral} and~\eqref{eq:secondIntegral}
of the main text respectively ---
we derive these first, then generalize to higher-orders in $q^\mu$
by taking derivatives with respect to the shifted impact parameter
$\tilde{b}^\mu$.

Our starting point for ${\cal J}^\mu$ is
\begin{align}\label{eq:easyInt}
  4\pi\int_q \dd(q\cdot v_1)
  e^{
    -i q\cdot \ltb
  }\frac{q^\mu}{q^2}
  =
  -i\frac{ P_1^{\mu \nu} \ltb_\nu
  }{
    \tbmo^3
  }\,,
\end{align}
which is easily derived by specializing to the rest frame of massive body 1 ---
$P_1^{\mu\nu}\tilde{b}_\nu$ and $\tbmo$ \eqref{eq:modBDef} are the covariant
``uplifts'' of $\tilde{\mathbf{b}}^i$ and $|\tilde{\mathbf{b}}|$ from this frame.
Using
\begin{align}
  \int_\omega e^{-i\omega\tau} \frac{f(\omega)}{\omega+i\epsilon}
  =
  -i\int_{-\infty}^{\tau} \text{d}\tau'
  \int_\omega e^{-i\omega \tau'} f(\omega)
\end{align}
the $\jty^\mu$ integral can be re-written as
\begin{align}
  \jty^\mu =
  -4 \pi i
  \int_{-\infty}^{u_2} \d u_2'
  \int_q \dd(q\cdot v_1)
  e^{
    -i q\cdot \ltb'
  }\frac{q^\mu}{q^2}
  \,,
\end{align}
where $\ltb'^\mu=b^\mu+u_2'v_2^\mu-u_1v_1^\mu$.
Inserting \eqref{eq:easyInt} and performing the one-dimensional
$u_2'$ integration produces Eq.~\eqref{eq:firstIntegral}.

In addition to $v_1^\mu$ the $\ity^{\mu\nu}$ integral is also orthogonal to
$\ltb^\mu$, i.e. $\ltb_\mu\ity^{\mu\nu}=v_{1,\mu}\ity^{\mu\nu}=0$.
This follows from
\begin{align}
  \ltb_\mu \ity^{\mu\nu}=
  i |\ltb|
  \frac{\partial}{\partial|\ltb|}
  \ity^\nu=0\,,
\end{align}
where the first equality is derived from the ${\cal I}$-type
integrals definition \eqref{eq:iInts}.
The integrand of $\ity^\mu$ is dimensionless in $q^\mu$,
so its integrated form depends only on dimensionless combinations of $\ltb^\mu$ ---
hence the second equality.
$\ity^{\mu\nu}$ therefore lives in a two-dimensional subspace
orthogonal to $v^\mu_1$ and $\ltb^\mu$, and we make an ansatz:
\begin{align}\label{eq:ansatz}
  \ity^{\mu \nu}
  =
  c_1 K_1^{\mu \nu}
  +
  c_2 (v_2\cdot K_1)^{(\mu} (\rho\cdot K_1)^{\nu)}
  \, .
\end{align}
$K_1^{\mu\nu}$ was defined in Eq.~\eqref{eq:kTensor} as the four-dimensional
projector into this subspace.
We solve for the coefficients by contracting $\ity^{\mu \nu}$
with $v_2^\mu$ and/or $\rho^\mu$,
evaluating the resulting scalar integrals to obtain
\begin{align}\label{eq:iTypeIntegral}
  &
  \rho_\mu v_{2\nu} \ity^{\mu\nu}
  =
  -\frac{1}{\tbmo}\,,
  &
  \rho_\mu \rho_\nu \ity^{\mu\nu}
  =
  \frac{v_2\cdot K_1\cdot\rho
  }{
    (\gamma^2-1)|b|^2\tbmo
  }\,.
\end{align}
These allow us to fix $c_1$ and $c_2$,
and we recover Eq.~\eqref{eq:secondIntegral}.

By differentiating these integrals with respect to
$\ltb^\mu$ one can pull down additional factors of $q^\mu$.
For the ${\cal J}$-type integrals this procedure is unambiguous;
special care should be taken for the ${\cal I}$-type integrals
as $\ltb^\mu$ is constrained by $\rho \cdot \ltb =0$.
However, provided one always works in the three-dimensional subspace
defined by $P_1^{\mu\nu}$ then one overcomes this problem,
as all contractions involve $P_1^{\mu\nu}$ and $\rho\cdot P_1 \cdot \ltb\neq0$.

\sec{Radiated energy and angular momentum}
In Ref.~\cite{Jakobsen:2021smu} we used the spin-less Bremsstrahlung
waveform to compute expressions for the radiated energy and angular momentum,
so here we extend these to include spin.
The relevant starting points are the same \cite{Damour:2020tta,Bonga:2018gzr}:
\begin{align}
	P^{\mu}_{\text{rad}}&=
	\frac1{32\pi G}\int\!\d u \d\sigma [\dot{f}_{ij}]^{2}\rho^{\mu}\label{eq:momdef}\,,
	\quad \text{where } f=f_{ij}\epsilon^{ij}
	\\
	J^{\text{rad}}_{ij}&=\frac1{8\pi G}\int\!\d u \d\sigma \left (f_{k[i}\dot{f}_{j]k} -\frac{1}{2} x_{[i}\partial_{j]}f_{kl}\dot{f}_{kl}\right )\, ,
\end{align}
with $\dot{f}_{ij}:=\partial_{u}f_{ij}$ and
$\d\sigma=\sin\theta\d\theta\d\phi$ is the unit sphere measure.
Here we have introduced a spherical polar coordinate system via
\be
\mathbf{\hat{x}}=
\mathbf{\hat{e}}_1 \cos\theta +\sin\theta\big(\mathbf{\hat{e}}_2\cos\phi +\mathbf{\hat{e}}_3\sin\phi \big)\, ,
\ee
which defines the angles $\theta$ and $\phi$ towards the observer;
$\mathbf{\hat{e}}_i$ are Cartesian spatial unit vectors
(with Latin indices $i,j,\ldots$).
Without loss of generality we assume that $\bv_2$ and $\bv_3$
are orthogonal to the initial velocities $v^\mu_1$ and $v^\nu_2$,
and $\mathbf{b}=|b|\,\bv_2$ where $b^\mu=(0,\mathbf{b})$.
The waveform $f_{ij}(u,\theta,\phi)$ is conveniently
decomposed on a basis of transverse-traceless polarization tensors:
\begin{equation}\label{eq:polBasis}
f_{ij}=f_+ (e_+)_{ij}+f_\times(e_\times)_{ij} \,,
\end{equation}
where $f_{+,\times}=\frac12(e_{+,\times})_{ij}f_{ij}$
and the polarization tensors are explicitly given as
\begin{align}
e_+^{ij}=\hat{\bm\theta}^i \hat{\bm\theta}^j - \hat{\bm\phi}^i \hat{\bm\phi}^j\,,\quad
e_\times^{ij}=\hat{\bm\theta}^i \hat{\bm\phi}^j+\hat{\bm\phi}^i \hat{\bm\theta}^j\, .
\end{align}
The two angular vectors orthogonal to $\mathbf{\hat{x}}$
are $\hat{\bm\theta}:=\partial_\theta\mathbf{\hat{x}}$
and $\hat{\bm\phi}:=(\sin\theta)^{-1}\partial_\phi \mathbf{\hat{x}}$.

Given our starting point of a fully Lorentz-covariant
expression for the waveform $f_{ij}$,
we can make different choices of inertial frame for intermediate expressions.
There are two of particular interest to us:
the rest frame of the first massive body,
and the center-of-mass (c.o.m.) frame.
In either case, we decompose the velocities $v_i^\mu$ and Pauli-Lubanski
spin vectors $a_i^\mu$; defined via
$\cS_i^{\mu\nu}={\eps^{\mu\nu}}_{\rho\sigma}v_i^\rho a_i^\sigma$; as
\begin{align}
	v_i^\mu=\left(\begin{matrix}
		\gamma_i \\
		\gamma_i\mathbf{v}_i
	\end{matrix}\right)\,, &&
	a_i^\mu=\left(\begin{matrix}
		\gamma_i(\mathbf{v}_i\cdot\mathbf{a}_i) \\
		\mathbf{a}_i+\sfrac{\gamma_i^2}{1+\gamma_i}
		(\mathbf{v}_i\cdot\mathbf{a}_i)\mathbf{v}_i
	\end{matrix}\right)\,,
\end{align}
where $\mathbf{v}_i\parallel\bv_2$.
These choices manifestly ensure that
$a_i\cdot v_i=0$, $v_i^2=1$, and $a_i^2=-\mathbf{a}_i^2$.
Note that $\mathbf{a_{i}}$ always denotes the spin vector of the $i$th body in its restframe.
In the first rest frame $\mathbf{v}_1=\mathbf{0}\implies\gamma_1=1$,
$\mathbf{v}_2=\mathbf{v}\implies\gamma_2=\gamma$;
in the c.o.m.~frame $\mathbf{v}_1=v_1\bv_1$ and $\mathbf{v}_2=-v_2\bv_1$, where
\begin{align}
	v_i=\frac{p_\infty}{E_i}\,, && \gamma_i=\frac{E_i}{m_i}\,,
\end{align}
and $E_i=\sqrt{m_i^2+p_\infty^2}$.
The initial momenta are $p_1^\mu=m_1v_1^\mu=(E_1,p_\infty,0,0)$
and $p_2^\mu=m_2v_2^\mu=(E_2,-p_\infty,0,0)$.
The c.o.m.~momentum $p_\infty$ is
\begin{align}
	p_\infty=\frac{m_1m_2\sqrt{\gamma^2-1}}{\sqrt{m_1^2+m_2^2+2\gamma m_1m_2}}\,.
\end{align}
When working in the c.o.m.~frame we prefer to
express intermediate results in terms of $\gamma_i$ and $v_i$, then use
\begin{equation}
	v=\frac{v_1+v_2}{1+v_1v_2}
\end{equation}
to reassemble final expressions in terms of $\gamma$ and $v$.

We begin with the radiated angular momentum $J^{\text{rad}}_{ij}$,
which contributes at leading PM order $G^2$.
There are two non-zero components:
$J^{\text{rad}}_{zx}$ and $J^{\text{rad}}_{xy}$.
As $f^{(1)}$ \eqref{eq:res1PM} is static the $u$-integration
is trivially performed by expressing $J^{\text{rad}}_{zx}$ and $J^{\text{rad}}_{xy}$
in terms of the wave memories
$\Delta{f}_{+,\times}:=
\left.f_{+,\times}\right|_{u=\infty}-\left.f_{+,\times}\right|_{u=-\infty}$:
\begin{align}
	J^{\text{rad}}_{xy}+iJ^{\text{rad}}_{zx}&=
	\frac1{8\pi}\int\!\d\sigma\,e^{-i\phi}\Bigl [
		i\frac{f_{+}^{(1)}\Delta f_{\times}}{\sin\theta}
		-\partial_{\theta}f_{+}^{(1)}\frac{\Delta f_{+}}2
		\Bigr ]\nn\\
	&\qquad+\cO(G^3)\,.
\end{align}
The result after integration is
Eq.~\eqref{eq:radAngMom}. It  holds in both the
rest frame of the first body and the c.o.m. frame:
in the former case $\left.J^{\rm init}_{xy}\right|_{\cS=0}=m_2\sqrt{\gamma^2-1}\,|b|$;
in the latter $\left.J^{\rm init}_{xy}\right|_{\cS=0}=p_\infty|b|$.

The radiated four-momentum $P^{\mu}_{\text{rad}}$
\eqref{eq:momdef} contributes to leading PM order $G^3$.
In the center-of-mass frame the radiated energy is
$E_{\text{rad,CoM}}=v_{\text{CoM}}^\mu P^{\text{rad}}_\mu$, where
\begin{equation}
  v_{\text{CoM}}=\frac{m_1 v_1+m_2 v_2}{\sqrt{m_1^2+m_2^2+2\gamma m_1 m_2}}
  \ .
\end{equation}
Due to the multi-scale nature of the waveform $f_{ij}$ it is difficult
to perform the time and solid-angle integrations in Eq.~\eqref{eq:momdef} directly;
however, in a low velocity expansion we suceeded and the result is stated in Eq.~\eqref{eq:radEnergy}. 

\bibliography{../bib/wqft_spin}

\begin{thebibliography}{134}%
\makeatletter
\providecommand \@ifxundefined [1]{%
 \@ifx{#1\undefined}
}%
\providecommand \@ifnum [1]{%
 \ifnum #1\expandafter \@firstoftwo
 \else \expandafter \@secondoftwo
 \fi
}%
\providecommand \@ifx [1]{%
 \ifx #1\expandafter \@firstoftwo
 \else \expandafter \@secondoftwo
 \fi
}%
\providecommand \natexlab [1]{#1}%
\providecommand \enquote  [1]{``#1''}%
\providecommand \bibnamefont  [1]{#1}%
\providecommand \bibfnamefont [1]{#1}%
\providecommand \citenamefont [1]{#1}%
\providecommand \href@noop [0]{\@secondoftwo}%
\providecommand \href [0]{\begingroup \@sanitize@url \@href}%
\providecommand \@href[1]{\@@startlink{#1}\@@href}%
\providecommand \@@href[1]{\endgroup#1\@@endlink}%
\providecommand \@sanitize@url [0]{\catcode `\\12\catcode `\$12\catcode
  `\&12\catcode `\#12\catcode `\^12\catcode `\_12\catcode `\%12\relax}%
\providecommand \@@startlink[1]{}%
\providecommand \@@endlink[0]{}%
\providecommand \url  [0]{\begingroup\@sanitize@url \@url }%
\providecommand \@url [1]{\endgroup\@href {#1}{\urlprefix }}%
\providecommand \urlprefix  [0]{URL }%
\providecommand \Eprint [0]{\href }%
\providecommand \doibase [0]{http://dx.doi.org/}%
\providecommand \selectlanguage [0]{\@gobble}%
\providecommand \bibinfo  [0]{\@secondoftwo}%
\providecommand \bibfield  [0]{\@secondoftwo}%
\providecommand \translation [1]{[#1]}%
\providecommand \BibitemOpen [0]{}%
\providecommand \bibitemStop [0]{}%
\providecommand \bibitemNoStop [0]{.\EOS\space}%
\providecommand \EOS [0]{\spacefactor3000\relax}%
\providecommand \BibitemShut  [1]{\csname bibitem#1\endcsname}%
\let\auto@bib@innerbib\@empty
\bibitem [{\citenamefont {Abbott}\ \emph {et~al.}(2016)\citenamefont {Abbott}
  \emph {et~al.}}]{Abbott:2016blz}%
  \BibitemOpen
  \bibfield  {author} {\bibinfo {author} {\bibfnamefont {B.P.}\ \bibnamefont
  {Abbott}} \emph {et~al.} (\bibinfo {collaboration} {LIGO Scientific,
  Virgo}),\ }\bibfield  {title} {\enquote {\bibinfo {title} {{Observation of
  Gravitational Waves from a Binary Black Hole Merger}},}\ }\href {\doibase
  10.1103/PhysRevLett.116.061102} {\bibfield  {journal} {\bibinfo  {journal}
  {Phys. Rev. Lett.}\ }\textbf {\bibinfo {volume} {116}},\ \bibinfo {pages}
  {061102} (\bibinfo {year} {2016})},\ \Eprint
  {http://arxiv.org/abs/1602.03837} {arXiv:1602.03837 [gr-qc]} \BibitemShut
  {NoStop}%
\bibitem [{\citenamefont {Abbott}\ \emph
  {et~al.}(2019{\natexlab{a}})\citenamefont {Abbott} \emph
  {et~al.}}]{LIGOScientific:2018mvr}%
  \BibitemOpen
  \bibfield  {author} {\bibinfo {author} {\bibfnamefont {B.P.}\ \bibnamefont
  {Abbott}} \emph {et~al.} (\bibinfo {collaboration} {LIGO Scientific,
  Virgo}),\ }\bibfield  {title} {\enquote {\bibinfo {title} {{GWTC-1: A
  Gravitational-Wave Transient Catalog of Compact Binary Mergers Observed by
  LIGO and Virgo during the First and Second Observing Runs}},}\ }\href
  {\doibase 10.1103/PhysRevX.9.031040} {\bibfield  {journal} {\bibinfo
  {journal} {Phys. Rev. X}\ }\textbf {\bibinfo {volume} {9}},\ \bibinfo {pages}
  {031040} (\bibinfo {year} {2019}{\natexlab{a}})},\ \Eprint
  {http://arxiv.org/abs/1811.12907} {arXiv:1811.12907 [astro-ph.HE]}
  \BibitemShut {NoStop}%
\bibitem [{\citenamefont {Abbott}\ \emph
  {et~al.}(2020{\natexlab{a}})\citenamefont {Abbott} \emph
  {et~al.}}]{Abbott:2020niy}%
  \BibitemOpen
  \bibfield  {author} {\bibinfo {author} {\bibfnamefont {R.}~\bibnamefont
  {Abbott}} \emph {et~al.} (\bibinfo {collaboration} {LIGO Scientific,
  Virgo}),\ }\bibfield  {title} {\enquote {\bibinfo {title} {{GWTC-2: Compact
  Binary Coalescences Observed by LIGO and Virgo During the First Half of the
  Third Observing Run}},}\ }\href@noop {} {\  (\bibinfo {year}
  {2020}{\natexlab{a}})},\ \Eprint {http://arxiv.org/abs/2010.14527}
  {arXiv:2010.14527 [gr-qc]} \BibitemShut {NoStop}%
\bibitem [{\citenamefont {Abbott}\ \emph
  {et~al.}(2020{\natexlab{b}})\citenamefont {Abbott} \emph
  {et~al.}}]{Abbott:2020gyp}%
  \BibitemOpen
  \bibfield  {author} {\bibinfo {author} {\bibfnamefont {R.}~\bibnamefont
  {Abbott}} \emph {et~al.} (\bibinfo {collaboration} {LIGO Scientific,
  Virgo}),\ }\bibfield  {title} {\enquote {\bibinfo {title} {{Population
  Properties of Compact Objects from the Second LIGO-Virgo Gravitational-Wave
  Transient Catalog}},}\ }\href@noop {} {\  (\bibinfo {year}
  {2020}{\natexlab{b}})},\ \Eprint {http://arxiv.org/abs/2010.14533}
  {arXiv:2010.14533 [astro-ph.HE]} \BibitemShut {NoStop}%
\bibitem [{\citenamefont {Abbott}\ \emph
  {et~al.}(2020{\natexlab{c}})\citenamefont {Abbott} \emph
  {et~al.}}]{Abbott:2020jks}%
  \BibitemOpen
  \bibfield  {author} {\bibinfo {author} {\bibfnamefont {R.}~\bibnamefont
  {Abbott}} \emph {et~al.} (\bibinfo {collaboration} {LIGO Scientific,
  Virgo}),\ }\bibfield  {title} {\enquote {\bibinfo {title} {{Tests of General
  Relativity with Binary Black Holes from the second LIGO-Virgo
  Gravitational-Wave Transient Catalog}},}\ }\href@noop {} {\  (\bibinfo {year}
  {2020}{\natexlab{c}})},\ \Eprint {http://arxiv.org/abs/2010.14529}
  {arXiv:2010.14529 [gr-qc]} \BibitemShut {NoStop}%
\bibitem [{\citenamefont {Abbott}\ \emph {et~al.}(2021)\citenamefont {Abbott}
  \emph {et~al.}}]{Abbott:2019yzh}%
  \BibitemOpen
  \bibfield  {author} {\bibinfo {author} {\bibfnamefont {B.~P.}\ \bibnamefont
  {Abbott}} \emph {et~al.} (\bibinfo {collaboration} {LIGO Scientific,
  Virgo}),\ }\bibfield  {title} {\enquote {\bibinfo {title} {{A
  Gravitational-wave Measurement of the Hubble Constant Following the Second
  Observing Run of Advanced LIGO and Virgo}},}\ }\href {\doibase
  10.3847/1538-4357/abdcb7} {\bibfield  {journal} {\bibinfo  {journal}
  {Astrophys. J.}\ }\textbf {\bibinfo {volume} {909}},\ \bibinfo {pages} {218}
  (\bibinfo {year} {2021})},\ \Eprint {http://arxiv.org/abs/1908.06060}
  {arXiv:1908.06060 [astro-ph.CO]} \BibitemShut {NoStop}%
\bibitem [{\citenamefont {Abbott}\ \emph {et~al.}(2018)\citenamefont {Abbott}
  \emph {et~al.}}]{Abbott:2018exr}%
  \BibitemOpen
  \bibfield  {author} {\bibinfo {author} {\bibfnamefont {B.~P.}\ \bibnamefont
  {Abbott}} \emph {et~al.} (\bibinfo {collaboration} {LIGO Scientific,
  Virgo}),\ }\bibfield  {title} {\enquote {\bibinfo {title} {{GW170817:
  Measurements of neutron star radii and equation of state}},}\ }\href
  {\doibase 10.1103/PhysRevLett.121.161101} {\bibfield  {journal} {\bibinfo
  {journal} {Phys. Rev. Lett.}\ }\textbf {\bibinfo {volume} {121}},\ \bibinfo
  {pages} {161101} (\bibinfo {year} {2018})},\ \Eprint
  {http://arxiv.org/abs/1805.11581} {arXiv:1805.11581 [gr-qc]} \BibitemShut
  {NoStop}%
\bibitem [{\citenamefont {Aasi}\ \emph {et~al.}(2015)\citenamefont {Aasi} \emph
  {et~al.}}]{TheLIGOScientific:2014jea}%
  \BibitemOpen
  \bibfield  {author} {\bibinfo {author} {\bibfnamefont {J.}~\bibnamefont
  {Aasi}} \emph {et~al.} (\bibinfo {collaboration} {LIGO Scientific}),\
  }\bibfield  {title} {\enquote {\bibinfo {title} {{Advanced LIGO}},}\ }\href
  {\doibase 10.1088/0264-9381/32/7/074001} {\bibfield  {journal} {\bibinfo
  {journal} {Class. Quant. Grav.}\ }\textbf {\bibinfo {volume} {32}},\ \bibinfo
  {pages} {074001} (\bibinfo {year} {2015})},\ \Eprint
  {http://arxiv.org/abs/1411.4547} {arXiv:1411.4547 [gr-qc]} \BibitemShut
  {NoStop}%
\bibitem [{\citenamefont {Acernese}\ \emph {et~al.}(2015)\citenamefont
  {Acernese} \emph {et~al.}}]{TheVirgo:2014hva}%
  \BibitemOpen
  \bibfield  {author} {\bibinfo {author} {\bibfnamefont {F.}~\bibnamefont
  {Acernese}} \emph {et~al.} (\bibinfo {collaboration} {VIRGO}),\ }\bibfield
  {title} {\enquote {\bibinfo {title} {{Advanced Virgo: a second-generation
  interferometric gravitational wave detector}},}\ }\href {\doibase
  10.1088/0264-9381/32/2/024001} {\bibfield  {journal} {\bibinfo  {journal}
  {Class. Quant. Grav.}\ }\textbf {\bibinfo {volume} {32}},\ \bibinfo {pages}
  {024001} (\bibinfo {year} {2015})},\ \Eprint {http://arxiv.org/abs/1408.3978}
  {arXiv:1408.3978 [gr-qc]} \BibitemShut {NoStop}%
\bibitem [{\citenamefont {Aso}\ \emph {et~al.}(2013)\citenamefont {Aso},
  \citenamefont {Michimura}, \citenamefont {Somiya}, \citenamefont {Ando},
  \citenamefont {Miyakawa}, \citenamefont {Sekiguchi}, \citenamefont
  {Tatsumi},\ and\ \citenamefont {Yamamoto}}]{Aso:2013eba}%
  \BibitemOpen
  \bibfield  {author} {\bibinfo {author} {\bibfnamefont {Yoichi}\ \bibnamefont
  {Aso}}, \bibinfo {author} {\bibfnamefont {Yuta}\ \bibnamefont {Michimura}},
  \bibinfo {author} {\bibfnamefont {Kentaro}\ \bibnamefont {Somiya}}, \bibinfo
  {author} {\bibfnamefont {Masaki}\ \bibnamefont {Ando}}, \bibinfo {author}
  {\bibfnamefont {Osamu}\ \bibnamefont {Miyakawa}}, \bibinfo {author}
  {\bibfnamefont {Takanori}\ \bibnamefont {Sekiguchi}}, \bibinfo {author}
  {\bibfnamefont {Daisuke}\ \bibnamefont {Tatsumi}}, \ and\ \bibinfo {author}
  {\bibfnamefont {Hiroaki}\ \bibnamefont {Yamamoto}} (\bibinfo {collaboration}
  {KAGRA}),\ }\bibfield  {title} {\enquote {\bibinfo {title} {{Interferometer
  design of the KAGRA gravitational wave detector}},}\ }\href {\doibase
  10.1103/PhysRevD.88.043007} {\bibfield  {journal} {\bibinfo  {journal} {Phys.
  Rev. D}\ }\textbf {\bibinfo {volume} {88}},\ \bibinfo {pages} {043007}
  (\bibinfo {year} {2013})},\ \Eprint {http://arxiv.org/abs/1306.6747}
  {arXiv:1306.6747 [gr-qc]} \BibitemShut {NoStop}%
\bibitem [{\citenamefont {{P\"urrer, Michael and Haster,
  Carl-Johan}}(2020)}]{Purrer:2019jcp}%
  \BibitemOpen
  \bibfield  {author} {\bibinfo {author} {\bibnamefont {{P\"urrer, Michael and
  Haster, Carl-Johan}}},\ }\bibfield  {title} {\enquote {\bibinfo {title}
  {{Gravitational waveform accuracy requirements for future ground-based
  detectors}},}\ }\href {\doibase 10.1103/PhysRevResearch.2.023151} {\bibfield
  {journal} {\bibinfo  {journal} {Phys. Rev. Res.}\ }\textbf {\bibinfo {volume}
  {2}},\ \bibinfo {pages} {023151} (\bibinfo {year} {2020})},\ \Eprint
  {http://arxiv.org/abs/1912.10055} {arXiv:1912.10055 [gr-qc]} \BibitemShut
  {NoStop}%
\bibitem [{\citenamefont {Buonanno}\ and\ \citenamefont
  {Damour}(1999)}]{Buonanno:1998gg}%
  \BibitemOpen
  \bibfield  {author} {\bibinfo {author} {\bibfnamefont {A.}~\bibnamefont
  {Buonanno}}\ and\ \bibinfo {author} {\bibfnamefont {T.}~\bibnamefont
  {Damour}},\ }\bibfield  {title} {\enquote {\bibinfo {title} {{Effective
  one-body approach to general relativistic two-body dynamics}},}\ }\href
  {\doibase 10.1103/PhysRevD.59.084006} {\bibfield  {journal} {\bibinfo
  {journal} {Phys. Rev. D}\ }\textbf {\bibinfo {volume} {59}},\ \bibinfo
  {pages} {084006} (\bibinfo {year} {1999})},\ \Eprint
  {http://arxiv.org/abs/gr-qc/9811091} {arXiv:gr-qc/9811091} \BibitemShut
  {NoStop}%
\bibitem [{\citenamefont {Ajith}\ \emph {et~al.}(2007)\citenamefont {Ajith}
  \emph {et~al.}}]{Ajith:2007qp}%
  \BibitemOpen
  \bibfield  {author} {\bibinfo {author} {\bibfnamefont {Parameswaran}\
  \bibnamefont {Ajith}} \emph {et~al.},\ }\bibfield  {title} {\enquote
  {\bibinfo {title} {{Phenomenological template family for black-hole
  coalescence waveforms}},}\ }\href {\doibase 10.1088/0264-9381/24/19/S31}
  {\bibfield  {journal} {\bibinfo  {journal} {Class. Quant. Grav.}\ }\textbf
  {\bibinfo {volume} {24}},\ \bibinfo {pages} {S689--S700} (\bibinfo {year}
  {2007})},\ \Eprint {http://arxiv.org/abs/0704.3764} {arXiv:0704.3764 [gr-qc]}
  \BibitemShut {NoStop}%
\bibitem [{\citenamefont {van~de Meent}\ and\ \citenamefont
  {Pfeiffer}(2020)}]{vandeMeent:2020xgc}%
  \BibitemOpen
  \bibfield  {author} {\bibinfo {author} {\bibfnamefont {Maarten}\ \bibnamefont
  {van~de Meent}}\ and\ \bibinfo {author} {\bibfnamefont {Harald~P.}\
  \bibnamefont {Pfeiffer}},\ }\bibfield  {title} {\enquote {\bibinfo {title}
  {{Intermediate mass-ratio black hole binaries: Applicability of small
  mass-ratio perturbation theory}},}\ }\href {\doibase
  10.1103/PhysRevLett.125.181101} {\bibfield  {journal} {\bibinfo  {journal}
  {Phys. Rev. Lett.}\ }\textbf {\bibinfo {volume} {125}},\ \bibinfo {pages}
  {181101} (\bibinfo {year} {2020})},\ \Eprint
  {http://arxiv.org/abs/2006.12036} {arXiv:2006.12036 [gr-qc]} \BibitemShut
  {NoStop}%
\bibitem [{\citenamefont {Ramos-Buades}\ \emph {et~al.}(2020)\citenamefont
  {Ramos-Buades}, \citenamefont {Husa}, \citenamefont {Pratten}, \citenamefont
  {Estell\'es}, \citenamefont {Garc\'\i{}a-Quir\'os}, \citenamefont
  {Mateu-Lucena}, \citenamefont {Colleoni},\ and\ \citenamefont
  {Jaume}}]{Ramos-Buades:2019uvh}%
  \BibitemOpen
  \bibfield  {author} {\bibinfo {author} {\bibfnamefont {Antoni}\ \bibnamefont
  {Ramos-Buades}}, \bibinfo {author} {\bibfnamefont {Sascha}\ \bibnamefont
  {Husa}}, \bibinfo {author} {\bibfnamefont {Geraint}\ \bibnamefont {Pratten}},
  \bibinfo {author} {\bibfnamefont {H\'ector}\ \bibnamefont {Estell\'es}},
  \bibinfo {author} {\bibfnamefont {Cecilio}\ \bibnamefont
  {Garc\'\i{}a-Quir\'os}}, \bibinfo {author} {\bibfnamefont {Maite}\
  \bibnamefont {Mateu-Lucena}}, \bibinfo {author} {\bibfnamefont {Marta}\
  \bibnamefont {Colleoni}}, \ and\ \bibinfo {author} {\bibfnamefont {Rafel}\
  \bibnamefont {Jaume}},\ }\bibfield  {title} {\enquote {\bibinfo {title}
  {{First survey of spinning eccentric black hole mergers: Numerical relativity
  simulations, hybrid waveforms, and parameter estimation}},}\ }\href {\doibase
  10.1103/PhysRevD.101.083015} {\bibfield  {journal} {\bibinfo  {journal}
  {Phys. Rev. D}\ }\textbf {\bibinfo {volume} {101}},\ \bibinfo {pages}
  {083015} (\bibinfo {year} {2020})},\ \Eprint
  {http://arxiv.org/abs/1909.11011} {arXiv:1909.11011 [gr-qc]} \BibitemShut
  {NoStop}%
\bibitem [{\citenamefont {Chiaramello}\ and\ \citenamefont
  {Nagar}(2020)}]{Chiaramello:2020ehz}%
  \BibitemOpen
  \bibfield  {author} {\bibinfo {author} {\bibfnamefont {Danilo}\ \bibnamefont
  {Chiaramello}}\ and\ \bibinfo {author} {\bibfnamefont {Alessandro}\
  \bibnamefont {Nagar}},\ }\bibfield  {title} {\enquote {\bibinfo {title}
  {{Faithful analytical effective-one-body waveform model for spin-aligned,
  moderately eccentric, coalescing black hole binaries}},}\ }\href {\doibase
  10.1103/PhysRevD.101.101501} {\bibfield  {journal} {\bibinfo  {journal}
  {Phys. Rev. D}\ }\textbf {\bibinfo {volume} {101}},\ \bibinfo {pages}
  {101501} (\bibinfo {year} {2020})},\ \Eprint
  {http://arxiv.org/abs/2001.11736} {arXiv:2001.11736 [gr-qc]} \BibitemShut
  {NoStop}%
\bibitem [{\citenamefont {Nagar}\ \emph {et~al.}(2021)\citenamefont {Nagar},
  \citenamefont {Bonino},\ and\ \citenamefont {Rettegno}}]{Nagar:2021gss}%
  \BibitemOpen
  \bibfield  {author} {\bibinfo {author} {\bibfnamefont {Alessandro}\
  \bibnamefont {Nagar}}, \bibinfo {author} {\bibfnamefont {Alice}\ \bibnamefont
  {Bonino}}, \ and\ \bibinfo {author} {\bibfnamefont {Piero}\ \bibnamefont
  {Rettegno}},\ }\bibfield  {title} {\enquote {\bibinfo {title} {{Effective
  one-body multipolar waveform model for spin-aligned, quasicircular,
  eccentric, hyperbolic black hole binaries}},}\ }\href {\doibase
  10.1103/PhysRevD.103.104021} {\bibfield  {journal} {\bibinfo  {journal}
  {Phys. Rev. D}\ }\textbf {\bibinfo {volume} {103}},\ \bibinfo {pages}
  {104021} (\bibinfo {year} {2021})},\ \Eprint
  {http://arxiv.org/abs/2101.08624} {arXiv:2101.08624 [gr-qc]} \BibitemShut
  {NoStop}%
\bibitem [{\citenamefont {Liu}\ \emph {et~al.}(2021{\natexlab{a}})\citenamefont
  {Liu}, \citenamefont {Cao},\ and\ \citenamefont {Zhu}}]{Liu:2021pkr}%
  \BibitemOpen
  \bibfield  {author} {\bibinfo {author} {\bibfnamefont {Xiaolin}\ \bibnamefont
  {Liu}}, \bibinfo {author} {\bibfnamefont {Zhoujian}\ \bibnamefont {Cao}}, \
  and\ \bibinfo {author} {\bibfnamefont {Zong-Hong}\ \bibnamefont {Zhu}},\
  }\bibfield  {title} {\enquote {\bibinfo {title} {{A higher-multipole
  gravitational waveform model for an eccentric binary black holes based on the
  effective-one-body-numerical-relativity formalism}},}\ }\href@noop {} {\
  (\bibinfo {year} {2021}{\natexlab{a}})},\ \Eprint
  {http://arxiv.org/abs/2102.08614} {arXiv:2102.08614 [gr-qc]} \BibitemShut
  {NoStop}%
\bibitem [{\citenamefont {Khalil}\ \emph {et~al.}(2021)\citenamefont {Khalil},
  \citenamefont {Buonanno}, \citenamefont {Steinhoff},\ and\ \citenamefont
  {Vines}}]{Khalil:2021txt}%
  \BibitemOpen
  \bibfield  {author} {\bibinfo {author} {\bibfnamefont {Mohammed}\
  \bibnamefont {Khalil}}, \bibinfo {author} {\bibfnamefont {Alessandra}\
  \bibnamefont {Buonanno}}, \bibinfo {author} {\bibfnamefont {Jan}\
  \bibnamefont {Steinhoff}}, \ and\ \bibinfo {author} {\bibfnamefont {Justin}\
  \bibnamefont {Vines}},\ }\bibfield  {title} {\enquote {\bibinfo {title}
  {{Radiation-reaction force and multipolar waveforms for eccentric,
  spin-aligned binaries in the effective-one-body formalism}},}\ }\href@noop {}
  {\  (\bibinfo {year} {2021})},\ \Eprint {http://arxiv.org/abs/2104.11705}
  {arXiv:2104.11705 [gr-qc]} \BibitemShut {NoStop}%
\bibitem [{\citenamefont {Hinderer}\ and\ \citenamefont
  {Babak}(2017)}]{Hinderer:2017jcs}%
  \BibitemOpen
  \bibfield  {author} {\bibinfo {author} {\bibfnamefont {Tanja}\ \bibnamefont
  {Hinderer}}\ and\ \bibinfo {author} {\bibfnamefont {Stanislav}\ \bibnamefont
  {Babak}},\ }\bibfield  {title} {\enquote {\bibinfo {title} {{Foundations of
  an effective-one-body model for coalescing binaries on eccentric orbits}},}\
  }\href {\doibase 10.1103/PhysRevD.96.104048} {\bibfield  {journal} {\bibinfo
  {journal} {Phys. Rev. D}\ }\textbf {\bibinfo {volume} {96}},\ \bibinfo
  {pages} {104048} (\bibinfo {year} {2017})},\ \Eprint
  {http://arxiv.org/abs/1707.08426} {arXiv:1707.08426 [gr-qc]} \BibitemShut
  {NoStop}%
\bibitem [{\citenamefont {Islam}\ \emph {et~al.}(2021)\citenamefont {Islam},
  \citenamefont {Varma}, \citenamefont {Lodman}, \citenamefont {Field},
  \citenamefont {Khanna}, \citenamefont {Scheel}, \citenamefont {Pfeiffer},
  \citenamefont {Gerosa},\ and\ \citenamefont {Kidder}}]{Islam:2021mha}%
  \BibitemOpen
  \bibfield  {author} {\bibinfo {author} {\bibfnamefont {Tousif}\ \bibnamefont
  {Islam}}, \bibinfo {author} {\bibfnamefont {Vijay}\ \bibnamefont {Varma}},
  \bibinfo {author} {\bibfnamefont {Jackie}\ \bibnamefont {Lodman}}, \bibinfo
  {author} {\bibfnamefont {Scott~E.}\ \bibnamefont {Field}}, \bibinfo {author}
  {\bibfnamefont {Gaurav}\ \bibnamefont {Khanna}}, \bibinfo {author}
  {\bibfnamefont {Mark~A.}\ \bibnamefont {Scheel}}, \bibinfo {author}
  {\bibfnamefont {Harald~P.}\ \bibnamefont {Pfeiffer}}, \bibinfo {author}
  {\bibfnamefont {Davide}\ \bibnamefont {Gerosa}}, \ and\ \bibinfo {author}
  {\bibfnamefont {Lawrence~E.}\ \bibnamefont {Kidder}},\ }\bibfield  {title}
  {\enquote {\bibinfo {title} {{Eccentric binary black hole surrogate models
  for the gravitational waveform and remnant properties: comparable mass,
  nonspinning case}},}\ }\href {\doibase 10.1103/PhysRevD.103.064022}
  {\bibfield  {journal} {\bibinfo  {journal} {Phys. Rev. D}\ }\textbf {\bibinfo
  {volume} {103}},\ \bibinfo {pages} {064022} (\bibinfo {year} {2021})},\
  \Eprint {http://arxiv.org/abs/2101.11798} {arXiv:2101.11798 [gr-qc]}
  \BibitemShut {NoStop}%
\bibitem [{\citenamefont {Samsing}(2018)}]{Samsing:2017xmd}%
  \BibitemOpen
  \bibfield  {author} {\bibinfo {author} {\bibfnamefont {Johan}\ \bibnamefont
  {Samsing}},\ }\bibfield  {title} {\enquote {\bibinfo {title} {{Eccentric
  Black Hole Mergers Forming in Globular Clusters}},}\ }\href {\doibase
  10.1103/PhysRevD.97.103014} {\bibfield  {journal} {\bibinfo  {journal} {Phys.
  Rev. D}\ }\textbf {\bibinfo {volume} {97}},\ \bibinfo {pages} {103014}
  (\bibinfo {year} {2018})},\ \Eprint {http://arxiv.org/abs/1711.07452}
  {arXiv:1711.07452 [astro-ph.HE]} \BibitemShut {NoStop}%
\bibitem [{\citenamefont {Rodriguez}\ \emph {et~al.}(2018)\citenamefont
  {Rodriguez}, \citenamefont {Amaro-Seoane}, \citenamefont {Chatterjee},\ and\
  \citenamefont {Rasio}}]{Rodriguez:2017pec}%
  \BibitemOpen
  \bibfield  {author} {\bibinfo {author} {\bibfnamefont {Carl~L.}\ \bibnamefont
  {Rodriguez}}, \bibinfo {author} {\bibfnamefont {Pau}\ \bibnamefont
  {Amaro-Seoane}}, \bibinfo {author} {\bibfnamefont {Sourav}\ \bibnamefont
  {Chatterjee}}, \ and\ \bibinfo {author} {\bibfnamefont {Frederic~A.}\
  \bibnamefont {Rasio}},\ }\bibfield  {title} {\enquote {\bibinfo {title}
  {{Post-Newtonian Dynamics in Dense Star Clusters: Highly-Eccentric,
  Highly-Spinning, and Repeated Binary Black Hole Mergers}},}\ }\href {\doibase
  10.1103/PhysRevLett.120.151101} {\bibfield  {journal} {\bibinfo  {journal}
  {Phys. Rev. Lett.}\ }\textbf {\bibinfo {volume} {120}},\ \bibinfo {pages}
  {151101} (\bibinfo {year} {2018})},\ \Eprint
  {http://arxiv.org/abs/1712.04937} {arXiv:1712.04937 [astro-ph.HE]}
  \BibitemShut {NoStop}%
\bibitem [{\citenamefont {Gond\'an}\ and\ \citenamefont
  {Kocsis}(2020)}]{Gondan:2020svr}%
  \BibitemOpen
  \bibfield  {author} {\bibinfo {author} {\bibfnamefont {L\'aszl\'o}\
  \bibnamefont {Gond\'an}}\ and\ \bibinfo {author} {\bibfnamefont {Bence}\
  \bibnamefont {Kocsis}},\ }\bibfield  {title} {\enquote {\bibinfo {title}
  {{High Eccentricities and High Masses Characterize Gravitational-wave
  Captures in Galactic Nuclei as Seen by Earth-based Detectors}},}\ }\href@noop
  {} {\  (\bibinfo {year} {2020})},\ \Eprint {http://arxiv.org/abs/2011.02507}
  {arXiv:2011.02507 [astro-ph.HE]} \BibitemShut {NoStop}%
\bibitem [{\citenamefont {Abbott}\ \emph
  {et~al.}(2019{\natexlab{b}})\citenamefont {Abbott} \emph
  {et~al.}}]{LIGOScientific:2018jsj}%
  \BibitemOpen
  \bibfield  {author} {\bibinfo {author} {\bibfnamefont {B.~P.}\ \bibnamefont
  {Abbott}} \emph {et~al.} (\bibinfo {collaboration} {LIGO Scientific,
  Virgo}),\ }\bibfield  {title} {\enquote {\bibinfo {title} {{Binary Black Hole
  Population Properties Inferred from the First and Second Observing Runs of
  Advanced LIGO and Advanced Virgo}},}\ }\href {\doibase
  10.3847/2041-8213/ab3800} {\bibfield  {journal} {\bibinfo  {journal}
  {Astrophys. J. Lett.}\ }\textbf {\bibinfo {volume} {882}},\ \bibinfo {pages}
  {L24} (\bibinfo {year} {2019}{\natexlab{b}})},\ \Eprint
  {http://arxiv.org/abs/1811.12940} {arXiv:1811.12940 [astro-ph.HE]}
  \BibitemShut {NoStop}%
\bibitem [{\citenamefont {Kocsis}\ \emph {et~al.}(2006)\citenamefont {Kocsis},
  \citenamefont {Gaspar},\ and\ \citenamefont {Marka}}]{Kocsis:2006hq}%
  \BibitemOpen
  \bibfield  {author} {\bibinfo {author} {\bibfnamefont {Bence}\ \bibnamefont
  {Kocsis}}, \bibinfo {author} {\bibfnamefont {Merse~E.}\ \bibnamefont
  {Gaspar}}, \ and\ \bibinfo {author} {\bibfnamefont {Szabolcs}\ \bibnamefont
  {Marka}},\ }\bibfield  {title} {\enquote {\bibinfo {title} {{Detection rate
  estimates of gravity-waves emitted during parabolic encounters of stellar
  black holes in globular clusters}},}\ }\href {\doibase 10.1086/505641}
  {\bibfield  {journal} {\bibinfo  {journal} {Astrophys. J.}\ }\textbf
  {\bibinfo {volume} {648}},\ \bibinfo {pages} {411--429} (\bibinfo {year}
  {2006})},\ \Eprint {http://arxiv.org/abs/astro-ph/0603441}
  {arXiv:astro-ph/0603441} \BibitemShut {NoStop}%
\bibitem [{\citenamefont {Mukherjee}\ \emph {et~al.}(2020)\citenamefont
  {Mukherjee}, \citenamefont {Mitra},\ and\ \citenamefont
  {Chatterjee}}]{Mukherjee:2020hnm}%
  \BibitemOpen
  \bibfield  {author} {\bibinfo {author} {\bibfnamefont {Sajal}\ \bibnamefont
  {Mukherjee}}, \bibinfo {author} {\bibfnamefont {Sanjit}\ \bibnamefont
  {Mitra}}, \ and\ \bibinfo {author} {\bibfnamefont {Sourav}\ \bibnamefont
  {Chatterjee}},\ }\bibfield  {title} {\enquote {\bibinfo {title}
  {{Detectability of hyperbolic encounters of compact stars with ground-based
  gravitational waves detectors}},}\ }\href@noop {} {\  (\bibinfo {year}
  {2020})},\ \Eprint {http://arxiv.org/abs/2010.00916} {arXiv:2010.00916
  [gr-qc]} \BibitemShut {NoStop}%
\bibitem [{\citenamefont {Zevin}\ \emph {et~al.}(2019)\citenamefont {Zevin},
  \citenamefont {Samsing}, \citenamefont {Rodriguez}, \citenamefont {Haster},\
  and\ \citenamefont {Ramirez-Ruiz}}]{Zevin:2018kzq}%
  \BibitemOpen
  \bibfield  {author} {\bibinfo {author} {\bibfnamefont {Michael}\ \bibnamefont
  {Zevin}}, \bibinfo {author} {\bibfnamefont {Johan}\ \bibnamefont {Samsing}},
  \bibinfo {author} {\bibfnamefont {Carl}\ \bibnamefont {Rodriguez}}, \bibinfo
  {author} {\bibfnamefont {Carl-Johan}\ \bibnamefont {Haster}}, \ and\ \bibinfo
  {author} {\bibfnamefont {Enrico}\ \bibnamefont {Ramirez-Ruiz}},\ }\bibfield
  {title} {\enquote {\bibinfo {title} {{Eccentric Black Hole Mergers in Dense
  Star Clusters: The Role of Binary\textendash{}Binary Encounters}},}\ }\href
  {\doibase 10.3847/1538-4357/aaf6ec} {\bibfield  {journal} {\bibinfo
  {journal} {Astrophys. J.}\ }\textbf {\bibinfo {volume} {871}},\ \bibinfo
  {pages} {91} (\bibinfo {year} {2019})},\ \Eprint
  {http://arxiv.org/abs/1810.00901} {arXiv:1810.00901 [astro-ph.HE]}
  \BibitemShut {NoStop}%
\bibitem [{\citenamefont {Gamba}\ \emph {et~al.}(2021)\citenamefont {Gamba},
  \citenamefont {Breschi}, \citenamefont {Carullo}, \citenamefont {Rettegno},
  \citenamefont {Albanesi}, \citenamefont {Bernuzzi},\ and\ \citenamefont
  {Nagar}}]{Gamba:2021gap}%
  \BibitemOpen
  \bibfield  {author} {\bibinfo {author} {\bibfnamefont {Rossella}\
  \bibnamefont {Gamba}}, \bibinfo {author} {\bibfnamefont {Matteo}\
  \bibnamefont {Breschi}}, \bibinfo {author} {\bibfnamefont {Gregorio}\
  \bibnamefont {Carullo}}, \bibinfo {author} {\bibfnamefont {Piero}\
  \bibnamefont {Rettegno}}, \bibinfo {author} {\bibfnamefont {Simone}\
  \bibnamefont {Albanesi}}, \bibinfo {author} {\bibfnamefont {Sebastiano}\
  \bibnamefont {Bernuzzi}}, \ and\ \bibinfo {author} {\bibfnamefont
  {Alessandro}\ \bibnamefont {Nagar}},\ }\bibfield  {title} {\enquote {\bibinfo
  {title} {{GW190521: A dynamical capture of two black holes}},}\ }\href@noop
  {} {\  (\bibinfo {year} {2021})},\ \Eprint {http://arxiv.org/abs/2106.05575}
  {arXiv:2106.05575 [gr-qc]} \BibitemShut {NoStop}%
\bibitem [{\citenamefont {Zackay}\ \emph {et~al.}(2019)\citenamefont {Zackay},
  \citenamefont {Venumadhav}, \citenamefont {Dai}, \citenamefont {Roulet},\
  and\ \citenamefont {Zaldarriaga}}]{Zackay:2019tzo}%
  \BibitemOpen
  \bibfield  {author} {\bibinfo {author} {\bibfnamefont {Barak}\ \bibnamefont
  {Zackay}}, \bibinfo {author} {\bibfnamefont {Tejaswi}\ \bibnamefont
  {Venumadhav}}, \bibinfo {author} {\bibfnamefont {Liang}\ \bibnamefont {Dai}},
  \bibinfo {author} {\bibfnamefont {Javier}\ \bibnamefont {Roulet}}, \ and\
  \bibinfo {author} {\bibfnamefont {Matias}\ \bibnamefont {Zaldarriaga}},\
  }\bibfield  {title} {\enquote {\bibinfo {title} {{Highly spinning and aligned
  binary black hole merger in the Advanced LIGO first observing run}},}\ }\href
  {\doibase 10.1103/PhysRevD.100.023007} {\bibfield  {journal} {\bibinfo
  {journal} {Phys. Rev. D}\ }\textbf {\bibinfo {volume} {100}},\ \bibinfo
  {pages} {023007} (\bibinfo {year} {2019})},\ \Eprint
  {http://arxiv.org/abs/1902.10331} {arXiv:1902.10331 [astro-ph.HE]}
  \BibitemShut {NoStop}%
\bibitem [{\citenamefont {Huang}\ \emph {et~al.}(2020)\citenamefont {Huang},
  \citenamefont {Haster}, \citenamefont {Vitale}, \citenamefont {Zimmerman},
  \citenamefont {Roulet}, \citenamefont {Venumadhav}, \citenamefont {Zackay},
  \citenamefont {Dai},\ and\ \citenamefont {Zaldarriaga}}]{Huang:2020ysn}%
  \BibitemOpen
  \bibfield  {author} {\bibinfo {author} {\bibfnamefont {Yiwen}\ \bibnamefont
  {Huang}}, \bibinfo {author} {\bibfnamefont {Carl-Johan}\ \bibnamefont
  {Haster}}, \bibinfo {author} {\bibfnamefont {Salvatore}\ \bibnamefont
  {Vitale}}, \bibinfo {author} {\bibfnamefont {Aaron}\ \bibnamefont
  {Zimmerman}}, \bibinfo {author} {\bibfnamefont {Javier}\ \bibnamefont
  {Roulet}}, \bibinfo {author} {\bibfnamefont {Tejaswi}\ \bibnamefont
  {Venumadhav}}, \bibinfo {author} {\bibfnamefont {Barak}\ \bibnamefont
  {Zackay}}, \bibinfo {author} {\bibfnamefont {Liang}\ \bibnamefont {Dai}}, \
  and\ \bibinfo {author} {\bibfnamefont {Matias}\ \bibnamefont {Zaldarriaga}},\
  }\bibfield  {title} {\enquote {\bibinfo {title} {{Source properties of the
  lowest signal-to-noise-ratio binary black hole detections}},}\ }\href
  {\doibase 10.1103/PhysRevD.102.103024} {\bibfield  {journal} {\bibinfo
  {journal} {Phys. Rev. D}\ }\textbf {\bibinfo {volume} {102}},\ \bibinfo
  {pages} {103024} (\bibinfo {year} {2020})},\ \Eprint
  {http://arxiv.org/abs/2003.04513} {arXiv:2003.04513 [gr-qc]} \BibitemShut
  {NoStop}%
\bibitem [{\citenamefont {{Thorne}}\ and\ \citenamefont
  {{Kovacs}}(1975)}]{1975ApJ...200..245T}%
  \BibitemOpen
  \bibfield  {author} {\bibinfo {author} {\bibfnamefont {K.~S.}\ \bibnamefont
  {{Thorne}}}\ and\ \bibinfo {author} {\bibfnamefont {S.~J.}\ \bibnamefont
  {{Kovacs}}},\ }\bibfield  {title} {\enquote {\bibinfo {title} {{The
  generation of gravitational waves. I. Weak-field sources.}}}\ }\href
  {\doibase 10.1086/153783} {\bibfield  {journal} {\bibinfo  {journal} {\apj}\
  }\textbf {\bibinfo {volume} {200}},\ \bibinfo {pages} {245--262} (\bibinfo
  {year} {1975})}\BibitemShut {NoStop}%
\bibitem [{\citenamefont {{Crowley}}\ and\ \citenamefont
  {{Thorne}}(1977)}]{1977ApJ...215..624C}%
  \BibitemOpen
  \bibfield  {author} {\bibinfo {author} {\bibfnamefont {R.~J.}\ \bibnamefont
  {{Crowley}}}\ and\ \bibinfo {author} {\bibfnamefont {K.~S.}\ \bibnamefont
  {{Thorne}}},\ }\bibfield  {title} {\enquote {\bibinfo {title} {{The
  generation of gravitational waves. II. The postlinear formation
  revisited.}}}\ }\href {\doibase 10.1086/155397} {\bibfield  {journal}
  {\bibinfo  {journal} {\apj}\ }\textbf {\bibinfo {volume} {215}},\ \bibinfo
  {pages} {624--635} (\bibinfo {year} {1977})}\BibitemShut {NoStop}%
\bibitem [{\citenamefont {Kovacs}\ and\ \citenamefont
  {Thorne}(1977)}]{Kovacs:1977uw}%
  \BibitemOpen
  \bibfield  {author} {\bibinfo {author} {\bibfnamefont {S.J.}\ \bibnamefont
  {Kovacs}}\ and\ \bibinfo {author} {\bibfnamefont {K.S.}\ \bibnamefont
  {Thorne}},\ }\bibfield  {title} {\enquote {\bibinfo {title} {{The Generation
  of Gravitational Waves. 3. Derivation of Bremsstrahlung Formulas}},}\ }\href
  {\doibase 10.1086/155576} {\bibfield  {journal} {\bibinfo  {journal}
  {Astrophys. J.}\ }\textbf {\bibinfo {volume} {217}},\ \bibinfo {pages}
  {252--280} (\bibinfo {year} {1977})}\BibitemShut {NoStop}%
\bibitem [{\citenamefont {Kovacs}\ and\ \citenamefont
  {Thorne}(1978)}]{Kovacs:1978eu}%
  \BibitemOpen
  \bibfield  {author} {\bibinfo {author} {\bibfnamefont {S.J.}\ \bibnamefont
  {Kovacs}}\ and\ \bibinfo {author} {\bibfnamefont {K.S.}\ \bibnamefont
  {Thorne}},\ }\bibfield  {title} {\enquote {\bibinfo {title} {{The Generation
  of Gravitational Waves. 4. Bremsstrahlung}},}\ }\href {\doibase
  10.1086/156350} {\bibfield  {journal} {\bibinfo  {journal} {Astrophys. J.}\
  }\textbf {\bibinfo {volume} {224}},\ \bibinfo {pages} {62--85} (\bibinfo
  {year} {1978})}\BibitemShut {NoStop}%
\bibitem [{\citenamefont {Jakobsen}\ \emph
  {et~al.}(2021{\natexlab{a}})\citenamefont {Jakobsen}, \citenamefont {Mogull},
  \citenamefont {Plefka},\ and\ \citenamefont {Steinhoff}}]{Jakobsen:2021smu}%
  \BibitemOpen
  \bibfield  {author} {\bibinfo {author} {\bibfnamefont {Gustav~Uhre}\
  \bibnamefont {Jakobsen}}, \bibinfo {author} {\bibfnamefont {Gustav}\
  \bibnamefont {Mogull}}, \bibinfo {author} {\bibfnamefont {Jan}\ \bibnamefont
  {Plefka}}, \ and\ \bibinfo {author} {\bibfnamefont {Jan}\ \bibnamefont
  {Steinhoff}},\ }\bibfield  {title} {\enquote {\bibinfo {title} {{Classical
  Gravitational Bremsstrahlung from a Worldline Quantum Field Theory}},}\
  }\href {\doibase 10.1103/PhysRevLett.126.201103} {\bibfield  {journal}
  {\bibinfo  {journal} {Phys. Rev. Lett.}\ }\textbf {\bibinfo {volume} {126}},\
  \bibinfo {pages} {201103} (\bibinfo {year} {2021}{\natexlab{a}})},\ \Eprint
  {http://arxiv.org/abs/2101.12688} {arXiv:2101.12688 [gr-qc]} \BibitemShut
  {NoStop}%
\bibitem [{\citenamefont {Mogull}\ \emph {et~al.}(2021)\citenamefont {Mogull},
  \citenamefont {Plefka},\ and\ \citenamefont {Steinhoff}}]{Mogull:2020sak}%
  \BibitemOpen
  \bibfield  {author} {\bibinfo {author} {\bibfnamefont {Gustav}\ \bibnamefont
  {Mogull}}, \bibinfo {author} {\bibfnamefont {Jan}\ \bibnamefont {Plefka}}, \
  and\ \bibinfo {author} {\bibfnamefont {Jan}\ \bibnamefont {Steinhoff}},\
  }\bibfield  {title} {\enquote {\bibinfo {title} {{Classical black hole
  scattering from a worldline quantum field theory}},}\ }\href {\doibase
  10.1007/JHEP02(2021)048} {\bibfield  {journal} {\bibinfo  {journal} {JHEP}\
  }\textbf {\bibinfo {volume} {02}},\ \bibinfo {pages} {048} (\bibinfo {year}
  {2021})},\ \Eprint {http://arxiv.org/abs/2010.02865} {arXiv:2010.02865
  [hep-th]} \BibitemShut {NoStop}%
\bibitem [{\citenamefont {Howe}\ \emph {et~al.}(1988)\citenamefont {Howe},
  \citenamefont {Penati}, \citenamefont {Pernici},\ and\ \citenamefont
  {Townsend}}]{Howe:1988ft}%
  \BibitemOpen
  \bibfield  {author} {\bibinfo {author} {\bibfnamefont {Paul~S.}\ \bibnamefont
  {Howe}}, \bibinfo {author} {\bibfnamefont {Silvia}\ \bibnamefont {Penati}},
  \bibinfo {author} {\bibfnamefont {Mario}\ \bibnamefont {Pernici}}, \ and\
  \bibinfo {author} {\bibfnamefont {Paul~K.}\ \bibnamefont {Townsend}},\
  }\bibfield  {title} {\enquote {\bibinfo {title} {{Wave Equations for
  Arbitrary Spin From Quantization of the Extended Supersymmetric Spinning
  Particle}},}\ }\href {\doibase 10.1016/0370-2693(88)91358-5} {\bibfield
  {journal} {\bibinfo  {journal} {Phys. Lett. B}\ }\textbf {\bibinfo {volume}
  {215}},\ \bibinfo {pages} {555--558} (\bibinfo {year} {1988})}\BibitemShut
  {NoStop}%
\bibitem [{\citenamefont {Gibbons}\ \emph {et~al.}(1993)\citenamefont
  {Gibbons}, \citenamefont {Rietdijk},\ and\ \citenamefont {van
  Holten}}]{Gibbons:1993ap}%
  \BibitemOpen
  \bibfield  {author} {\bibinfo {author} {\bibfnamefont {G.~W.}\ \bibnamefont
  {Gibbons}}, \bibinfo {author} {\bibfnamefont {R.~H.}\ \bibnamefont
  {Rietdijk}}, \ and\ \bibinfo {author} {\bibfnamefont {J.~W.}\ \bibnamefont
  {van Holten}},\ }\bibfield  {title} {\enquote {\bibinfo {title} {{SUSY in the
  sky}},}\ }\href {\doibase 10.1016/0550-3213(93)90472-2} {\bibfield  {journal}
  {\bibinfo  {journal} {Nucl. Phys. B}\ }\textbf {\bibinfo {volume} {404}},\
  \bibinfo {pages} {42--64} (\bibinfo {year} {1993})},\ \Eprint
  {http://arxiv.org/abs/hep-th/9303112} {arXiv:hep-th/9303112} \BibitemShut
  {NoStop}%
\bibitem [{\citenamefont {Bastianelli}\ \emph
  {et~al.}(2005{\natexlab{a}})\citenamefont {Bastianelli}, \citenamefont
  {Benincasa},\ and\ \citenamefont {Giombi}}]{Bastianelli:2005vk}%
  \BibitemOpen
  \bibfield  {author} {\bibinfo {author} {\bibfnamefont {Fiorenzo}\
  \bibnamefont {Bastianelli}}, \bibinfo {author} {\bibfnamefont {Paolo}\
  \bibnamefont {Benincasa}}, \ and\ \bibinfo {author} {\bibfnamefont {Simone}\
  \bibnamefont {Giombi}},\ }\bibfield  {title} {\enquote {\bibinfo {title}
  {{Worldline approach to vector and antisymmetric tensor fields}},}\ }\href
  {\doibase 10.1088/1126-6708/2005/04/010} {\bibfield  {journal} {\bibinfo
  {journal} {JHEP}\ }\textbf {\bibinfo {volume} {04}},\ \bibinfo {pages} {010}
  (\bibinfo {year} {2005}{\natexlab{a}})},\ \Eprint
  {http://arxiv.org/abs/hep-th/0503155} {arXiv:hep-th/0503155} \BibitemShut
  {NoStop}%
\bibitem [{\citenamefont {Bastianelli}\ \emph
  {et~al.}(2005{\natexlab{b}})\citenamefont {Bastianelli}, \citenamefont
  {Benincasa},\ and\ \citenamefont {Giombi}}]{Bastianelli:2005uy}%
  \BibitemOpen
  \bibfield  {author} {\bibinfo {author} {\bibfnamefont {Fiorenzo}\
  \bibnamefont {Bastianelli}}, \bibinfo {author} {\bibfnamefont {Paolo}\
  \bibnamefont {Benincasa}}, \ and\ \bibinfo {author} {\bibfnamefont {Simone}\
  \bibnamefont {Giombi}},\ }\bibfield  {title} {\enquote {\bibinfo {title}
  {{Worldline approach to vector and antisymmetric tensor fields. II.}}}\
  }\href {\doibase 10.1088/1126-6708/2005/10/114} {\bibfield  {journal}
  {\bibinfo  {journal} {JHEP}\ }\textbf {\bibinfo {volume} {10}},\ \bibinfo
  {pages} {114} (\bibinfo {year} {2005}{\natexlab{b}})},\ \Eprint
  {http://arxiv.org/abs/hep-th/0510010} {arXiv:hep-th/0510010} \BibitemShut
  {NoStop}%
\bibitem [{\citenamefont {Porto}(2006)}]{Porto:2005ac}%
  \BibitemOpen
  \bibfield  {author} {\bibinfo {author} {\bibfnamefont {Rafael~A.}\
  \bibnamefont {Porto}},\ }\bibfield  {title} {\enquote {\bibinfo {title}
  {{Post-Newtonian corrections to the motion of spinning bodies in NRGR}},}\
  }\href {\doibase 10.1103/PhysRevD.73.104031} {\bibfield  {journal} {\bibinfo
  {journal} {Phys. Rev. D}\ }\textbf {\bibinfo {volume} {73}},\ \bibinfo
  {pages} {104031} (\bibinfo {year} {2006})},\ \Eprint
  {http://arxiv.org/abs/gr-qc/0511061} {arXiv:gr-qc/0511061} \BibitemShut
  {NoStop}%
\bibitem [{\citenamefont {Levi}\ and\ \citenamefont
  {Steinhoff}(2015)}]{Levi:2015msa}%
  \BibitemOpen
  \bibfield  {author} {\bibinfo {author} {\bibfnamefont {Michele}\ \bibnamefont
  {Levi}}\ and\ \bibinfo {author} {\bibfnamefont {Jan}\ \bibnamefont
  {Steinhoff}},\ }\bibfield  {title} {\enquote {\bibinfo {title} {{Spinning
  gravitating objects in the effective field theory in the post-Newtonian
  scheme}},}\ }\href {\doibase 10.1007/JHEP09(2015)219} {\bibfield  {journal}
  {\bibinfo  {journal} {JHEP}\ }\textbf {\bibinfo {volume} {09}},\ \bibinfo
  {pages} {219} (\bibinfo {year} {2015})},\ \Eprint
  {http://arxiv.org/abs/1501.04956} {arXiv:1501.04956 [gr-qc]} \BibitemShut
  {NoStop}%
\bibitem [{\citenamefont {Goldberger}\ and\ \citenamefont
  {Rothstein}(2006{\natexlab{a}})}]{Goldberger:2004jt}%
  \BibitemOpen
  \bibfield  {author} {\bibinfo {author} {\bibfnamefont {Walter~D.}\
  \bibnamefont {Goldberger}}\ and\ \bibinfo {author} {\bibfnamefont {Ira~Z.}\
  \bibnamefont {Rothstein}},\ }\bibfield  {title} {\enquote {\bibinfo {title}
  {{An Effective field theory of gravity for extended objects}},}\ }\href
  {\doibase 10.1103/PhysRevD.73.104029} {\bibfield  {journal} {\bibinfo
  {journal} {Phys. Rev. D}\ }\textbf {\bibinfo {volume} {73}},\ \bibinfo
  {pages} {104029} (\bibinfo {year} {2006}{\natexlab{a}})},\ \Eprint
  {http://arxiv.org/abs/hep-th/0409156} {arXiv:hep-th/0409156} \BibitemShut
  {NoStop}%
\bibitem [{\citenamefont {Goldberger}\ and\ \citenamefont
  {Rothstein}(2006{\natexlab{b}})}]{Goldberger:2006bd}%
  \BibitemOpen
  \bibfield  {author} {\bibinfo {author} {\bibfnamefont {Walter~D.}\
  \bibnamefont {Goldberger}}\ and\ \bibinfo {author} {\bibfnamefont {Ira~Z.}\
  \bibnamefont {Rothstein}},\ }\bibfield  {title} {\enquote {\bibinfo {title}
  {{Towers of Gravitational Theories}},}\ }\href {\doibase
  10.1142/S0218271806009698} {\bibfield  {journal} {\bibinfo  {journal} {Gen.
  Rel. Grav.}\ }\textbf {\bibinfo {volume} {38}},\ \bibinfo {pages}
  {1537--1546} (\bibinfo {year} {2006}{\natexlab{b}})},\ \Eprint
  {http://arxiv.org/abs/hep-th/0605238} {arXiv:hep-th/0605238} \BibitemShut
  {NoStop}%
\bibitem [{\citenamefont {Goldberger}\ and\ \citenamefont
  {Ross}(2010)}]{Goldberger:2009qd}%
  \BibitemOpen
  \bibfield  {author} {\bibinfo {author} {\bibfnamefont {Walter~D.}\
  \bibnamefont {Goldberger}}\ and\ \bibinfo {author} {\bibfnamefont {Andreas}\
  \bibnamefont {Ross}},\ }\bibfield  {title} {\enquote {\bibinfo {title}
  {{Gravitational radiative corrections from effective field theory}},}\ }\href
  {\doibase 10.1103/PhysRevD.81.124015} {\bibfield  {journal} {\bibinfo
  {journal} {Phys. Rev. D}\ }\textbf {\bibinfo {volume} {81}},\ \bibinfo
  {pages} {124015} (\bibinfo {year} {2010})},\ \Eprint
  {http://arxiv.org/abs/0912.4254} {arXiv:0912.4254 [gr-qc]} \BibitemShut
  {NoStop}%
\bibitem [{\citenamefont {Porto}(2016)}]{Porto:2016pyg}%
  \BibitemOpen
  \bibfield  {author} {\bibinfo {author} {\bibfnamefont {Rafael~A.}\
  \bibnamefont {Porto}},\ }\bibfield  {title} {\enquote {\bibinfo {title} {{The
  effective field theorist\textquoteright{}s approach to gravitational
  dynamics}},}\ }\href {\doibase 10.1016/j.physrep.2016.04.003} {\bibfield
  {journal} {\bibinfo  {journal} {Phys. Rept.}\ }\textbf {\bibinfo {volume}
  {633}},\ \bibinfo {pages} {1--104} (\bibinfo {year} {2016})},\ \Eprint
  {http://arxiv.org/abs/1601.04914} {arXiv:1601.04914 [hep-th]} \BibitemShut
  {NoStop}%
\bibitem [{\citenamefont {Levi}(2020)}]{Levi:2018nxp}%
  \BibitemOpen
  \bibfield  {author} {\bibinfo {author} {\bibfnamefont {Mich\`ele}\
  \bibnamefont {Levi}},\ }\bibfield  {title} {\enquote {\bibinfo {title}
  {{Effective Field Theories of Post-Newtonian Gravity: A comprehensive
  review}},}\ }\href {\doibase 10.1088/1361-6633/ab12bc} {\bibfield  {journal}
  {\bibinfo  {journal} {Rept. Prog. Phys.}\ }\textbf {\bibinfo {volume} {83}},\
  \bibinfo {pages} {075901} (\bibinfo {year} {2020})},\ \Eprint
  {http://arxiv.org/abs/1807.01699} {arXiv:1807.01699 [hep-th]} \BibitemShut
  {NoStop}%
\bibitem [{\citenamefont {Goldberger}\ \emph {et~al.}(2018)\citenamefont
  {Goldberger}, \citenamefont {Li},\ and\ \citenamefont
  {Prabhu}}]{Goldberger:2017ogt}%
  \BibitemOpen
  \bibfield  {author} {\bibinfo {author} {\bibfnamefont {Walter~D.}\
  \bibnamefont {Goldberger}}, \bibinfo {author} {\bibfnamefont {Jingping}\
  \bibnamefont {Li}}, \ and\ \bibinfo {author} {\bibfnamefont {Siddharth~G.}\
  \bibnamefont {Prabhu}},\ }\bibfield  {title} {\enquote {\bibinfo {title}
  {{Spinning particles, axion radiation, and the classical double copy}},}\
  }\href {\doibase 10.1103/PhysRevD.97.105018} {\bibfield  {journal} {\bibinfo
  {journal} {Phys. Rev. D}\ }\textbf {\bibinfo {volume} {97}},\ \bibinfo
  {pages} {105018} (\bibinfo {year} {2018})},\ \Eprint
  {http://arxiv.org/abs/1712.09250} {arXiv:1712.09250 [hep-th]} \BibitemShut
  {NoStop}%
\bibitem [{\citenamefont {Goldberger}\ and\ \citenamefont
  {Ridgway}(2017)}]{Goldberger:2016iau}%
  \BibitemOpen
  \bibfield  {author} {\bibinfo {author} {\bibfnamefont {Walter~D.}\
  \bibnamefont {Goldberger}}\ and\ \bibinfo {author} {\bibfnamefont
  {Alexander~K.}\ \bibnamefont {Ridgway}},\ }\bibfield  {title} {\enquote
  {\bibinfo {title} {{Radiation and the classical double copy for color
  charges}},}\ }\href {\doibase 10.1103/PhysRevD.95.125010} {\bibfield
  {journal} {\bibinfo  {journal} {Phys. Rev.}\ }\textbf {\bibinfo {volume}
  {D95}},\ \bibinfo {pages} {125010} (\bibinfo {year} {2017})},\ \Eprint
  {http://arxiv.org/abs/1611.03493} {arXiv:1611.03493 [hep-th]} \BibitemShut
  {NoStop}%
\bibitem [{\citenamefont {Shen}(2018)}]{Shen:2018ebu}%
  \BibitemOpen
  \bibfield  {author} {\bibinfo {author} {\bibfnamefont {Chia-Hsien}\
  \bibnamefont {Shen}},\ }\bibfield  {title} {\enquote {\bibinfo {title}
  {{Gravitational Radiation from Color-Kinematics Duality}},}\ }\href {\doibase
  10.1007/JHEP11(2018)162} {\bibfield  {journal} {\bibinfo  {journal} {JHEP}\
  }\textbf {\bibinfo {volume} {11}},\ \bibinfo {pages} {162} (\bibinfo {year}
  {2018})},\ \Eprint {http://arxiv.org/abs/1806.07388} {arXiv:1806.07388
  [hep-th]} \BibitemShut {NoStop}%
\bibitem [{\citenamefont {Liu}\ \emph {et~al.}(2021{\natexlab{b}})\citenamefont
  {Liu}, \citenamefont {Porto},\ and\ \citenamefont {Yang}}]{Liu:2021zxr}%
  \BibitemOpen
  \bibfield  {author} {\bibinfo {author} {\bibfnamefont {Zhengwen}\
  \bibnamefont {Liu}}, \bibinfo {author} {\bibfnamefont {Rafael~A.}\
  \bibnamefont {Porto}}, \ and\ \bibinfo {author} {\bibfnamefont {Zixin}\
  \bibnamefont {Yang}},\ }\bibfield  {title} {\enquote {\bibinfo {title} {{Spin
  Effects in the Effective Field Theory Approach to Post-Minkowskian
  Conservative Dynamics}},}\ }\href {\doibase 10.1007/JHEP06(2021)012}
  {\bibfield  {journal} {\bibinfo  {journal} {JHEP}\ }\textbf {\bibinfo
  {volume} {06}},\ \bibinfo {pages} {012} (\bibinfo {year}
  {2021}{\natexlab{b}})},\ \Eprint {http://arxiv.org/abs/2102.10059}
  {arXiv:2102.10059 [hep-th]} \BibitemShut {NoStop}%
\bibitem [{\citenamefont {K\"alin}\ and\ \citenamefont
  {Porto}(2020)}]{Kalin:2020mvi}%
  \BibitemOpen
  \bibfield  {author} {\bibinfo {author} {\bibfnamefont {Gregor}\ \bibnamefont
  {K\"alin}}\ and\ \bibinfo {author} {\bibfnamefont {Rafael~A.}\ \bibnamefont
  {Porto}},\ }\bibfield  {title} {\enquote {\bibinfo {title} {{Post-Minkowskian
  Effective Field Theory for Conservative Binary Dynamics}},}\ }\href {\doibase
  10.1007/JHEP11(2020)106} {\bibfield  {journal} {\bibinfo  {journal} {JHEP}\
  }\textbf {\bibinfo {volume} {11}},\ \bibinfo {pages} {106} (\bibinfo {year}
  {2020})},\ \Eprint {http://arxiv.org/abs/2006.01184} {arXiv:2006.01184
  [hep-th]} \BibitemShut {NoStop}%
\bibitem [{\citenamefont {Bern}\ \emph
  {et~al.}(2020{\natexlab{a}})\citenamefont {Bern}, \citenamefont {Luna},
  \citenamefont {Roiban}, \citenamefont {Shen},\ and\ \citenamefont
  {Zeng}}]{Bern:2020buy}%
  \BibitemOpen
  \bibfield  {author} {\bibinfo {author} {\bibfnamefont {Zvi}\ \bibnamefont
  {Bern}}, \bibinfo {author} {\bibfnamefont {Andres}\ \bibnamefont {Luna}},
  \bibinfo {author} {\bibfnamefont {Radu}\ \bibnamefont {Roiban}}, \bibinfo
  {author} {\bibfnamefont {Chia-Hsien}\ \bibnamefont {Shen}}, \ and\ \bibinfo
  {author} {\bibfnamefont {Mao}\ \bibnamefont {Zeng}},\ }\bibfield  {title}
  {\enquote {\bibinfo {title} {{Spinning Black Hole Binary Dynamics, Scattering
  Amplitudes and Effective Field Theory}},}\ }\href@noop {} {\  (\bibinfo
  {year} {2020}{\natexlab{a}})},\ \Eprint {http://arxiv.org/abs/2005.03071}
  {arXiv:2005.03071 [hep-th]} \BibitemShut {NoStop}%
\bibitem [{\citenamefont {Kosmopoulos}\ and\ \citenamefont
  {Luna}(2021)}]{Kosmopoulos:2021zoq}%
  \BibitemOpen
  \bibfield  {author} {\bibinfo {author} {\bibfnamefont {Dimitrios}\
  \bibnamefont {Kosmopoulos}}\ and\ \bibinfo {author} {\bibfnamefont {Andres}\
  \bibnamefont {Luna}},\ }\bibfield  {title} {\enquote {\bibinfo {title}
  {{Quadratic-in-spin Hamiltonian at $ \mathcal{O} $(G$^{2}$) from scattering
  amplitudes}},}\ }\href {\doibase 10.1007/JHEP07(2021)037} {\bibfield
  {journal} {\bibinfo  {journal} {JHEP}\ }\textbf {\bibinfo {volume} {07}},\
  \bibinfo {pages} {037} (\bibinfo {year} {2021})},\ \Eprint
  {http://arxiv.org/abs/2102.10137} {arXiv:2102.10137 [hep-th]} \BibitemShut
  {NoStop}%
\bibitem [{\citenamefont {Porto}\ \emph {et~al.}(2011)\citenamefont {Porto},
  \citenamefont {Ross},\ and\ \citenamefont {Rothstein}}]{Porto:2010zg}%
  \BibitemOpen
  \bibfield  {author} {\bibinfo {author} {\bibfnamefont {Rafael~A.}\
  \bibnamefont {Porto}}, \bibinfo {author} {\bibfnamefont {Andreas}\
  \bibnamefont {Ross}}, \ and\ \bibinfo {author} {\bibfnamefont {Ira~Z.}\
  \bibnamefont {Rothstein}},\ }\bibfield  {title} {\enquote {\bibinfo {title}
  {{Spin induced multipole moments for the gravitational wave flux from binary
  inspirals to third Post-Newtonian order}},}\ }\href {\doibase
  10.1088/1475-7516/2011/03/009} {\bibfield  {journal} {\bibinfo  {journal}
  {JCAP}\ }\textbf {\bibinfo {volume} {1103}},\ \bibinfo {pages} {009}
  (\bibinfo {year} {2011})},\ \Eprint {http://arxiv.org/abs/1007.1312}
  {arXiv:1007.1312 [gr-qc]} \BibitemShut {NoStop}%
\bibitem [{\citenamefont {Porto}\ \emph {et~al.}(2012)\citenamefont {Porto},
  \citenamefont {Ross},\ and\ \citenamefont {Rothstein}}]{Porto:2012as}%
  \BibitemOpen
  \bibfield  {author} {\bibinfo {author} {\bibfnamefont {Rafael~A.}\
  \bibnamefont {Porto}}, \bibinfo {author} {\bibfnamefont {Andreas}\
  \bibnamefont {Ross}}, \ and\ \bibinfo {author} {\bibfnamefont {Ira~Z.}\
  \bibnamefont {Rothstein}},\ }\bibfield  {title} {\enquote {\bibinfo {title}
  {{Spin induced multipole moments for the gravitational wave amplitude from
  binary inspirals to 2.5 Post-Newtonian order}},}\ }\href {\doibase
  10.1088/1475-7516/2012/09/028} {\bibfield  {journal} {\bibinfo  {journal}
  {JCAP}\ }\textbf {\bibinfo {volume} {1209}},\ \bibinfo {pages} {028}
  (\bibinfo {year} {2012})},\ \Eprint {http://arxiv.org/abs/1203.2962}
  {arXiv:1203.2962 [gr-qc]} \BibitemShut {NoStop}%
\bibitem [{\citenamefont {Maia}\ \emph
  {et~al.}(2017{\natexlab{a}})\citenamefont {Maia}, \citenamefont {Galley},
  \citenamefont {Leibovich},\ and\ \citenamefont {Porto}}]{Maia:2017gxn}%
  \BibitemOpen
  \bibfield  {author} {\bibinfo {author} {\bibfnamefont {Natalia~T.}\
  \bibnamefont {Maia}}, \bibinfo {author} {\bibfnamefont {Chad~R.}\
  \bibnamefont {Galley}}, \bibinfo {author} {\bibfnamefont {Adam~K.}\
  \bibnamefont {Leibovich}}, \ and\ \bibinfo {author} {\bibfnamefont
  {Rafael~A.}\ \bibnamefont {Porto}},\ }\bibfield  {title} {\enquote {\bibinfo
  {title} {{Radiation reaction for spinning bodies in effective field theory I:
  Spin-orbit effects}},}\ }\href {\doibase 10.1103/PhysRevD.96.084064}
  {\bibfield  {journal} {\bibinfo  {journal} {Phys. Rev.}\ }\textbf {\bibinfo
  {volume} {D96}},\ \bibinfo {pages} {084064} (\bibinfo {year}
  {2017}{\natexlab{a}})},\ \Eprint {http://arxiv.org/abs/1705.07934}
  {arXiv:1705.07934 [gr-qc]} \BibitemShut {NoStop}%
\bibitem [{\citenamefont {Maia}\ \emph
  {et~al.}(2017{\natexlab{b}})\citenamefont {Maia}, \citenamefont {Galley},
  \citenamefont {Leibovich},\ and\ \citenamefont {Porto}}]{Maia:2017yok}%
  \BibitemOpen
  \bibfield  {author} {\bibinfo {author} {\bibfnamefont {Natalia~T.}\
  \bibnamefont {Maia}}, \bibinfo {author} {\bibfnamefont {Chad~R.}\
  \bibnamefont {Galley}}, \bibinfo {author} {\bibfnamefont {Adam~K.}\
  \bibnamefont {Leibovich}}, \ and\ \bibinfo {author} {\bibfnamefont
  {Rafael~A.}\ \bibnamefont {Porto}},\ }\bibfield  {title} {\enquote {\bibinfo
  {title} {{Radiation reaction for spinning bodies in effective field theory
  II: Spin-spin effects}},}\ }\href {\doibase 10.1103/PhysRevD.96.084065}
  {\bibfield  {journal} {\bibinfo  {journal} {Phys. Rev.}\ }\textbf {\bibinfo
  {volume} {D96}},\ \bibinfo {pages} {084065} (\bibinfo {year}
  {2017}{\natexlab{b}})},\ \Eprint {http://arxiv.org/abs/1705.07938}
  {arXiv:1705.07938 [gr-qc]} \BibitemShut {NoStop}%
\bibitem [{\citenamefont {Cho}\ \emph {et~al.}(2021)\citenamefont {Cho},
  \citenamefont {Pardo},\ and\ \citenamefont {Porto}}]{Cho:2021mqw}%
  \BibitemOpen
  \bibfield  {author} {\bibinfo {author} {\bibfnamefont {Gihyuk}\ \bibnamefont
  {Cho}}, \bibinfo {author} {\bibfnamefont {Brian}\ \bibnamefont {Pardo}}, \
  and\ \bibinfo {author} {\bibfnamefont {Rafael~A.}\ \bibnamefont {Porto}},\
  }\bibfield  {title} {\enquote {\bibinfo {title} {{Gravitational radiation
  from inspiralling compact objects: Spin-spin effects completed at the
  next-to-leading post-Newtonian order}},}\ }\href@noop {} {\  (\bibinfo {year}
  {2021})},\ \Eprint {http://arxiv.org/abs/2103.14612} {arXiv:2103.14612
  [gr-qc]} \BibitemShut {NoStop}%
\bibitem [{\citenamefont {Mishra}\ \emph {et~al.}(2016)\citenamefont {Mishra},
  \citenamefont {Kela}, \citenamefont {Arun},\ and\ \citenamefont
  {Faye}}]{Mishra:2016whh}%
  \BibitemOpen
  \bibfield  {author} {\bibinfo {author} {\bibfnamefont {Chandra~Kant}\
  \bibnamefont {Mishra}}, \bibinfo {author} {\bibfnamefont {Aditya}\
  \bibnamefont {Kela}}, \bibinfo {author} {\bibfnamefont {K.~G.}\ \bibnamefont
  {Arun}}, \ and\ \bibinfo {author} {\bibfnamefont {Guillaume}\ \bibnamefont
  {Faye}},\ }\bibfield  {title} {\enquote {\bibinfo {title} {{Ready-to-use
  post-Newtonian gravitational waveforms for binary black holes with
  nonprecessing spins: An update}},}\ }\href {\doibase
  10.1103/PhysRevD.93.084054} {\bibfield  {journal} {\bibinfo  {journal} {Phys.
  Rev.}\ }\textbf {\bibinfo {volume} {D93}},\ \bibinfo {pages} {084054}
  (\bibinfo {year} {2016})},\ \Eprint {http://arxiv.org/abs/1601.05588}
  {arXiv:1601.05588 [gr-qc]} \BibitemShut {NoStop}%
\bibitem [{\citenamefont {Buonanno}\ \emph {et~al.}(2013)\citenamefont
  {Buonanno}, \citenamefont {Faye},\ and\ \citenamefont
  {Hinderer}}]{Buonanno:2012rv}%
  \BibitemOpen
  \bibfield  {author} {\bibinfo {author} {\bibfnamefont {Alessandra}\
  \bibnamefont {Buonanno}}, \bibinfo {author} {\bibfnamefont {Guillaume}\
  \bibnamefont {Faye}}, \ and\ \bibinfo {author} {\bibfnamefont {Tanja}\
  \bibnamefont {Hinderer}},\ }\bibfield  {title} {\enquote {\bibinfo {title}
  {{Spin effects on gravitational waves from inspiraling compact binaries at
  second post-Newtonian order}},}\ }\href {\doibase 10.1103/PhysRevD.87.044009}
  {\bibfield  {journal} {\bibinfo  {journal} {Phys. Rev.}\ }\textbf {\bibinfo
  {volume} {D87}},\ \bibinfo {pages} {044009} (\bibinfo {year} {2013})},\
  \Eprint {http://arxiv.org/abs/1209.6349} {arXiv:1209.6349 [gr-qc]}
  \BibitemShut {NoStop}%
\bibitem [{\citenamefont {Vines}(2018)}]{Vines:2017hyw}%
  \BibitemOpen
  \bibfield  {author} {\bibinfo {author} {\bibfnamefont {Justin}\ \bibnamefont
  {Vines}},\ }\bibfield  {title} {\enquote {\bibinfo {title} {{Scattering of
  two spinning black holes in post-Minkowskian gravity, to all orders in spin,
  and effective-one-body mappings}},}\ }\href {\doibase
  10.1088/1361-6382/aaa3a8} {\bibfield  {journal} {\bibinfo  {journal} {Class.
  Quant. Grav.}\ }\textbf {\bibinfo {volume} {35}},\ \bibinfo {pages} {084002}
  (\bibinfo {year} {2018})},\ \Eprint {http://arxiv.org/abs/1709.06016}
  {arXiv:1709.06016 [gr-qc]} \BibitemShut {NoStop}%
\bibitem [{\citenamefont {Bini}\ and\ \citenamefont
  {Damour}(2017)}]{Bini:2017xzy}%
  \BibitemOpen
  \bibfield  {author} {\bibinfo {author} {\bibfnamefont {Donato}\ \bibnamefont
  {Bini}}\ and\ \bibinfo {author} {\bibfnamefont {Thibault}\ \bibnamefont
  {Damour}},\ }\bibfield  {title} {\enquote {\bibinfo {title} {{Gravitational
  spin-orbit coupling in binary systems, post-Minkowskian approximation and
  effective one-body theory}},}\ }\href {\doibase 10.1103/PhysRevD.96.104038}
  {\bibfield  {journal} {\bibinfo  {journal} {Phys. Rev.}\ }\textbf {\bibinfo
  {volume} {D96}},\ \bibinfo {pages} {104038} (\bibinfo {year} {2017})},\
  \Eprint {http://arxiv.org/abs/1709.00590} {arXiv:1709.00590 [gr-qc]}
  \BibitemShut {NoStop}%
\bibitem [{\citenamefont {Bini}\ and\ \citenamefont
  {Damour}(2018)}]{Bini:2018ywr}%
  \BibitemOpen
  \bibfield  {author} {\bibinfo {author} {\bibfnamefont {Donato}\ \bibnamefont
  {Bini}}\ and\ \bibinfo {author} {\bibfnamefont {Thibault}\ \bibnamefont
  {Damour}},\ }\bibfield  {title} {\enquote {\bibinfo {title} {{Gravitational
  spin-orbit coupling in binary systems at the second post-Minkowskian
  approximation}},}\ }\href {\doibase 10.1103/PhysRevD.98.044036} {\bibfield
  {journal} {\bibinfo  {journal} {Phys. Rev.}\ }\textbf {\bibinfo {volume}
  {D98}},\ \bibinfo {pages} {044036} (\bibinfo {year} {2018})},\ \Eprint
  {http://arxiv.org/abs/1805.10809} {arXiv:1805.10809 [gr-qc]} \BibitemShut
  {NoStop}%
\bibitem [{\citenamefont {Guevara}(2019)}]{Guevara:2017csg}%
  \BibitemOpen
  \bibfield  {author} {\bibinfo {author} {\bibfnamefont {Alfredo}\ \bibnamefont
  {Guevara}},\ }\bibfield  {title} {\enquote {\bibinfo {title} {{Holomorphic
  Classical Limit for Spin Effects in Gravitational and Electromagnetic
  Scattering}},}\ }\href {\doibase 10.1007/JHEP04(2019)033} {\bibfield
  {journal} {\bibinfo  {journal} {JHEP}\ }\textbf {\bibinfo {volume} {04}},\
  \bibinfo {pages} {033} (\bibinfo {year} {2019})},\ \Eprint
  {http://arxiv.org/abs/1706.02314} {arXiv:1706.02314 [hep-th]} \BibitemShut
  {NoStop}%
\bibitem [{\citenamefont {Vines}\ \emph {et~al.}(2019)\citenamefont {Vines},
  \citenamefont {Steinhoff},\ and\ \citenamefont {Buonanno}}]{Vines:2018gqi}%
  \BibitemOpen
  \bibfield  {author} {\bibinfo {author} {\bibfnamefont {Justin}\ \bibnamefont
  {Vines}}, \bibinfo {author} {\bibfnamefont {Jan}\ \bibnamefont {Steinhoff}},
  \ and\ \bibinfo {author} {\bibfnamefont {Alessandra}\ \bibnamefont
  {Buonanno}},\ }\bibfield  {title} {\enquote {\bibinfo {title}
  {{Spinning-black-hole scattering and the test-black-hole limit at second
  post-Minkowskian order}},}\ }\href {\doibase 10.1103/PhysRevD.99.064054}
  {\bibfield  {journal} {\bibinfo  {journal} {Phys. Rev. D}\ }\textbf {\bibinfo
  {volume} {99}},\ \bibinfo {pages} {064054} (\bibinfo {year} {2019})},\
  \Eprint {http://arxiv.org/abs/1812.00956} {arXiv:1812.00956 [gr-qc]}
  \BibitemShut {NoStop}%
\bibitem [{\citenamefont {Guevara}\ \emph
  {et~al.}(2019{\natexlab{a}})\citenamefont {Guevara}, \citenamefont
  {Ochirov},\ and\ \citenamefont {Vines}}]{Guevara:2018wpp}%
  \BibitemOpen
  \bibfield  {author} {\bibinfo {author} {\bibfnamefont {Alfredo}\ \bibnamefont
  {Guevara}}, \bibinfo {author} {\bibfnamefont {Alexander}\ \bibnamefont
  {Ochirov}}, \ and\ \bibinfo {author} {\bibfnamefont {Justin}\ \bibnamefont
  {Vines}},\ }\bibfield  {title} {\enquote {\bibinfo {title} {{Scattering of
  Spinning Black Holes from Exponentiated Soft Factors}},}\ }\href {\doibase
  10.1007/JHEP09(2019)056} {\bibfield  {journal} {\bibinfo  {journal} {JHEP}\
  }\textbf {\bibinfo {volume} {09}},\ \bibinfo {pages} {056} (\bibinfo {year}
  {2019}{\natexlab{a}})},\ \Eprint {http://arxiv.org/abs/1812.06895}
  {arXiv:1812.06895 [hep-th]} \BibitemShut {NoStop}%
\bibitem [{\citenamefont {Chung}\ \emph {et~al.}(2019)\citenamefont {Chung},
  \citenamefont {Huang}, \citenamefont {Kim},\ and\ \citenamefont
  {Lee}}]{Chung:2018kqs}%
  \BibitemOpen
  \bibfield  {author} {\bibinfo {author} {\bibfnamefont {Ming-Zhi}\
  \bibnamefont {Chung}}, \bibinfo {author} {\bibfnamefont {Yu-Tin}\
  \bibnamefont {Huang}}, \bibinfo {author} {\bibfnamefont {Jung-Wook}\
  \bibnamefont {Kim}}, \ and\ \bibinfo {author} {\bibfnamefont {Sangmin}\
  \bibnamefont {Lee}},\ }\bibfield  {title} {\enquote {\bibinfo {title} {{The
  simplest massive S-matrix: from minimal coupling to Black Holes}},}\ }\href
  {\doibase 10.1007/JHEP04(2019)156} {\bibfield  {journal} {\bibinfo  {journal}
  {JHEP}\ }\textbf {\bibinfo {volume} {04}},\ \bibinfo {pages} {156} (\bibinfo
  {year} {2019})},\ \Eprint {http://arxiv.org/abs/1812.08752} {arXiv:1812.08752
  [hep-th]} \BibitemShut {NoStop}%
\bibitem [{\citenamefont {Guevara}\ \emph
  {et~al.}(2019{\natexlab{b}})\citenamefont {Guevara}, \citenamefont
  {Ochirov},\ and\ \citenamefont {Vines}}]{Guevara:2019fsj}%
  \BibitemOpen
  \bibfield  {author} {\bibinfo {author} {\bibfnamefont {Alfredo}\ \bibnamefont
  {Guevara}}, \bibinfo {author} {\bibfnamefont {Alexander}\ \bibnamefont
  {Ochirov}}, \ and\ \bibinfo {author} {\bibfnamefont {Justin}\ \bibnamefont
  {Vines}},\ }\bibfield  {title} {\enquote {\bibinfo {title} {{Black-hole
  scattering with general spin directions from minimal-coupling amplitudes}},}\
  }\href {\doibase 10.1103/PhysRevD.100.104024} {\bibfield  {journal} {\bibinfo
   {journal} {Phys. Rev. D}\ }\textbf {\bibinfo {volume} {100}},\ \bibinfo
  {pages} {104024} (\bibinfo {year} {2019}{\natexlab{b}})},\ \Eprint
  {http://arxiv.org/abs/1906.10071} {arXiv:1906.10071 [hep-th]} \BibitemShut
  {NoStop}%
\bibitem [{\citenamefont {Chung}\ \emph {et~al.}(2020)\citenamefont {Chung},
  \citenamefont {Huang},\ and\ \citenamefont {Kim}}]{Chung:2019duq}%
  \BibitemOpen
  \bibfield  {author} {\bibinfo {author} {\bibfnamefont {Ming-Zhi}\
  \bibnamefont {Chung}}, \bibinfo {author} {\bibfnamefont {Yu-Tin}\
  \bibnamefont {Huang}}, \ and\ \bibinfo {author} {\bibfnamefont {Jung-Wook}\
  \bibnamefont {Kim}},\ }\bibfield  {title} {\enquote {\bibinfo {title}
  {{Classical potential for general spinning bodies}},}\ }\href {\doibase
  10.1007/JHEP09(2020)074} {\bibfield  {journal} {\bibinfo  {journal} {JHEP}\
  }\textbf {\bibinfo {volume} {09}},\ \bibinfo {pages} {074} (\bibinfo {year}
  {2020})},\ \Eprint {http://arxiv.org/abs/1908.08463} {arXiv:1908.08463
  [hep-th]} \BibitemShut {NoStop}%
\bibitem [{\citenamefont {Damgaard}\ \emph {et~al.}(2019)\citenamefont
  {Damgaard}, \citenamefont {Haddad},\ and\ \citenamefont
  {Helset}}]{Damgaard:2019lfh}%
  \BibitemOpen
  \bibfield  {author} {\bibinfo {author} {\bibfnamefont {Poul~H.}\ \bibnamefont
  {Damgaard}}, \bibinfo {author} {\bibfnamefont {Kays}\ \bibnamefont {Haddad}},
  \ and\ \bibinfo {author} {\bibfnamefont {Andreas}\ \bibnamefont {Helset}},\
  }\bibfield  {title} {\enquote {\bibinfo {title} {{Heavy Black Hole Effective
  Theory}},}\ }\href {\doibase 10.1007/JHEP11(2019)070} {\bibfield  {journal}
  {\bibinfo  {journal} {JHEP}\ }\textbf {\bibinfo {volume} {11}},\ \bibinfo
  {pages} {070} (\bibinfo {year} {2019})},\ \Eprint
  {http://arxiv.org/abs/1908.10308} {arXiv:1908.10308 [hep-ph]} \BibitemShut
  {NoStop}%
\bibitem [{\citenamefont {Aoude}\ \emph {et~al.}(2020)\citenamefont {Aoude},
  \citenamefont {Haddad},\ and\ \citenamefont {Helset}}]{Aoude:2020onz}%
  \BibitemOpen
  \bibfield  {author} {\bibinfo {author} {\bibfnamefont {Rafael}\ \bibnamefont
  {Aoude}}, \bibinfo {author} {\bibfnamefont {Kays}\ \bibnamefont {Haddad}}, \
  and\ \bibinfo {author} {\bibfnamefont {Andreas}\ \bibnamefont {Helset}},\
  }\bibfield  {title} {\enquote {\bibinfo {title} {{On-shell heavy particle
  effective theories}},}\ }\href {\doibase 10.1007/JHEP05(2020)051} {\bibfield
  {journal} {\bibinfo  {journal} {JHEP}\ }\textbf {\bibinfo {volume} {05}},\
  \bibinfo {pages} {051} (\bibinfo {year} {2020})},\ \Eprint
  {http://arxiv.org/abs/2001.09164} {arXiv:2001.09164 [hep-th]} \BibitemShut
  {NoStop}%
\bibitem [{\citenamefont {Guevara}\ \emph {et~al.}(2020)\citenamefont
  {Guevara}, \citenamefont {Maybee}, \citenamefont {Ochirov}, \citenamefont
  {O'Connell},\ and\ \citenamefont {Vines}}]{Guevara:2020xjx}%
  \BibitemOpen
  \bibfield  {author} {\bibinfo {author} {\bibfnamefont {Alfredo}\ \bibnamefont
  {Guevara}}, \bibinfo {author} {\bibfnamefont {Ben}\ \bibnamefont {Maybee}},
  \bibinfo {author} {\bibfnamefont {Alexander}\ \bibnamefont {Ochirov}},
  \bibinfo {author} {\bibfnamefont {Donal}\ \bibnamefont {O'Connell}}, \ and\
  \bibinfo {author} {\bibfnamefont {Justin}\ \bibnamefont {Vines}},\ }\bibfield
   {title} {\enquote {\bibinfo {title} {{A worldsheet for Kerr}},}\ }\href@noop
  {} {\  (\bibinfo {year} {2020})},\ \Eprint {http://arxiv.org/abs/2012.11570}
  {arXiv:2012.11570 [hep-th]} \BibitemShut {NoStop}%
\bibitem [{\citenamefont {Jakobsen}\ \emph
  {et~al.}(2021{\natexlab{b}})\citenamefont {Jakobsen}, \citenamefont {Mogull},
  \citenamefont {Plefka},\ and\ \citenamefont {Steinhoff}}]{Jakobsen:2021zvh}%
  \BibitemOpen
  \bibfield  {author} {\bibinfo {author} {\bibfnamefont {Gustav~Uhre}\
  \bibnamefont {Jakobsen}}, \bibinfo {author} {\bibfnamefont {Gustav}\
  \bibnamefont {Mogull}}, \bibinfo {author} {\bibfnamefont {Jan}\ \bibnamefont
  {Plefka}}, \ and\ \bibinfo {author} {\bibfnamefont {Jan}\ \bibnamefont
  {Steinhoff}},\ }\bibfield  {title} {\enquote {\bibinfo {title} {{SUSY in the
  Sky with Gravitons}},}\ }\href@noop {} {\  (\bibinfo {year}
  {2021}{\natexlab{b}})},\ \Eprint {http://arxiv.org/abs/2109.04465}
  {arXiv:2109.04465 [hep-th]} \BibitemShut {NoStop}%
\bibitem [{\citenamefont {Vines}\ \emph {et~al.}(2016)\citenamefont {Vines},
  \citenamefont {Kunst}, \citenamefont {Steinhoff},\ and\ \citenamefont
  {Hinderer}}]{Vines:2016unv}%
  \BibitemOpen
  \bibfield  {author} {\bibinfo {author} {\bibfnamefont {Justin}\ \bibnamefont
  {Vines}}, \bibinfo {author} {\bibfnamefont {Daniela}\ \bibnamefont {Kunst}},
  \bibinfo {author} {\bibfnamefont {Jan}\ \bibnamefont {Steinhoff}}, \ and\
  \bibinfo {author} {\bibfnamefont {Tanja}\ \bibnamefont {Hinderer}},\
  }\bibfield  {title} {\enquote {\bibinfo {title} {{Canonical Hamiltonian for
  an extended test body in curved spacetime: To quadratic order in spin}},}\
  }\href {\doibase 10.1103/PhysRevD.93.103008} {\bibfield  {journal} {\bibinfo
  {journal} {Phys. Rev. D}\ }\textbf {\bibinfo {volume} {93}},\ \bibinfo
  {pages} {103008} (\bibinfo {year} {2016})},\ \Eprint
  {http://arxiv.org/abs/1601.07529} {arXiv:1601.07529 [gr-qc]} \BibitemShut
  {NoStop}%
\bibitem [{\citenamefont {Porto}\ and\ \citenamefont
  {Rothstein}(2008)}]{Porto:2008jj}%
  \BibitemOpen
  \bibfield  {author} {\bibinfo {author} {\bibfnamefont {Rafael~A}\
  \bibnamefont {Porto}}\ and\ \bibinfo {author} {\bibfnamefont {Ira~Z.}\
  \bibnamefont {Rothstein}},\ }\bibfield  {title} {\enquote {\bibinfo {title}
  {{Next to Leading Order Spin(1)Spin(1) Effects in the Motion of Inspiralling
  Compact Binaries}},}\ }\href {\doibase 10.1103/PhysRevD.78.044013} {\bibfield
   {journal} {\bibinfo  {journal} {Phys. Rev. D}\ }\textbf {\bibinfo {volume}
  {78}},\ \bibinfo {pages} {044013} (\bibinfo {year} {2008})},\ \bibinfo {note}
  {[Erratum: Phys.Rev.D 81, 029905 (2010)]},\ \Eprint
  {http://arxiv.org/abs/0804.0260} {arXiv:0804.0260 [gr-qc]} \BibitemShut
  {NoStop}%
\bibitem [{\citenamefont {Mathisson}(1937)}]{Mathisson:1937zz}%
  \BibitemOpen
  \bibfield  {author} {\bibinfo {author} {\bibfnamefont {Myron}\ \bibnamefont
  {Mathisson}},\ }\bibfield  {title} {\enquote {\bibinfo {title} {{Neue
  Mechanik materieller Systeme}},}\ }\href@noop {} {\bibfield  {journal}
  {\bibinfo  {journal} {Acta Phys. Polon.}\ }\textbf {\bibinfo {volume} {6}},\
  \bibinfo {pages} {163--2900} (\bibinfo {year} {1937})}\BibitemShut {NoStop}%
\bibitem [{\citenamefont {Papapetrou}(1951)}]{Papapetrou:1951pa}%
  \BibitemOpen
  \bibfield  {author} {\bibinfo {author} {\bibfnamefont {Achille}\ \bibnamefont
  {Papapetrou}},\ }\bibfield  {title} {\enquote {\bibinfo {title} {{Spinning
  test particles in general relativity. 1.}}}\ }\href {\doibase
  10.1098/rspa.1951.0200} {\bibfield  {journal} {\bibinfo  {journal} {Proc.
  Roy. Soc. Lond. A}\ }\textbf {\bibinfo {volume} {209}},\ \bibinfo {pages}
  {248--258} (\bibinfo {year} {1951})}\BibitemShut {NoStop}%
\bibitem [{\citenamefont {Dixon}(1970)}]{Dixon:1970zza}%
  \BibitemOpen
  \bibfield  {author} {\bibinfo {author} {\bibfnamefont {W.~G.}\ \bibnamefont
  {Dixon}},\ }\bibfield  {title} {\enquote {\bibinfo {title} {{Dynamics of
  extended bodies in general relativity. I. Momentum and angular momentum}},}\
  }\href {\doibase 10.1098/rspa.1970.0020} {\bibfield  {journal} {\bibinfo
  {journal} {Proc. Roy. Soc. Lond. A}\ }\textbf {\bibinfo {volume} {314}},\
  \bibinfo {pages} {499--527} (\bibinfo {year} {1970})}\BibitemShut {NoStop}%
\bibitem [{\citenamefont {Herrmann}\ \emph
  {et~al.}(2021{\natexlab{a}})\citenamefont {Herrmann}, \citenamefont
  {Parra-Martinez}, \citenamefont {Ruf},\ and\ \citenamefont
  {Zeng}}]{Herrmann:2021lqe}%
  \BibitemOpen
  \bibfield  {author} {\bibinfo {author} {\bibfnamefont {Enrico}\ \bibnamefont
  {Herrmann}}, \bibinfo {author} {\bibfnamefont {Julio}\ \bibnamefont
  {Parra-Martinez}}, \bibinfo {author} {\bibfnamefont {Michael~S.}\
  \bibnamefont {Ruf}}, \ and\ \bibinfo {author} {\bibfnamefont {Mao}\
  \bibnamefont {Zeng}},\ }\bibfield  {title} {\enquote {\bibinfo {title}
  {{Gravitational Bremsstrahlung from Reverse Unitarity}},}\ }\href {\doibase
  10.1103/PhysRevLett.126.201602} {\bibfield  {journal} {\bibinfo  {journal}
  {Phys. Rev. Lett.}\ }\textbf {\bibinfo {volume} {126}},\ \bibinfo {pages}
  {201602} (\bibinfo {year} {2021}{\natexlab{a}})},\ \Eprint
  {http://arxiv.org/abs/2101.07255} {arXiv:2101.07255 [hep-th]} \BibitemShut
  {NoStop}%
\bibitem [{\citenamefont {Herrmann}\ \emph
  {et~al.}(2021{\natexlab{b}})\citenamefont {Herrmann}, \citenamefont
  {Parra-Martinez}, \citenamefont {Ruf},\ and\ \citenamefont
  {Zeng}}]{Herrmann:2021tct}%
  \BibitemOpen
  \bibfield  {author} {\bibinfo {author} {\bibfnamefont {Enrico}\ \bibnamefont
  {Herrmann}}, \bibinfo {author} {\bibfnamefont {Julio}\ \bibnamefont
  {Parra-Martinez}}, \bibinfo {author} {\bibfnamefont {Michael~S.}\
  \bibnamefont {Ruf}}, \ and\ \bibinfo {author} {\bibfnamefont {Mao}\
  \bibnamefont {Zeng}},\ }\bibfield  {title} {\enquote {\bibinfo {title}
  {{Radiative classical gravitational observables at $ \mathcal{O} $(G$^{3}$)
  from scattering amplitudes}},}\ }\href {\doibase 10.1007/JHEP10(2021)148}
  {\bibfield  {journal} {\bibinfo  {journal} {JHEP}\ }\textbf {\bibinfo
  {volume} {10}},\ \bibinfo {pages} {148} (\bibinfo {year}
  {2021}{\natexlab{b}})},\ \Eprint {http://arxiv.org/abs/2104.03957}
  {arXiv:2104.03957 [hep-th]} \BibitemShut {NoStop}%
\bibitem [{\citenamefont {Kosower}\ \emph {et~al.}(2019)\citenamefont
  {Kosower}, \citenamefont {Maybee},\ and\ \citenamefont
  {O'Connell}}]{Kosower:2018adc}%
  \BibitemOpen
  \bibfield  {author} {\bibinfo {author} {\bibfnamefont {David~A.}\
  \bibnamefont {Kosower}}, \bibinfo {author} {\bibfnamefont {Ben}\ \bibnamefont
  {Maybee}}, \ and\ \bibinfo {author} {\bibfnamefont {Donal}\ \bibnamefont
  {O'Connell}},\ }\bibfield  {title} {\enquote {\bibinfo {title} {{Amplitudes,
  Observables, and Classical Scattering}},}\ }\href {\doibase
  10.1007/JHEP02(2019)137} {\bibfield  {journal} {\bibinfo  {journal} {JHEP}\
  }\textbf {\bibinfo {volume} {02}},\ \bibinfo {pages} {137} (\bibinfo {year}
  {2019})},\ \Eprint {http://arxiv.org/abs/1811.10950} {arXiv:1811.10950
  [hep-th]} \BibitemShut {NoStop}%
\bibitem [{\citenamefont {Maybee}\ \emph {et~al.}(2019)\citenamefont {Maybee},
  \citenamefont {O'Connell},\ and\ \citenamefont {Vines}}]{Maybee:2019jus}%
  \BibitemOpen
  \bibfield  {author} {\bibinfo {author} {\bibfnamefont {Ben}\ \bibnamefont
  {Maybee}}, \bibinfo {author} {\bibfnamefont {Donal}\ \bibnamefont
  {O'Connell}}, \ and\ \bibinfo {author} {\bibfnamefont {Justin}\ \bibnamefont
  {Vines}},\ }\bibfield  {title} {\enquote {\bibinfo {title} {{Observables and
  amplitudes for spinning particles and black holes}},}\ }\href {\doibase
  10.1007/JHEP12(2019)156} {\bibfield  {journal} {\bibinfo  {journal} {JHEP}\
  }\textbf {\bibinfo {volume} {12}},\ \bibinfo {pages} {156} (\bibinfo {year}
  {2019})},\ \Eprint {http://arxiv.org/abs/1906.09260} {arXiv:1906.09260
  [hep-th]} \BibitemShut {NoStop}%
\bibitem [{\citenamefont {Di~Vecchia}\ \emph
  {et~al.}(2020{\natexlab{a}})\citenamefont {Di~Vecchia}, \citenamefont
  {Heissenberg}, \citenamefont {Russo},\ and\ \citenamefont
  {Veneziano}}]{DiVecchia:2020ymx}%
  \BibitemOpen
  \bibfield  {author} {\bibinfo {author} {\bibfnamefont {Paolo}\ \bibnamefont
  {Di~Vecchia}}, \bibinfo {author} {\bibfnamefont {Carlo}\ \bibnamefont
  {Heissenberg}}, \bibinfo {author} {\bibfnamefont {Rodolfo}\ \bibnamefont
  {Russo}}, \ and\ \bibinfo {author} {\bibfnamefont {Gabriele}\ \bibnamefont
  {Veneziano}},\ }\bibfield  {title} {\enquote {\bibinfo {title} {{Universality
  of ultra-relativistic gravitational scattering}},}\ }\href {\doibase
  10.1016/j.physletb.2020.135924} {\bibfield  {journal} {\bibinfo  {journal}
  {Phys. Lett. B}\ }\textbf {\bibinfo {volume} {811}},\ \bibinfo {pages}
  {135924} (\bibinfo {year} {2020}{\natexlab{a}})},\ \Eprint
  {http://arxiv.org/abs/2008.12743} {arXiv:2008.12743 [hep-th]} \BibitemShut
  {NoStop}%
\bibitem [{\citenamefont {Damour}(2020{\natexlab{a}})}]{Damour:2020tta}%
  \BibitemOpen
  \bibfield  {author} {\bibinfo {author} {\bibfnamefont {Thibault}\
  \bibnamefont {Damour}},\ }\bibfield  {title} {\enquote {\bibinfo {title}
  {{Radiative contribution to classical gravitational scattering at the third
  order in $G$}},}\ }\href {\doibase 10.1103/PhysRevD.102.124008} {\bibfield
  {journal} {\bibinfo  {journal} {Phys. Rev. D}\ }\textbf {\bibinfo {volume}
  {102}},\ \bibinfo {pages} {124008} (\bibinfo {year} {2020}{\natexlab{a}})},\
  \Eprint {http://arxiv.org/abs/2010.01641} {arXiv:2010.01641 [gr-qc]}
  \BibitemShut {NoStop}%
\bibitem [{\citenamefont {Bern}\ \emph {et~al.}(1994)\citenamefont {Bern},
  \citenamefont {Dixon}, \citenamefont {Dunbar},\ and\ \citenamefont
  {Kosower}}]{Bern:1994zx}%
  \BibitemOpen
  \bibfield  {author} {\bibinfo {author} {\bibfnamefont {Zvi}\ \bibnamefont
  {Bern}}, \bibinfo {author} {\bibfnamefont {Lance~J.}\ \bibnamefont {Dixon}},
  \bibinfo {author} {\bibfnamefont {David~C.}\ \bibnamefont {Dunbar}}, \ and\
  \bibinfo {author} {\bibfnamefont {David~A.}\ \bibnamefont {Kosower}},\
  }\bibfield  {title} {\enquote {\bibinfo {title} {{One loop n point gauge
  theory amplitudes, unitarity and collinear limits}},}\ }\href {\doibase
  10.1016/0550-3213(94)90179-1} {\bibfield  {journal} {\bibinfo  {journal}
  {Nucl. Phys.}\ }\textbf {\bibinfo {volume} {B425}},\ \bibinfo {pages}
  {217--260} (\bibinfo {year} {1994})},\ \Eprint
  {http://arxiv.org/abs/hep-ph/9403226} {arXiv:hep-ph/9403226 [hep-ph]}
  \BibitemShut {NoStop}%
\bibitem [{\citenamefont {Bern}\ \emph {et~al.}(1995)\citenamefont {Bern},
  \citenamefont {Dixon}, \citenamefont {Dunbar},\ and\ \citenamefont
  {Kosower}}]{Bern:1994cg}%
  \BibitemOpen
  \bibfield  {author} {\bibinfo {author} {\bibfnamefont {Zvi}\ \bibnamefont
  {Bern}}, \bibinfo {author} {\bibfnamefont {Lance~J.}\ \bibnamefont {Dixon}},
  \bibinfo {author} {\bibfnamefont {David~C.}\ \bibnamefont {Dunbar}}, \ and\
  \bibinfo {author} {\bibfnamefont {David~A.}\ \bibnamefont {Kosower}},\
  }\bibfield  {title} {\enquote {\bibinfo {title} {{Fusing gauge theory tree
  amplitudes into loop amplitudes}},}\ }\href {\doibase
  10.1016/0550-3213(94)00488-Z} {\bibfield  {journal} {\bibinfo  {journal}
  {Nucl. Phys.}\ }\textbf {\bibinfo {volume} {B435}},\ \bibinfo {pages}
  {59--101} (\bibinfo {year} {1995})},\ \Eprint
  {http://arxiv.org/abs/hep-ph/9409265} {arXiv:hep-ph/9409265 [hep-ph]}
  \BibitemShut {NoStop}%
\bibitem [{\citenamefont {Britto}\ \emph {et~al.}(2005)\citenamefont {Britto},
  \citenamefont {Cachazo},\ and\ \citenamefont {Feng}}]{Britto:2004nc}%
  \BibitemOpen
  \bibfield  {author} {\bibinfo {author} {\bibfnamefont {Ruth}\ \bibnamefont
  {Britto}}, \bibinfo {author} {\bibfnamefont {Freddy}\ \bibnamefont
  {Cachazo}}, \ and\ \bibinfo {author} {\bibfnamefont {Bo}~\bibnamefont
  {Feng}},\ }\bibfield  {title} {\enquote {\bibinfo {title} {{Generalized
  unitarity and one-loop amplitudes in N=4 super-Yang-Mills}},}\ }\href
  {\doibase 10.1016/j.nuclphysb.2005.07.014} {\bibfield  {journal} {\bibinfo
  {journal} {Nucl. Phys.}\ }\textbf {\bibinfo {volume} {B725}},\ \bibinfo
  {pages} {275--305} (\bibinfo {year} {2005})},\ \Eprint
  {http://arxiv.org/abs/hep-th/0412103} {arXiv:hep-th/0412103 [hep-th]}
  \BibitemShut {NoStop}%
\bibitem [{\citenamefont {Bern}\ \emph {et~al.}(2008)\citenamefont {Bern},
  \citenamefont {Carrasco},\ and\ \citenamefont {Johansson}}]{Bern:2008qj}%
  \BibitemOpen
  \bibfield  {author} {\bibinfo {author} {\bibfnamefont {Z.}~\bibnamefont
  {Bern}}, \bibinfo {author} {\bibfnamefont {J.~J.~M.}\ \bibnamefont
  {Carrasco}}, \ and\ \bibinfo {author} {\bibfnamefont {Henrik}\ \bibnamefont
  {Johansson}},\ }\bibfield  {title} {\enquote {\bibinfo {title} {{New
  Relations for Gauge-Theory Amplitudes}},}\ }\href {\doibase
  10.1103/PhysRevD.78.085011} {\bibfield  {journal} {\bibinfo  {journal} {Phys.
  Rev.}\ }\textbf {\bibinfo {volume} {D78}},\ \bibinfo {pages} {085011}
  (\bibinfo {year} {2008})},\ \Eprint {http://arxiv.org/abs/0805.3993}
  {arXiv:0805.3993 [hep-ph]} \BibitemShut {NoStop}%
\bibitem [{\citenamefont {Bern}\ \emph {et~al.}(2010)\citenamefont {Bern},
  \citenamefont {Carrasco},\ and\ \citenamefont {Johansson}}]{Bern:2010ue}%
  \BibitemOpen
  \bibfield  {author} {\bibinfo {author} {\bibfnamefont {Zvi}\ \bibnamefont
  {Bern}}, \bibinfo {author} {\bibfnamefont {John Joseph~M.}\ \bibnamefont
  {Carrasco}}, \ and\ \bibinfo {author} {\bibfnamefont {Henrik}\ \bibnamefont
  {Johansson}},\ }\bibfield  {title} {\enquote {\bibinfo {title} {{Perturbative
  Quantum Gravity as a Double Copy of Gauge Theory}},}\ }\href {\doibase
  10.1103/PhysRevLett.105.061602} {\bibfield  {journal} {\bibinfo  {journal}
  {Phys. Rev. Lett.}\ }\textbf {\bibinfo {volume} {105}},\ \bibinfo {pages}
  {061602} (\bibinfo {year} {2010})},\ \Eprint {http://arxiv.org/abs/1004.0476}
  {arXiv:1004.0476 [hep-th]} \BibitemShut {NoStop}%
\bibitem [{\citenamefont {Bern}\ \emph {et~al.}(2012)\citenamefont {Bern},
  \citenamefont {Carrasco}, \citenamefont {Dixon}, \citenamefont {Johansson},\
  and\ \citenamefont {Roiban}}]{Bern:2012uf}%
  \BibitemOpen
  \bibfield  {author} {\bibinfo {author} {\bibfnamefont {Z.}~\bibnamefont
  {Bern}}, \bibinfo {author} {\bibfnamefont {J.~J.~M.}\ \bibnamefont
  {Carrasco}}, \bibinfo {author} {\bibfnamefont {L.~J.}\ \bibnamefont {Dixon}},
  \bibinfo {author} {\bibfnamefont {H.}~\bibnamefont {Johansson}}, \ and\
  \bibinfo {author} {\bibfnamefont {R.}~\bibnamefont {Roiban}},\ }\bibfield
  {title} {\enquote {\bibinfo {title} {{Simplifying Multiloop Integrands and
  Ultraviolet Divergences of Gauge Theory and Gravity Amplitudes}},}\ }\href
  {\doibase 10.1103/PhysRevD.85.105014} {\bibfield  {journal} {\bibinfo
  {journal} {Phys. Rev.}\ }\textbf {\bibinfo {volume} {D85}},\ \bibinfo {pages}
  {105014} (\bibinfo {year} {2012})},\ \Eprint {http://arxiv.org/abs/1201.5366}
  {arXiv:1201.5366 [hep-th]} \BibitemShut {NoStop}%
\bibitem [{\citenamefont {Bern}\ \emph {et~al.}(2017)\citenamefont {Bern},
  \citenamefont {Carrasco}, \citenamefont {Chen}, \citenamefont {Johansson},
  \citenamefont {Roiban},\ and\ \citenamefont {Zeng}}]{Bern:2017ucb}%
  \BibitemOpen
  \bibfield  {author} {\bibinfo {author} {\bibfnamefont {Zvi}\ \bibnamefont
  {Bern}}, \bibinfo {author} {\bibfnamefont {John Joseph~M.}\ \bibnamefont
  {Carrasco}}, \bibinfo {author} {\bibfnamefont {Wei-Ming}\ \bibnamefont
  {Chen}}, \bibinfo {author} {\bibfnamefont {Henrik}\ \bibnamefont
  {Johansson}}, \bibinfo {author} {\bibfnamefont {Radu}\ \bibnamefont
  {Roiban}}, \ and\ \bibinfo {author} {\bibfnamefont {Mao}\ \bibnamefont
  {Zeng}},\ }\bibfield  {title} {\enquote {\bibinfo {title} {{Five-loop
  four-point integrand of $N=8$ supergravity as a generalized double copy}},}\
  }\href {\doibase 10.1103/PhysRevD.96.126012} {\bibfield  {journal} {\bibinfo
  {journal} {Phys. Rev.}\ }\textbf {\bibinfo {volume} {D96}},\ \bibinfo {pages}
  {126012} (\bibinfo {year} {2017})},\ \Eprint
  {http://arxiv.org/abs/1708.06807} {arXiv:1708.06807 [hep-th]} \BibitemShut
  {NoStop}%
\bibitem [{\citenamefont {Bern}\ \emph {et~al.}(2018)\citenamefont {Bern},
  \citenamefont {Carrasco}, \citenamefont {Chen}, \citenamefont {Edison},
  \citenamefont {Johansson}, \citenamefont {Parra-Martinez}, \citenamefont
  {Roiban},\ and\ \citenamefont {Zeng}}]{Bern:2018jmv}%
  \BibitemOpen
  \bibfield  {author} {\bibinfo {author} {\bibfnamefont {Zvi}\ \bibnamefont
  {Bern}}, \bibinfo {author} {\bibfnamefont {John~Joseph}\ \bibnamefont
  {Carrasco}}, \bibinfo {author} {\bibfnamefont {Wei-Ming}\ \bibnamefont
  {Chen}}, \bibinfo {author} {\bibfnamefont {Alex}\ \bibnamefont {Edison}},
  \bibinfo {author} {\bibfnamefont {Henrik}\ \bibnamefont {Johansson}},
  \bibinfo {author} {\bibfnamefont {Julio}\ \bibnamefont {Parra-Martinez}},
  \bibinfo {author} {\bibfnamefont {Radu}\ \bibnamefont {Roiban}}, \ and\
  \bibinfo {author} {\bibfnamefont {Mao}\ \bibnamefont {Zeng}},\ }\bibfield
  {title} {\enquote {\bibinfo {title} {{Ultraviolet Properties of $\mathcal N =
  8$ Supergravity at Five Loops}},}\ }\href {\doibase
  10.1103/PhysRevD.98.086021} {\bibfield  {journal} {\bibinfo  {journal} {Phys.
  Rev.}\ }\textbf {\bibinfo {volume} {D98}},\ \bibinfo {pages} {086021}
  (\bibinfo {year} {2018})},\ \Eprint {http://arxiv.org/abs/1804.09311}
  {arXiv:1804.09311 [hep-th]} \BibitemShut {NoStop}%
\bibitem [{\citenamefont {Bern}\ \emph
  {et~al.}(2019{\natexlab{a}})\citenamefont {Bern}, \citenamefont {Carrasco},
  \citenamefont {Chiodaroli}, \citenamefont {Johansson},\ and\ \citenamefont
  {Roiban}}]{Bern:2019prr}%
  \BibitemOpen
  \bibfield  {author} {\bibinfo {author} {\bibfnamefont {Zvi}\ \bibnamefont
  {Bern}}, \bibinfo {author} {\bibfnamefont {John~Joseph}\ \bibnamefont
  {Carrasco}}, \bibinfo {author} {\bibfnamefont {Marco}\ \bibnamefont
  {Chiodaroli}}, \bibinfo {author} {\bibfnamefont {Henrik}\ \bibnamefont
  {Johansson}}, \ and\ \bibinfo {author} {\bibfnamefont {Radu}\ \bibnamefont
  {Roiban}},\ }\bibfield  {title} {\enquote {\bibinfo {title} {{The Duality
  Between Color and Kinematics and its Applications}},}\ }\href@noop {} {\
  (\bibinfo {year} {2019}{\natexlab{a}})},\ \Eprint
  {http://arxiv.org/abs/1909.01358} {arXiv:1909.01358 [hep-th]} \BibitemShut
  {NoStop}%
\bibitem [{\citenamefont {Bern}\ \emph
  {et~al.}(2019{\natexlab{b}})\citenamefont {Bern}, \citenamefont {Cheung},
  \citenamefont {Roiban}, \citenamefont {Shen}, \citenamefont {Solon},\ and\
  \citenamefont {Zeng}}]{Bern:2019crd}%
  \BibitemOpen
  \bibfield  {author} {\bibinfo {author} {\bibfnamefont {Zvi}\ \bibnamefont
  {Bern}}, \bibinfo {author} {\bibfnamefont {Clifford}\ \bibnamefont {Cheung}},
  \bibinfo {author} {\bibfnamefont {Radu}\ \bibnamefont {Roiban}}, \bibinfo
  {author} {\bibfnamefont {Chia-Hsien}\ \bibnamefont {Shen}}, \bibinfo {author}
  {\bibfnamefont {Mikhail~P.}\ \bibnamefont {Solon}}, \ and\ \bibinfo {author}
  {\bibfnamefont {Mao}\ \bibnamefont {Zeng}},\ }\bibfield  {title} {\enquote
  {\bibinfo {title} {{Black Hole Binary Dynamics from the Double Copy and
  Effective Theory}},}\ }\href {\doibase 10.1007/JHEP10(2019)206} {\bibfield
  {journal} {\bibinfo  {journal} {JHEP}\ }\textbf {\bibinfo {volume} {10}},\
  \bibinfo {pages} {206} (\bibinfo {year} {2019}{\natexlab{b}})},\ \Eprint
  {http://arxiv.org/abs/1908.01493} {arXiv:1908.01493 [hep-th]} \BibitemShut
  {NoStop}%
\bibitem [{\citenamefont {Parra-Martinez}\ \emph {et~al.}(2020)\citenamefont
  {Parra-Martinez}, \citenamefont {Ruf},\ and\ \citenamefont
  {Zeng}}]{Parra-Martinez:2020dzs}%
  \BibitemOpen
  \bibfield  {author} {\bibinfo {author} {\bibfnamefont {Julio}\ \bibnamefont
  {Parra-Martinez}}, \bibinfo {author} {\bibfnamefont {Michael~S.}\
  \bibnamefont {Ruf}}, \ and\ \bibinfo {author} {\bibfnamefont {Mao}\
  \bibnamefont {Zeng}},\ }\bibfield  {title} {\enquote {\bibinfo {title}
  {{Extremal black hole scattering at $\mathcal{O}(G^3)$: graviton dominance,
  eikonal exponentiation, and differential equations}},}\ }\href {\doibase
  10.1007/JHEP11(2020)023} {\bibfield  {journal} {\bibinfo  {journal} {JHEP}\
  }\textbf {\bibinfo {volume} {11}},\ \bibinfo {pages} {023} (\bibinfo {year}
  {2020})},\ \Eprint {http://arxiv.org/abs/2005.04236} {arXiv:2005.04236
  [hep-th]} \BibitemShut {NoStop}%
\bibitem [{\citenamefont {Bern}\ \emph
  {et~al.}(2021{\natexlab{a}})\citenamefont {Bern}, \citenamefont
  {Parra-Martinez}, \citenamefont {Roiban}, \citenamefont {Ruf}, \citenamefont
  {Shen}, \citenamefont {Solon},\ and\ \citenamefont {Zeng}}]{Bern:2021dqo}%
  \BibitemOpen
  \bibfield  {author} {\bibinfo {author} {\bibfnamefont {Zvi}\ \bibnamefont
  {Bern}}, \bibinfo {author} {\bibfnamefont {Julio}\ \bibnamefont
  {Parra-Martinez}}, \bibinfo {author} {\bibfnamefont {Radu}\ \bibnamefont
  {Roiban}}, \bibinfo {author} {\bibfnamefont {Michael~S.}\ \bibnamefont
  {Ruf}}, \bibinfo {author} {\bibfnamefont {Chia-Hsien}\ \bibnamefont {Shen}},
  \bibinfo {author} {\bibfnamefont {Mikhail~P.}\ \bibnamefont {Solon}}, \ and\
  \bibinfo {author} {\bibfnamefont {Mao}\ \bibnamefont {Zeng}},\ }\bibfield
  {title} {\enquote {\bibinfo {title} {{Scattering Amplitudes and Conservative
  Binary Dynamics at ${\cal O}(G^4)$}},}\ }\href@noop {} {\  (\bibinfo {year}
  {2021}{\natexlab{a}})},\ \Eprint {http://arxiv.org/abs/2101.07254}
  {arXiv:2101.07254 [hep-th]} \BibitemShut {NoStop}%
\bibitem [{\citenamefont {Dlapa}\ \emph {et~al.}(2021)\citenamefont {Dlapa},
  \citenamefont {K\"alin}, \citenamefont {Liu},\ and\ \citenamefont
  {Porto}}]{Dlapa:2021npj}%
  \BibitemOpen
  \bibfield  {author} {\bibinfo {author} {\bibfnamefont {Christoph}\
  \bibnamefont {Dlapa}}, \bibinfo {author} {\bibfnamefont {Gregor}\
  \bibnamefont {K\"alin}}, \bibinfo {author} {\bibfnamefont {Zhengwen}\
  \bibnamefont {Liu}}, \ and\ \bibinfo {author} {\bibfnamefont {Rafael~A.}\
  \bibnamefont {Porto}},\ }\bibfield  {title} {\enquote {\bibinfo {title}
  {{Dynamics of Binary Systems to Fourth Post-Minkowskian Order from the
  Effective Field Theory Approach}},}\ }\href@noop {} {\  (\bibinfo {year}
  {2021})},\ \Eprint {http://arxiv.org/abs/2106.08276} {arXiv:2106.08276
  [hep-th]} \BibitemShut {NoStop}%
\bibitem [{\citenamefont {Bern}\ \emph
  {et~al.}(2019{\natexlab{c}})\citenamefont {Bern}, \citenamefont {Cheung},
  \citenamefont {Roiban}, \citenamefont {Shen}, \citenamefont {Solon},\ and\
  \citenamefont {Zeng}}]{Bern:2019nnu}%
  \BibitemOpen
  \bibfield  {author} {\bibinfo {author} {\bibfnamefont {Zvi}\ \bibnamefont
  {Bern}}, \bibinfo {author} {\bibfnamefont {Clifford}\ \bibnamefont {Cheung}},
  \bibinfo {author} {\bibfnamefont {Radu}\ \bibnamefont {Roiban}}, \bibinfo
  {author} {\bibfnamefont {Chia-Hsien}\ \bibnamefont {Shen}}, \bibinfo {author}
  {\bibfnamefont {Mikhail~P.}\ \bibnamefont {Solon}}, \ and\ \bibinfo {author}
  {\bibfnamefont {Mao}\ \bibnamefont {Zeng}},\ }\bibfield  {title} {\enquote
  {\bibinfo {title} {{Scattering Amplitudes and the Conservative Hamiltonian
  for Binary Systems at Third Post-Minkowskian Order}},}\ }\href {\doibase
  10.1103/PhysRevLett.122.201603} {\bibfield  {journal} {\bibinfo  {journal}
  {Phys. Rev. Lett.}\ }\textbf {\bibinfo {volume} {122}},\ \bibinfo {pages}
  {201603} (\bibinfo {year} {2019}{\natexlab{c}})},\ \Eprint
  {http://arxiv.org/abs/1901.04424} {arXiv:1901.04424 [hep-th]} \BibitemShut
  {NoStop}%
\bibitem [{\citenamefont {Cheung}\ and\ \citenamefont
  {Solon}(2020{\natexlab{a}})}]{Cheung:2020gyp}%
  \BibitemOpen
  \bibfield  {author} {\bibinfo {author} {\bibfnamefont {Clifford}\
  \bibnamefont {Cheung}}\ and\ \bibinfo {author} {\bibfnamefont {Mikhail~P.}\
  \bibnamefont {Solon}},\ }\bibfield  {title} {\enquote {\bibinfo {title}
  {{Classical gravitational scattering at $ \mathcal{O} $(G$^{3}$) from Feynman
  diagrams}},}\ }\href {\doibase 10.1007/JHEP06(2020)144} {\bibfield  {journal}
  {\bibinfo  {journal} {JHEP}\ }\textbf {\bibinfo {volume} {06}},\ \bibinfo
  {pages} {144} (\bibinfo {year} {2020}{\natexlab{a}})},\ \Eprint
  {http://arxiv.org/abs/2003.08351} {arXiv:2003.08351 [hep-th]} \BibitemShut
  {NoStop}%
\bibitem [{\citenamefont {{K\"alin, Gregor and Liu, Zhengwen and Porto, Rafael
  A.}}(2020)}]{Kalin:2020fhe}%
  \BibitemOpen
  \bibfield  {author} {\bibinfo {author} {\bibnamefont {{K\"alin, Gregor and
  Liu, Zhengwen and Porto, Rafael A.}}},\ }\bibfield  {title} {\enquote
  {\bibinfo {title} {{Conservative Dynamics of Binary Systems to Third
  Post-Minkowskian Order from the Effective Field Theory Approach}},}\ }\href
  {\doibase 10.1103/PhysRevLett.125.261103} {\bibfield  {journal} {\bibinfo
  {journal} {Phys. Rev. Lett.}\ }\textbf {\bibinfo {volume} {125}},\ \bibinfo
  {pages} {261103} (\bibinfo {year} {2020})},\ \Eprint
  {http://arxiv.org/abs/2007.04977} {arXiv:2007.04977 [hep-th]} \BibitemShut
  {NoStop}%
\bibitem [{\citenamefont {Di~Vecchia}\ \emph
  {et~al.}(2021{\natexlab{a}})\citenamefont {Di~Vecchia}, \citenamefont
  {Heissenberg}, \citenamefont {Russo},\ and\ \citenamefont
  {Veneziano}}]{DiVecchia:2021bdo}%
  \BibitemOpen
  \bibfield  {author} {\bibinfo {author} {\bibfnamefont {Paolo}\ \bibnamefont
  {Di~Vecchia}}, \bibinfo {author} {\bibfnamefont {Carlo}\ \bibnamefont
  {Heissenberg}}, \bibinfo {author} {\bibfnamefont {Rodolfo}\ \bibnamefont
  {Russo}}, \ and\ \bibinfo {author} {\bibfnamefont {Gabriele}\ \bibnamefont
  {Veneziano}},\ }\bibfield  {title} {\enquote {\bibinfo {title} {{The Eikonal
  Approach to Gravitational Scattering and Radiation at $\mathcal O(G^3)$}},}\
  }\href@noop {} {\  (\bibinfo {year} {2021}{\natexlab{a}})},\ \Eprint
  {http://arxiv.org/abs/2104.03256} {arXiv:2104.03256 [hep-th]} \BibitemShut
  {NoStop}%
\bibitem [{\citenamefont {Bjerrum-Bohr}\ \emph
  {et~al.}(2021{\natexlab{a}})\citenamefont {Bjerrum-Bohr}, \citenamefont
  {Damgaard}, \citenamefont {Plant\'e},\ and\ \citenamefont
  {Vanhove}}]{Bjerrum-Bohr:2021din}%
  \BibitemOpen
  \bibfield  {author} {\bibinfo {author} {\bibfnamefont {N.~E.~J.}\
  \bibnamefont {Bjerrum-Bohr}}, \bibinfo {author} {\bibfnamefont {P.~H.}\
  \bibnamefont {Damgaard}}, \bibinfo {author} {\bibfnamefont {L.}~\bibnamefont
  {Plant\'e}}, \ and\ \bibinfo {author} {\bibfnamefont {P.}~\bibnamefont
  {Vanhove}},\ }\bibfield  {title} {\enquote {\bibinfo {title} {{The Amplitude
  for Classical Gravitational Scattering at Third Post-Minkowskian Order}},}\
  }\href@noop {} {\  (\bibinfo {year} {2021}{\natexlab{a}})},\ \Eprint
  {http://arxiv.org/abs/2105.05218} {arXiv:2105.05218 [hep-th]} \BibitemShut
  {NoStop}%
\bibitem [{\citenamefont {Bini}\ \emph {et~al.}(2020)\citenamefont {Bini},
  \citenamefont {Damour},\ and\ \citenamefont {Geralico}}]{Bini:2020flp}%
  \BibitemOpen
  \bibfield  {author} {\bibinfo {author} {\bibfnamefont {Donato}\ \bibnamefont
  {Bini}}, \bibinfo {author} {\bibfnamefont {Thibault}\ \bibnamefont {Damour}},
  \ and\ \bibinfo {author} {\bibfnamefont {Andrea}\ \bibnamefont {Geralico}},\
  }\bibfield  {title} {\enquote {\bibinfo {title} {{Scattering of tidally
  interacting bodies in post-Minkowskian gravity}},}\ }\href {\doibase
  10.1103/PhysRevD.101.044039} {\bibfield  {journal} {\bibinfo  {journal}
  {Phys. Rev.}\ }\textbf {\bibinfo {volume} {D101}},\ \bibinfo {pages} {044039}
  (\bibinfo {year} {2020})},\ \Eprint {http://arxiv.org/abs/2001.00352}
  {arXiv:2001.00352 [gr-qc]} \BibitemShut {NoStop}%
\bibitem [{\citenamefont {Cheung}\ and\ \citenamefont
  {Solon}(2020{\natexlab{b}})}]{Cheung:2020sdj}%
  \BibitemOpen
  \bibfield  {author} {\bibinfo {author} {\bibfnamefont {Clifford}\
  \bibnamefont {Cheung}}\ and\ \bibinfo {author} {\bibfnamefont {Mikhail~P.}\
  \bibnamefont {Solon}},\ }\bibfield  {title} {\enquote {\bibinfo {title}
  {{Tidal Effects in the Post-Minkowskian Expansion}},}\ }\href {\doibase
  10.1103/PhysRevLett.125.191601} {\bibfield  {journal} {\bibinfo  {journal}
  {Phys. Rev. Lett.}\ }\textbf {\bibinfo {volume} {125}},\ \bibinfo {pages}
  {191601} (\bibinfo {year} {2020}{\natexlab{b}})},\ \Eprint
  {http://arxiv.org/abs/2006.06665} {arXiv:2006.06665 [hep-th]} \BibitemShut
  {NoStop}%
\bibitem [{\citenamefont {Haddad}\ and\ \citenamefont
  {Helset}(2020)}]{Haddad:2020que}%
  \BibitemOpen
  \bibfield  {author} {\bibinfo {author} {\bibfnamefont {Kays}\ \bibnamefont
  {Haddad}}\ and\ \bibinfo {author} {\bibfnamefont {Andreas}\ \bibnamefont
  {Helset}},\ }\bibfield  {title} {\enquote {\bibinfo {title} {{Tidal effects
  in quantum field theory}},}\ }\href {\doibase 10.1007/JHEP12(2020)024}
  {\bibfield  {journal} {\bibinfo  {journal} {JHEP}\ }\textbf {\bibinfo
  {volume} {12}},\ \bibinfo {pages} {024} (\bibinfo {year} {2020})},\ \Eprint
  {http://arxiv.org/abs/2008.04920} {arXiv:2008.04920 [hep-th]} \BibitemShut
  {NoStop}%
\bibitem [{\citenamefont {K\"alin}\ \emph {et~al.}(2020)\citenamefont
  {K\"alin}, \citenamefont {Liu},\ and\ \citenamefont {Porto}}]{Kalin:2020lmz}%
  \BibitemOpen
  \bibfield  {author} {\bibinfo {author} {\bibfnamefont {Gregor}\ \bibnamefont
  {K\"alin}}, \bibinfo {author} {\bibfnamefont {Zhengwen}\ \bibnamefont {Liu}},
  \ and\ \bibinfo {author} {\bibfnamefont {Rafael~A.}\ \bibnamefont {Porto}},\
  }\bibfield  {title} {\enquote {\bibinfo {title} {{Conservative Tidal Effects
  in Compact Binary Systems to Next-to-Leading Post-Minkowskian Order}},}\
  }\href {\doibase 10.1103/PhysRevD.102.124025} {\bibfield  {journal} {\bibinfo
   {journal} {Phys. Rev. D}\ }\textbf {\bibinfo {volume} {102}},\ \bibinfo
  {pages} {124025} (\bibinfo {year} {2020})},\ \Eprint
  {http://arxiv.org/abs/2008.06047} {arXiv:2008.06047 [hep-th]} \BibitemShut
  {NoStop}%
\bibitem [{\citenamefont {Brandhuber}\ and\ \citenamefont
  {Travaglini}(2020)}]{Brandhuber:2019qpg}%
  \BibitemOpen
  \bibfield  {author} {\bibinfo {author} {\bibfnamefont {Andreas}\ \bibnamefont
  {Brandhuber}}\ and\ \bibinfo {author} {\bibfnamefont {Gabriele}\ \bibnamefont
  {Travaglini}},\ }\bibfield  {title} {\enquote {\bibinfo {title} {{On
  higher-derivative effects on the gravitational potential and particle
  bending}},}\ }\href {\doibase 10.1007/JHEP01(2020)010} {\bibfield  {journal}
  {\bibinfo  {journal} {JHEP}\ }\textbf {\bibinfo {volume} {01}},\ \bibinfo
  {pages} {010} (\bibinfo {year} {2020})},\ \Eprint
  {http://arxiv.org/abs/1905.05657} {arXiv:1905.05657 [hep-th]} \BibitemShut
  {NoStop}%
\bibitem [{\citenamefont {Accettulli~Huber}\ \emph
  {et~al.}(2020{\natexlab{a}})\citenamefont {Accettulli~Huber}, \citenamefont
  {Brandhuber}, \citenamefont {De~Angelis},\ and\ \citenamefont
  {Travaglini}}]{Huber:2019ugz}%
  \BibitemOpen
  \bibfield  {author} {\bibinfo {author} {\bibfnamefont {Manuel}\ \bibnamefont
  {Accettulli~Huber}}, \bibinfo {author} {\bibfnamefont {Andreas}\ \bibnamefont
  {Brandhuber}}, \bibinfo {author} {\bibfnamefont {Stefano}\ \bibnamefont
  {De~Angelis}}, \ and\ \bibinfo {author} {\bibfnamefont {Gabriele}\
  \bibnamefont {Travaglini}},\ }\bibfield  {title} {\enquote {\bibinfo {title}
  {{Note on the absence of $R^2$ corrections to Newton’s potential}},}\
  }\href {\doibase 10.1103/PhysRevD.101.046011} {\bibfield  {journal} {\bibinfo
   {journal} {Phys. Rev.}\ }\textbf {\bibinfo {volume} {D101}},\ \bibinfo
  {pages} {046011} (\bibinfo {year} {2020}{\natexlab{a}})},\ \Eprint
  {http://arxiv.org/abs/1911.10108} {arXiv:1911.10108 [hep-th]} \BibitemShut
  {NoStop}%
\bibitem [{\citenamefont {Accettulli~Huber}\ \emph
  {et~al.}(2020{\natexlab{b}})\citenamefont {Accettulli~Huber}, \citenamefont
  {Brandhuber}, \citenamefont {De~Angelis},\ and\ \citenamefont
  {Travaglini}}]{AccettulliHuber:2020oou}%
  \BibitemOpen
  \bibfield  {author} {\bibinfo {author} {\bibfnamefont {Manuel}\ \bibnamefont
  {Accettulli~Huber}}, \bibinfo {author} {\bibfnamefont {Andreas}\ \bibnamefont
  {Brandhuber}}, \bibinfo {author} {\bibfnamefont {Stefano}\ \bibnamefont
  {De~Angelis}}, \ and\ \bibinfo {author} {\bibfnamefont {Gabriele}\
  \bibnamefont {Travaglini}},\ }\bibfield  {title} {\enquote {\bibinfo {title}
  {{Eikonal phase matrix, deflection angle and time delay in effective field
  theories of gravity}},}\ }\href {\doibase 10.1103/PhysRevD.102.046014}
  {\bibfield  {journal} {\bibinfo  {journal} {Phys. Rev.}\ }\textbf {\bibinfo
  {volume} {D102}},\ \bibinfo {pages} {046014} (\bibinfo {year}
  {2020}{\natexlab{b}})},\ \Eprint {http://arxiv.org/abs/2006.02375}
  {arXiv:2006.02375 [hep-th]} \BibitemShut {NoStop}%
\bibitem [{\citenamefont {Bern}\ \emph
  {et~al.}(2021{\natexlab{b}})\citenamefont {Bern}, \citenamefont
  {Parra-Martinez}, \citenamefont {Roiban}, \citenamefont {Sawyer},\ and\
  \citenamefont {Shen}}]{Bern:2020uwk}%
  \BibitemOpen
  \bibfield  {author} {\bibinfo {author} {\bibfnamefont {Zvi}\ \bibnamefont
  {Bern}}, \bibinfo {author} {\bibfnamefont {Julio}\ \bibnamefont
  {Parra-Martinez}}, \bibinfo {author} {\bibfnamefont {Radu}\ \bibnamefont
  {Roiban}}, \bibinfo {author} {\bibfnamefont {Eric}\ \bibnamefont {Sawyer}}, \
  and\ \bibinfo {author} {\bibfnamefont {Chia-Hsien}\ \bibnamefont {Shen}},\
  }\bibfield  {title} {\enquote {\bibinfo {title} {{Leading Nonlinear Tidal
  Effects and Scattering Amplitudes}},}\ }\href {\doibase
  10.1007/JHEP05(2021)188} {\bibfield  {journal} {\bibinfo  {journal} {JHEP}\
  }\textbf {\bibinfo {volume} {05}},\ \bibinfo {pages} {188} (\bibinfo {year}
  {2021}{\natexlab{b}})},\ \Eprint {http://arxiv.org/abs/2010.08559}
  {arXiv:2010.08559 [hep-th]} \BibitemShut {NoStop}%
\bibitem [{\citenamefont {Cheung}\ \emph {et~al.}(2021)\citenamefont {Cheung},
  \citenamefont {Shah},\ and\ \citenamefont {Solon}}]{Cheung:2020gbf}%
  \BibitemOpen
  \bibfield  {author} {\bibinfo {author} {\bibfnamefont {Clifford}\
  \bibnamefont {Cheung}}, \bibinfo {author} {\bibfnamefont {Nabha}\
  \bibnamefont {Shah}}, \ and\ \bibinfo {author} {\bibfnamefont {Mikhail~P.}\
  \bibnamefont {Solon}},\ }\bibfield  {title} {\enquote {\bibinfo {title}
  {{Mining the Geodesic Equation for Scattering Data}},}\ }\href {\doibase
  10.1103/PhysRevD.103.024030} {\bibfield  {journal} {\bibinfo  {journal}
  {Phys. Rev.}\ }\textbf {\bibinfo {volume} {D103}},\ \bibinfo {pages} {024030}
  (\bibinfo {year} {2021})},\ \Eprint {http://arxiv.org/abs/2010.08568}
  {arXiv:2010.08568 [hep-th]} \BibitemShut {NoStop}%
\bibitem [{\citenamefont {Aoude}\ \emph {et~al.}(2021)\citenamefont {Aoude},
  \citenamefont {Haddad},\ and\ \citenamefont {Helset}}]{Aoude:2020ygw}%
  \BibitemOpen
  \bibfield  {author} {\bibinfo {author} {\bibfnamefont {Rafael}\ \bibnamefont
  {Aoude}}, \bibinfo {author} {\bibfnamefont {Kays}\ \bibnamefont {Haddad}}, \
  and\ \bibinfo {author} {\bibfnamefont {Andreas}\ \bibnamefont {Helset}},\
  }\bibfield  {title} {\enquote {\bibinfo {title} {{Tidal effects for spinning
  particles}},}\ }\href {\doibase 10.1007/JHEP03(2021)097} {\bibfield
  {journal} {\bibinfo  {journal} {JHEP}\ }\textbf {\bibinfo {volume} {03}},\
  \bibinfo {pages} {097} (\bibinfo {year} {2021})},\ \Eprint
  {http://arxiv.org/abs/2012.05256} {arXiv:2012.05256 [hep-th]} \BibitemShut
  {NoStop}%
\bibitem [{\citenamefont {Amati}\ \emph {et~al.}(1990)\citenamefont {Amati},
  \citenamefont {Ciafaloni},\ and\ \citenamefont {Veneziano}}]{Amati:1990xe}%
  \BibitemOpen
  \bibfield  {author} {\bibinfo {author} {\bibfnamefont {D.}~\bibnamefont
  {Amati}}, \bibinfo {author} {\bibfnamefont {M.}~\bibnamefont {Ciafaloni}}, \
  and\ \bibinfo {author} {\bibfnamefont {G.}~\bibnamefont {Veneziano}},\
  }\bibfield  {title} {\enquote {\bibinfo {title} {{Higher Order Gravitational
  Deflection and Soft Bremsstrahlung in Planckian Energy Superstring
  Collisions}},}\ }\href {\doibase 10.1016/0550-3213(90)90375-N} {\bibfield
  {journal} {\bibinfo  {journal} {Nucl. Phys.}\ }\textbf {\bibinfo {volume}
  {B347}},\ \bibinfo {pages} {550--580} (\bibinfo {year} {1990})}\BibitemShut
  {NoStop}%
\bibitem [{\citenamefont {Di~Vecchia}\ \emph {et~al.}(2019)\citenamefont
  {Di~Vecchia}, \citenamefont {Luna}, \citenamefont {Naculich}, \citenamefont
  {Russo}, \citenamefont {Veneziano},\ and\ \citenamefont
  {White}}]{DiVecchia:2019myk}%
  \BibitemOpen
  \bibfield  {author} {\bibinfo {author} {\bibfnamefont {Paolo}\ \bibnamefont
  {Di~Vecchia}}, \bibinfo {author} {\bibfnamefont {Andrés}\ \bibnamefont
  {Luna}}, \bibinfo {author} {\bibfnamefont {Stephen~G.}\ \bibnamefont
  {Naculich}}, \bibinfo {author} {\bibfnamefont {Rodolfo}\ \bibnamefont
  {Russo}}, \bibinfo {author} {\bibfnamefont {Gabriele}\ \bibnamefont
  {Veneziano}}, \ and\ \bibinfo {author} {\bibfnamefont {Chris~D.}\
  \bibnamefont {White}},\ }\bibfield  {title} {\enquote {\bibinfo {title} {{A
  tale of two exponentiations in ${\cal N}=8$ supergravity}},}\ }\href
  {\doibase 10.1016/j.physletb.2019.134927} {\bibfield  {journal} {\bibinfo
  {journal} {Phys. Lett.}\ }\textbf {\bibinfo {volume} {B798}},\ \bibinfo
  {pages} {134927} (\bibinfo {year} {2019})},\ \Eprint
  {http://arxiv.org/abs/1908.05603} {arXiv:1908.05603 [hep-th]} \BibitemShut
  {NoStop}%
\bibitem [{\citenamefont {Di~Vecchia}\ \emph
  {et~al.}(2020{\natexlab{b}})\citenamefont {Di~Vecchia}, \citenamefont
  {Naculich}, \citenamefont {Russo}, \citenamefont {Veneziano},\ and\
  \citenamefont {White}}]{DiVecchia:2019kta}%
  \BibitemOpen
  \bibfield  {author} {\bibinfo {author} {\bibfnamefont {Paolo}\ \bibnamefont
  {Di~Vecchia}}, \bibinfo {author} {\bibfnamefont {Stephen~G.}\ \bibnamefont
  {Naculich}}, \bibinfo {author} {\bibfnamefont {Rodolfo}\ \bibnamefont
  {Russo}}, \bibinfo {author} {\bibfnamefont {Gabriele}\ \bibnamefont
  {Veneziano}}, \ and\ \bibinfo {author} {\bibfnamefont {Chris~D.}\
  \bibnamefont {White}},\ }\bibfield  {title} {\enquote {\bibinfo {title} {{A
  tale of two exponentiations in $ \mathcal{N} $ = 8 supergravity at subleading
  level}},}\ }\href {\doibase 10.1007/JHEP03(2020)173} {\bibfield  {journal}
  {\bibinfo  {journal} {JHEP}\ }\textbf {\bibinfo {volume} {03}},\ \bibinfo
  {pages} {173} (\bibinfo {year} {2020}{\natexlab{b}})},\ \Eprint
  {http://arxiv.org/abs/1911.11716} {arXiv:1911.11716 [hep-th]} \BibitemShut
  {NoStop}%
\bibitem [{\citenamefont {Bern}\ \emph
  {et~al.}(2020{\natexlab{b}})\citenamefont {Bern}, \citenamefont {Ita},
  \citenamefont {Parra-Martinez},\ and\ \citenamefont {Ruf}}]{Bern:2020gjj}%
  \BibitemOpen
  \bibfield  {author} {\bibinfo {author} {\bibfnamefont {Zvi}\ \bibnamefont
  {Bern}}, \bibinfo {author} {\bibfnamefont {Harald}\ \bibnamefont {Ita}},
  \bibinfo {author} {\bibfnamefont {Julio}\ \bibnamefont {Parra-Martinez}}, \
  and\ \bibinfo {author} {\bibfnamefont {Michael~S.}\ \bibnamefont {Ruf}},\
  }\bibfield  {title} {\enquote {\bibinfo {title} {{Universality in the
  classical limit of massless gravitational scattering}},}\ }\href {\doibase
  10.1103/PhysRevLett.125.031601} {\bibfield  {journal} {\bibinfo  {journal}
  {Phys. Rev. Lett.}\ }\textbf {\bibinfo {volume} {125}},\ \bibinfo {pages}
  {031601} (\bibinfo {year} {2020}{\natexlab{b}})},\ \Eprint
  {http://arxiv.org/abs/2002.02459} {arXiv:2002.02459 [hep-th]} \BibitemShut
  {NoStop}%
\bibitem [{\citenamefont {Accettulli~Huber}\ \emph
  {et~al.}(2020{\natexlab{c}})\citenamefont {Accettulli~Huber}, \citenamefont
  {Brandhuber}, \citenamefont {De~Angelis},\ and\ \citenamefont
  {Travaglini}}]{Huber:2020xny}%
  \BibitemOpen
  \bibfield  {author} {\bibinfo {author} {\bibfnamefont {Manuel}\ \bibnamefont
  {Accettulli~Huber}}, \bibinfo {author} {\bibfnamefont {Andreas}\ \bibnamefont
  {Brandhuber}}, \bibinfo {author} {\bibfnamefont {Stefano}\ \bibnamefont
  {De~Angelis}}, \ and\ \bibinfo {author} {\bibfnamefont {Gabriele}\
  \bibnamefont {Travaglini}},\ }\bibfield  {title} {\enquote {\bibinfo {title}
  {{From amplitudes to gravitational radiation with cubic interactions and
  tidal effects}},}\ }\href@noop {} {\  (\bibinfo {year}
  {2020}{\natexlab{c}})},\ \Eprint {http://arxiv.org/abs/2012.06548}
  {arXiv:2012.06548 [hep-th]} \BibitemShut {NoStop}%
\bibitem [{\citenamefont {Di~Vecchia}\ \emph
  {et~al.}(2021{\natexlab{b}})\citenamefont {Di~Vecchia}, \citenamefont
  {Heissenberg}, \citenamefont {Russo},\ and\ \citenamefont
  {Veneziano}}]{DiVecchia:2021ndb}%
  \BibitemOpen
  \bibfield  {author} {\bibinfo {author} {\bibfnamefont {Paolo}\ \bibnamefont
  {Di~Vecchia}}, \bibinfo {author} {\bibfnamefont {Carlo}\ \bibnamefont
  {Heissenberg}}, \bibinfo {author} {\bibfnamefont {Rodolfo}\ \bibnamefont
  {Russo}}, \ and\ \bibinfo {author} {\bibfnamefont {Gabriele}\ \bibnamefont
  {Veneziano}},\ }\bibfield  {title} {\enquote {\bibinfo {title} {{Radiation
  Reaction from Soft Theorems}},}\ }\href@noop {} {\  (\bibinfo {year}
  {2021}{\natexlab{b}})},\ \Eprint {http://arxiv.org/abs/2101.05772}
  {arXiv:2101.05772 [hep-th]} \BibitemShut {NoStop}%
\bibitem [{\citenamefont {Bautista}\ and\ \citenamefont
  {Guevara}(2019)}]{Bautista:2019tdr}%
  \BibitemOpen
  \bibfield  {author} {\bibinfo {author} {\bibfnamefont {Yilber~Fabian}\
  \bibnamefont {Bautista}}\ and\ \bibinfo {author} {\bibfnamefont {Alfredo}\
  \bibnamefont {Guevara}},\ }\bibfield  {title} {\enquote {\bibinfo {title}
  {{From Scattering Amplitudes to Classical Physics: Universality, Double Copy
  and Soft Theorems}},}\ }\href@noop {} {\  (\bibinfo {year} {2019})},\ \Eprint
  {http://arxiv.org/abs/1903.12419} {arXiv:1903.12419 [hep-th]} \BibitemShut
  {NoStop}%
\bibitem [{\citenamefont {Laddha}\ and\ \citenamefont
  {Sen}(2018{\natexlab{a}})}]{Laddha:2018rle}%
  \BibitemOpen
  \bibfield  {author} {\bibinfo {author} {\bibfnamefont {Alok}\ \bibnamefont
  {Laddha}}\ and\ \bibinfo {author} {\bibfnamefont {Ashoke}\ \bibnamefont
  {Sen}},\ }\bibfield  {title} {\enquote {\bibinfo {title} {{Gravity Waves from
  Soft Theorem in General Dimensions}},}\ }\href {\doibase
  10.1007/JHEP09(2018)105} {\bibfield  {journal} {\bibinfo  {journal} {JHEP}\
  }\textbf {\bibinfo {volume} {09}},\ \bibinfo {pages} {105} (\bibinfo {year}
  {2018}{\natexlab{a}})},\ \Eprint {http://arxiv.org/abs/1801.07719}
  {arXiv:1801.07719 [hep-th]} \BibitemShut {NoStop}%
\bibitem [{\citenamefont {Laddha}\ and\ \citenamefont
  {Sen}(2018{\natexlab{b}})}]{Laddha:2018myi}%
  \BibitemOpen
  \bibfield  {author} {\bibinfo {author} {\bibfnamefont {Alok}\ \bibnamefont
  {Laddha}}\ and\ \bibinfo {author} {\bibfnamefont {Ashoke}\ \bibnamefont
  {Sen}},\ }\bibfield  {title} {\enquote {\bibinfo {title} {{Logarithmic Terms
  in the Soft Expansion in Four Dimensions}},}\ }\href {\doibase
  10.1007/JHEP10(2018)056} {\bibfield  {journal} {\bibinfo  {journal} {JHEP}\
  }\textbf {\bibinfo {volume} {10}},\ \bibinfo {pages} {056} (\bibinfo {year}
  {2018}{\natexlab{b}})},\ \Eprint {http://arxiv.org/abs/1804.09193}
  {arXiv:1804.09193 [hep-th]} \BibitemShut {NoStop}%
\bibitem [{\citenamefont {Sahoo}\ and\ \citenamefont
  {Sen}(2019)}]{Sahoo:2018lxl}%
  \BibitemOpen
  \bibfield  {author} {\bibinfo {author} {\bibfnamefont {Biswajit}\
  \bibnamefont {Sahoo}}\ and\ \bibinfo {author} {\bibfnamefont {Ashoke}\
  \bibnamefont {Sen}},\ }\bibfield  {title} {\enquote {\bibinfo {title}
  {{Classical and Quantum Results on Logarithmic Terms in the Soft Theorem in
  Four Dimensions}},}\ }\href {\doibase 10.1007/JHEP02(2019)086} {\bibfield
  {journal} {\bibinfo  {journal} {JHEP}\ }\textbf {\bibinfo {volume} {02}},\
  \bibinfo {pages} {086} (\bibinfo {year} {2019})},\ \Eprint
  {http://arxiv.org/abs/1808.03288} {arXiv:1808.03288 [hep-th]} \BibitemShut
  {NoStop}%
\bibitem [{\citenamefont {Laddha}\ and\ \citenamefont
  {Sen}(2020)}]{Laddha:2019yaj}%
  \BibitemOpen
  \bibfield  {author} {\bibinfo {author} {\bibfnamefont {Alok}\ \bibnamefont
  {Laddha}}\ and\ \bibinfo {author} {\bibfnamefont {Ashoke}\ \bibnamefont
  {Sen}},\ }\bibfield  {title} {\enquote {\bibinfo {title} {{Classical proof of
  the classical soft graviton theorem in $D>4$}},}\ }\href {\doibase
  10.1103/PhysRevD.101.084011} {\bibfield  {journal} {\bibinfo  {journal}
  {Phys. Rev.}\ }\textbf {\bibinfo {volume} {D101}},\ \bibinfo {pages} {084011}
  (\bibinfo {year} {2020})},\ \Eprint {http://arxiv.org/abs/1906.08288}
  {arXiv:1906.08288 [gr-qc]} \BibitemShut {NoStop}%
\bibitem [{\citenamefont {Saha}\ \emph {et~al.}(2020)\citenamefont {Saha},
  \citenamefont {Sahoo},\ and\ \citenamefont {Sen}}]{Saha:2019tub}%
  \BibitemOpen
  \bibfield  {author} {\bibinfo {author} {\bibfnamefont {Arnab~Priya}\
  \bibnamefont {Saha}}, \bibinfo {author} {\bibfnamefont {Biswajit}\
  \bibnamefont {Sahoo}}, \ and\ \bibinfo {author} {\bibfnamefont {Ashoke}\
  \bibnamefont {Sen}},\ }\bibfield  {title} {\enquote {\bibinfo {title} {{Proof
  of the classical soft graviton theorem in $D$ = 4}},}\ }\href {\doibase
  10.1007/JHEP06(2020)153} {\bibfield  {journal} {\bibinfo  {journal} {JHEP}\
  }\textbf {\bibinfo {volume} {06}},\ \bibinfo {pages} {153} (\bibinfo {year}
  {2020})},\ \Eprint {http://arxiv.org/abs/1912.06413} {arXiv:1912.06413
  [hep-th]} \BibitemShut {NoStop}%
\bibitem [{\citenamefont {A}\ \emph {et~al.}(2020)\citenamefont {A},
  \citenamefont {Ghosh}, \citenamefont {Laddha},\ and\ \citenamefont
  {Athira}}]{A:2020lub}%
  \BibitemOpen
  \bibfield  {author} {\bibinfo {author} {\bibfnamefont {Manu}\ \bibnamefont
  {A}}, \bibinfo {author} {\bibfnamefont {Debodirna}\ \bibnamefont {Ghosh}},
  \bibinfo {author} {\bibfnamefont {Alok}\ \bibnamefont {Laddha}}, \ and\
  \bibinfo {author} {\bibfnamefont {P.~V.}\ \bibnamefont {Athira}},\ }\bibfield
   {title} {\enquote {\bibinfo {title} {{Soft Radiation from Scattering
  Amplitudes Revisited}},}\ }\href@noop {} {\  (\bibinfo {year} {2020})},\
  \Eprint {http://arxiv.org/abs/2007.02077} {arXiv:2007.02077 [hep-th]}
  \BibitemShut {NoStop}%
\bibitem [{\citenamefont {Sahoo}(2020)}]{Sahoo:2020ryf}%
  \BibitemOpen
  \bibfield  {author} {\bibinfo {author} {\bibfnamefont {Biswajit}\
  \bibnamefont {Sahoo}},\ }\bibfield  {title} {\enquote {\bibinfo {title}
  {{Classical Sub-subleading Soft Photon and Soft Graviton Theorems in Four
  Spacetime Dimensions}},}\ }\href {\doibase 10.1007/JHEP12(2020)070}
  {\bibfield  {journal} {\bibinfo  {journal} {JHEP}\ }\textbf {\bibinfo
  {volume} {12}},\ \bibinfo {pages} {070} (\bibinfo {year} {2020})},\ \Eprint
  {http://arxiv.org/abs/2008.04376} {arXiv:2008.04376 [hep-th]} \BibitemShut
  {NoStop}%
\bibitem [{\citenamefont {Damour}(2020{\natexlab{b}})}]{Damour:2019lcq}%
  \BibitemOpen
  \bibfield  {author} {\bibinfo {author} {\bibfnamefont {Thibault}\
  \bibnamefont {Damour}},\ }\bibfield  {title} {\enquote {\bibinfo {title}
  {{Classical and quantum scattering in post-Minkowskian gravity}},}\ }\href
  {\doibase 10.1103/PhysRevD.102.024060} {\bibfield  {journal} {\bibinfo
  {journal} {Phys. Rev.}\ }\textbf {\bibinfo {volume} {D102}},\ \bibinfo
  {pages} {024060} (\bibinfo {year} {2020}{\natexlab{b}})},\ \Eprint
  {http://arxiv.org/abs/1912.02139} {arXiv:1912.02139 [gr-qc]} \BibitemShut
  {NoStop}%
\bibitem [{\citenamefont {Bjerrum-Bohr}\ \emph {et~al.}(2020)\citenamefont
  {Bjerrum-Bohr}, \citenamefont {Cristofoli},\ and\ \citenamefont
  {Damgaard}}]{Bjerrum-Bohr:2019kec}%
  \BibitemOpen
  \bibfield  {author} {\bibinfo {author} {\bibfnamefont {N.~E.~J.}\
  \bibnamefont {Bjerrum-Bohr}}, \bibinfo {author} {\bibfnamefont {Andrea}\
  \bibnamefont {Cristofoli}}, \ and\ \bibinfo {author} {\bibfnamefont
  {Poul~H.}\ \bibnamefont {Damgaard}},\ }\bibfield  {title} {\enquote {\bibinfo
  {title} {{Post-Minkowskian Scattering Angle in Einstein Gravity}},}\ }\href
  {\doibase 10.1007/JHEP08(2020)038} {\bibfield  {journal} {\bibinfo  {journal}
  {JHEP}\ }\textbf {\bibinfo {volume} {08}},\ \bibinfo {pages} {038} (\bibinfo
  {year} {2020})},\ \Eprint {http://arxiv.org/abs/1910.09366} {arXiv:1910.09366
  [hep-th]} \BibitemShut {NoStop}%
\bibitem [{\citenamefont {Bjerrum-Bohr}\ \emph
  {et~al.}(2021{\natexlab{b}})\citenamefont {Bjerrum-Bohr}, \citenamefont
  {Damgaard}, \citenamefont {Plant\'e},\ and\ \citenamefont
  {Vanhove}}]{Bjerrum-Bohr:2021vuf}%
  \BibitemOpen
  \bibfield  {author} {\bibinfo {author} {\bibfnamefont {N.~Emil~J.}\
  \bibnamefont {Bjerrum-Bohr}}, \bibinfo {author} {\bibfnamefont {Poul~H.}\
  \bibnamefont {Damgaard}}, \bibinfo {author} {\bibfnamefont {Ludovic}\
  \bibnamefont {Plant\'e}}, \ and\ \bibinfo {author} {\bibfnamefont {Pierre}\
  \bibnamefont {Vanhove}},\ }\bibfield  {title} {\enquote {\bibinfo {title}
  {{Classical Gravity from Loop Amplitudes}},}\ }\href@noop {} {\  (\bibinfo
  {year} {2021}{\natexlab{b}})},\ \Eprint {http://arxiv.org/abs/2104.04510}
  {arXiv:2104.04510 [hep-th]} \BibitemShut {NoStop}%
\bibitem [{\citenamefont {{K\"alin, Gregor and Porto, Rafael
  A.}}(2020{\natexlab{a}})}]{Kalin:2019rwq}%
  \BibitemOpen
  \bibfield  {author} {\bibinfo {author} {\bibnamefont {{K\"alin, Gregor and
  Porto, Rafael A.}}},\ }\bibfield  {title} {\enquote {\bibinfo {title} {{From
  Boundary Data to Bound States}},}\ }\href {\doibase 10.1007/JHEP01(2020)072}
  {\bibfield  {journal} {\bibinfo  {journal} {JHEP}\ }\textbf {\bibinfo
  {volume} {01}},\ \bibinfo {pages} {072} (\bibinfo {year}
  {2020}{\natexlab{a}})},\ \Eprint {http://arxiv.org/abs/1910.03008}
  {arXiv:1910.03008 [hep-th]} \BibitemShut {NoStop}%
\bibitem [{\citenamefont {{K\"alin, Gregor and Porto, Rafael
  A.}}(2020{\natexlab{b}})}]{Kalin:2019inp}%
  \BibitemOpen
  \bibfield  {author} {\bibinfo {author} {\bibnamefont {{K\"alin, Gregor and
  Porto, Rafael A.}}},\ }\bibfield  {title} {\enquote {\bibinfo {title} {{From
  boundary data to bound states. Part II. Scattering angle to dynamical
  invariants (with twist)}},}\ }\href {\doibase 10.1007/JHEP02(2020)120}
  {\bibfield  {journal} {\bibinfo  {journal} {JHEP}\ }\textbf {\bibinfo
  {volume} {02}},\ \bibinfo {pages} {120} (\bibinfo {year}
  {2020}{\natexlab{b}})},\ \Eprint {http://arxiv.org/abs/1911.09130}
  {arXiv:1911.09130 [hep-th]} \BibitemShut {NoStop}%
\bibitem [{\citenamefont {Bonga}\ and\ \citenamefont
  {Poisson}(2019)}]{Bonga:2018gzr}%
  \BibitemOpen
  \bibfield  {author} {\bibinfo {author} {\bibfnamefont {B\'eatrice}\
  \bibnamefont {Bonga}}\ and\ \bibinfo {author} {\bibfnamefont {Eric}\
  \bibnamefont {Poisson}},\ }\bibfield  {title} {\enquote {\bibinfo {title}
  {{Coulombic contribution to angular momentum flux in general relativity}},}\
  }\href {\doibase 10.1103/PhysRevD.99.064024} {\bibfield  {journal} {\bibinfo
  {journal} {Phys. Rev. D}\ }\textbf {\bibinfo {volume} {99}},\ \bibinfo
  {pages} {064024} (\bibinfo {year} {2019})},\ \Eprint
  {http://arxiv.org/abs/1808.01288} {arXiv:1808.01288 [gr-qc]} \BibitemShut
  {NoStop}%
\end{thebibliography}%

\end{document}